\documentclass[]{emulateapj}
\usepackage{natbib,amsmath,graphicx,longtable,amssymb,latexsym,epsf}

\pdfoutput=1

\def \nobsnir {13}
\def \nznir {11}
\def \nqpq {14}

\def \mzfg {z_{\rm fg}}
\def \zfg {$\mzfg$}
\def \mdvstat {\bar{\delta v}}
\def \dvstat {$\mdvstat$}
\def \mavgU {<U>}
\def \avgU {$\mavgU$}
\def \mxavg {<x_{\rm HI}>}
\def \xavg {$\mxavg$}
\def\smm{\sum\limits}
\def \mdelv{\Delta v_{90}}
\def \delv{$\mdelv$}
\def \kms  {\,km~s$^{-1}$}
\def \mkms  {{\rm km~s^{-1}}}
\def \mrphys {R_\perp}
\def \rphys {$\mrphys$}

\def \lya  {Ly$\alpha$}
\def \lyb  {Ly$\beta$}
\def \lyg  {Ly$\gamma$}
\def \msol      {{\rm\ M}_\odot}
\def \mnhi  {N_{\rm HI}}
\def \nhi  {$\mnhi$}
\def \mnhv  {n_{\rm H}}
\def \nhv  {$\mnhv$}
\def\psol#1#2#3#4{$\{ {\rm #1}^{#2}/{\rm #3}^{#4}\}$}
\newcommand{\cm}[1]{\, {\rm cm^{#1}}}
\def\N#1{{N({\rm #1})}}

\begin{document}

\title{Quasars Probing Quasars VIII. The Physical Properties of the Cool Circumgalactic Medium 
Surrounding $z\sim2\text{--}3$ Massive Galaxies Hosting Quasars} 

\author{
Marie Wingyee Lau\altaffilmark{1,2}, 
J. Xavier Prochaska\altaffilmark{2},
Joseph F. Hennawi\altaffilmark{3}}
\altaffiltext{1}{Email: lwymarie@ucolick.org}
\altaffiltext{2}{Department of Astronomy and Astrophysics, UCO/Lick Observatory, 
University of California, 1156 High Street, Santa Cruz, CA 95064, USA}
\altaffiltext{3}{Max-Planck-Institut f\"ur Astronomie, K\"onigstuhl 17, D-69115 Heidelberg, Germany}

\begin{abstract}
We characterize the physical properties of the cool $T\sim10^4$~K circumgalactic medium 
surrounding $z\sim$2--3 quasar host galaxies, which are predicted to evolve into present day 
massive ellipticals. Using a statistical sample of 14 quasar pairs with projected separation 
$<300\;{\rm kpc}$ and high dispersion, high S/N spectra, we find extreme kinematics with low metal 
ion lines typically spanning $\approx500\;{\rm km\;s^{-1}}$, exceeding any previously studied 
galactic population. The CGM is significantly enriched, even beyond the virial radius, with a 
median metallicity [M/H] $\approx-0.6$. The $\alpha$/Fe abundance ratio is enhanced, suggesting 
that halo gas is primarily enriched by core-collapse supernovae. The projected cool gas mass 
within the virial radius is 
estimated to be $1.9\times10^{11}\;{\rm M}_\odot\;(R_\perp/160\;{\rm kpc})^2$, accounting 
for $\approx 1/3$ of the galaxy halo baryonic budget. The ionization state of CGM gas 
increases with projected distance from the foreground quasars, contrary to expectation if the
quasar dominates the ionizing radiation flux. However, we also found peculiarities not exhibited
in the CGM of other galaxy populations. In one absorption system, we may be detecting unresolved
fluorescent Ly$\alpha$ emission, and another system shows strong \ion{N}{5} lines. Taken together
these anomalies suggest that transverse sightlines are at least in some cases possibly illuminated. 
We also discovered a peculiar case where detection of the \ion{C}{2} fine structure line implies an
electron density $>100\;{\rm cm^{-3}}$ and subparsec scale gas clumps.
\end{abstract}


\section{Introduction}
\label{sec:introduction}

The circumgalactic medium (CGM) is defined as the gaseous halo extending approximately 20--300~kpc 
from galaxies. 
It is the site of interplay between outflows from galaxies and accretion onto galaxies. 
Together these flows fuel, drain, heat, and enrich the CGM of dark matter halos. 
The impact of these processes on galaxy evolution and the enrichment of the intergalactic medium 
(IGM) remain open questions. 
%
The hot-phase CGM, or the intracluster medium, has been detected in X-rays at $z\lesssim1$ from 
the halos of massive galaxy clusters \citep[e.g.][]{KravtsovBorgani12,Sato+07,Mushotzky+96}. A 
warm-hot phase CGM traced by \ion{O}{6} is also observed for individual galaxy halos \citep{
Tumlinson+11,Peeples+14}.
The cool phase CGM, 
however, is typically too diffuse or low mass to directly detect 
far beyond the Local Group \citep[e.g.][]{Oosterloo+07,Oosterloo+05}. Instead, one is compelled to 
search in absorption using background sources whose sightlines pass close to foreground galaxies 
\citep[e.g.][]{BergeronBoisse91,Lanzetta+95,Chen+10,Prochaska+11}. 

Our understanding of the low-$z$ CGM is greatly advanced by the COS-Halos survey. The survey 
presented a statistical sample of high dispersion spectra that resolve the \ion{H}{1} Lyman series 
and many diagnostic metal ion transition lines of the CGM surrounding $L^*$ galaxies. 
The COS-Halos survey has demonstrated that, in addition to being a mediator for 
baryon recycling between galaxies and the intergalactic medium, the CGM also carries a significant 
portion of the baryonic budget of galaxy halos. Thus the CGM is essential in addressing the 
galaxy halo missing baryon problem \citep{Tumlinson+13,Werk+14}.

In the low-$z$ universe, outflows preferentially occur in star forming galaxies with appreciable 
star formation rates \citep{Rupke+05,Martin05,Rubin+14}. 
At higher redshifts, when the star formation density is higher, these galactic winds occur for 
galaxies of a wide range in mass 
\citep{Rubin+10,Steidel+10}. The cold inflows onto galaxies are also predicted to be more 
important at higher redshifts as they feed and regulate the higher star formation rates \citep{
Keres+05,Erb08}. In addition, at $z\sim2\text{--}3$, when both the universal star formation rate 
and active galactic nuclei activity peak, theories have predicted quasar driven outflows may 
couple with galaxy evolution by injecting heat into the CGM \citep{ScannapiecoOh04,Hopkins+08b}. 
Because of the difficulty in obtaining spectra of faint galaxies, previous studies on the CGM at 
high-$z$ have been largely confined to low dispersion, stacked spectra \citep{
Adelberger+05b,Steidel+10,Crighton+11}. Previous data obtained that are of sufficiently high 
quality for performing Voigt profile analysis have focused on modest samples of Lyman break 
galaxies \citep{Turner+14,Simcoe+06,Rudie+12,Crighton+13,Crighton+15}. 
 
It has been found the CGM of Lyman break galaxies exhibit strong enhancement in metal ion 
absorption out to $\approx200$~kpc 
relative to the IGM average both in the transverse direction and in the line-of-sight direction to 
the galaxies. Because Lyman break galaxies inhabit dark matter halos 
$\lesssim10^{12}\;{\rm M_\odot}$ \citep{Adelberger+05a}, the majority of them are not predicted to 
evolve into the present day, massive, red and dead, elliptical galaxies. To study mechanisms for 
maintaining or quenching massive galaxy formation, 
one would preferably perform a 
similar experiment using background sightlines that pass close to more massive galaxies. 

As a primary goal to assess the CGM of the most massive galaxies at $z\sim2\text{--}3$, we have 
performed the ``Quasars Probing Quasars'' (QPQ) survey~\footnote{http://www.qpqsurvey.org} to 
inform the processes of massive galaxy formation. Quasars are bright and easily observed at 
cosmological distances. In an ongoing survey we select closely projected quasar pairs from 
$\sim10^6$ quasars from SDSS, BOSS and 2dF surveys \citep{Bovy+11,Bovy+12}. We performed 
follow-up spectroscopy to confirm the pairs on 4~m class telescope including the 3.5~m telescope 
at Apache Point Observatory, the Mayall 4~m telescope at Kitt Peak National Observatory, the 
Multiple Mirror 6.5~m telescope, and the Calar Alto Observatory 3.5~m telescope. Our continuing 
effort to discover quasar pairs is described in \cite{Hennawi+06,Hennawi+10}. Detailed methodology 
of the QPQ experiment is described in \cite[][hereafter QPQ6]{QPQ6}. To date, we have confirmed 
$\approx700$ pairs to within 1~Mpc projected separation. In the 
series of QPQ papers, statistical inferences have generally been limited to results from low 
dispersion, stacked spectra, such as covering fractions and equivalent widths 
\citep[][hereafter QPQ2, QPQ5 and QPQ7]{QPQ2,QPQ5,QPQ7}. In the third paper of the QPQ series 
\citep[][hereafter QPQ3]{QPQ3}, we have reported a detailed analysis of the CGM surrounding one 
foreground quasar, using echelle resolution, high signal-to-noise ratio spectra. 
The previous QPQ studies, which suggested that properties of the CGM of QPQ are different from the 
LBGs, provoked questions about the physics of massive galaxy formation. 

The previous QPQ studies have yet to find definitive signatures of quasar feedback in the 
cool CGM, either. They provoked questions about the nature of quasar feedback. 
Quasars are the most luminous objects in the Universe and are thought to be powered by infall of 
matter onto a supermassive black hole at the center of a galaxy. 
On sub-kpc scales, quasars may ionize and accelerate dense clumps of material, manifested as broad 
absorption line features \citep{Weymann+91,Trump+06}. Because of the enormous energy liberated by 
quasars, quasar feedback is often invoked on larger scales as the energy source that quench star 
formation in massive galaxies \citep{Kimm+09,Lu+12} \citep[but see][]{Gabor+11}.

The degree to which a quasar can influence its host galaxy 
on kpc scales is less clear. The massive stars in the galaxy and the quasar may produce 
a significant flux of ionizing photons that would photoionize the surrounding gas on scales of at 
least tens of kpc. The line-of-sight proximity effect would suppress \ion{H}{1} absorption along 
the line-of-sight to a quasar because of the enhanced photoionization rate in its vicinity, but 
may yield a greater abundance of highly ionized gas, manifested in e.g. \ion{N}{5} and \ion{O}{6} 
\citep{Tripp+08,Simcoe+02,DOdorico+08}. 
The same may not hold for the transverse proximity effect, which is the expected suppression in 
Lyman-$\alpha$ forest opacity observed in another background sightline transverse to the quasar, 
caused by the ionizing flux of the foregound quasar \citep{GoncalvesSteidel+08}. 
This will not occur if the quasar emits anisotropically due to obscuration effects in 
AGN unification models, where the accreting black hole is centered within a torus of dust and gas 
\citep{Antonucci93}, or if the quasar emits episodically in short burst durations 
\citep{Croft04,Martini04,Hopkins+05}. In \citet[][hereafter QPQ1]{QPQ1}, QPQ2 and 
\citet[][hereafter QPQ4]{QPQ4}, we discussed the transverse proximity effect as it applies to 
optically thick absorbers in the quasar environment, and argued that most of the optically thick 
systems observed in background sightlines are likely not illuminated by the quasar.
These results are consistent with the results of 
complimentary work done on $z\sim1$ quasars using low dispersion spectra of projected 
quasar pairs \citep{Bowen+06,Farina+13,Farina+14,Johnson+15}.
They have revealed \ion{Mg}{2} absorption in the CGM along the background sightlines coincident 
with the foreground quasar's redshift, of strengths consistent with the CGM surrounding non-quasar 
host galaxies of similar masses. 

On the other hand, in the current galaxy formation paradigm, every massive galaxy has undergone a 
luminous quasar phase, making high redshift quasars the progenitors of dormant supermassive black 
holes found in the center of nearly all bulge dominated galaxies \citep{KormendyRichstone95}. 
Moreover, strong clustering of luminous quasars has been measured in various 
quasar surveys \citep[e.g.][]{Porciani+04,White+12}. The recent work \cite{White+12} found that at 
$z\approx2.4$, when star forming activity peaks, the projected autocorrelation function takes the 
form $\xi_{\rm QQ}=(r/r_0)^{-1}$, where the correlation length $r_0=8.4\;h^{-1}$~comoving~Mpc 
implies dark halo masses of $M_{\rm halo}\approx10^{12.5}\;{\rm M}_\odot$. Thus quasar hosts are 
the progenitors of the present day, massive, red and dead galaxies, whose physical processes that 
quench their star formation remain poorly constrained.

As we will frequently refer to other results from the Quasars Probing Quasars series, 
we briefly review the methodology and the results of each paper. In 
QPQ1 \citep{QPQ1} we introduced a novel technique of using projected quasar pairs to study 
the physical state of the gas in $z\sim2\text{--}3$ quasar environments.
Spectroscopic observations of the background quasar in each pair reveals the nature of the IGM 
transverse to the foreground quasar on scales of tens of kpc to several Mpc. We searched 149 
background quasar spectra for optically thick absorption in the vicinity of luminous foreground 
quasars, and found a high covering fraction to strong \ion{H}{1} absorbers. 
In QPQ2 \citep{QPQ2} we compared the statistics of this optically thick absorption in background 
sightlines near the redshift of the foreground quasars, to that observed along the line of sight 
to the foreground quasars. 
We found the clustering pattern of strong \ion{H}{1} systems around quasars to be highly 
anisotropic, and we argued that the foreground quasars anisotropically or intermittently emit 
their ionizing radiation. In QPQ3 \citep{QPQ3} we presented an echelle spectrum of a projected 
quasar pair, which resolved the velocity field and revealed the physical properties of the gas at 
$\approx100$~kpc from the foreground quasar. This gas shows extreme kinematics, an enrichment 
exceeding 1/10 solar metallicity, and has a temperature $T\approx10^4\;{\rm K}$. In QPQ4 
\citep{QPQ4} we simultaneously studied the quasar CGM in absorption and emission. We found that 
quasar powered Ly$\alpha$ fluorescence is generally absent from the absorbers observed in 
background sightlines, which implies the foreground quasars do not illuminate the surrounding gas.
In QPQ5 \citep{QPQ5} we used an enlarged sample of 74 closely projected quasar pair spectra to
study the CGM of quasar host galaxies. We reported a covering fraction
of $\approx60\%$ to optically thick, metal enriched gas within the virial radius
$\approx160$~kpc. In QPQ6 \citep{QPQ6}, with a sample enlarged to $\approx650$ 
quasar pairs, we confirmed the high incidence of optically thick gas in excess to IGM 
average extends to at least 1 physical Mpc transverse to the foreground quasars. 
The clustering found well exceeds CGM scales, which implies the gas may arise in large scale 
structures.
This enhanced \ion{H}{1} absorption measured exceeds that of other galaxy populations, consistent 
with quasars being hosted by massive dark matter halos. In QPQ7
\citep{QPQ7} we surveyed the incidence and absorption strength of metal line transitions. We 
found the cool CGM around $z\sim2$ quasars is the pinnacle amongst galaxies observed at all
epochs, regarding covering fraction and equivalent width of \ion{H}{1} Ly$\alpha$ and low ions. 

To summarize, the QPQ series suggests a massive, enriched and cool CGM surrounding massive 
galaxies at $z\sim2$, despite the presence of a luminous quasar whose ionizing flux is sufficient 
to suppress the local \ion{H}{1} Ly$\alpha$ opacity. 
Until recently, state-of-the-art cosmological \citep{Meiksin+15,Rahmati+15} and zoom-in \citep{
Fumagalli+14,FaucherGiguere+15} simulations of galaxy formation have had difficulties in 
reproducing the high covering fractions of optically thick gas seen in the QPQ work, even if one 
ignores quasar radiation. As explained in \cite{FaucherGiguere+16}, \cite{Rahmati+15} compared 
QPQ6 results with simulated halos that are typically more massive than the QPQ6 sample. We note 
the recent work \cite{FaucherGiguere+16} were able to reproduce high covering fractions of 
optically thick gas in massive halos, without invoking quasar feedback. 

Questions raised by previous QPQ studies can only be answered if we can map the kinematics, 
ionization structure, relative chemical abundance patterns, the presence or absence of a hot 
collisionally ionized phase, and the volume density and size of the absorbing clouds, using a 
statistical sample of high dispersion, high S/N spectra, which is the aim of this eighth paper in 
the series.
This manuscript is summarized as follows. In Section~\ref{sec:methodology}, we describe the 
spectral dataset that comprises QPQ8, including the criteria for selecting the subsample from the 
QPQ survey, the new observations and data reductions, and precise quasar redshift measurements. In 
Section~\ref{sec:analysis}, we present the metal ion and \ion{H}{1} absorption velocity profiles  
and their column density measurements, as well as modeling of the ionization state of the 
absorption systems. In Section~\ref{sec:results}, we constrain the kinematics, the relative 
chemical abundances, the surface density profiles of the CGM gas, the volume density and the 
linear size of the absorbers, and discuss peculiarities of the QPQ8 sample compared to 
expectations for the $z\sim2$ cool CGM surrounding quasars. In Section~\ref{sec:summary}, we 
conclude with our key findings. In the Appendix we describe our treatment of the self blended 
\ion{C}{4} doublet in our kinematic analysis, and then we present figures and tables for the 
absorption associated to each of the foreground quasars in the sample. This is a lengthy 
manuscript. The casual reader may wish to focus their attention on Section~\ref{sec:results} 
which discusses the results and implications. Throughout this manuscript we adopt a $\Lambda$CDM 
cosmology with $\Omega_M=0.26$, $\Omega_\Lambda=0.74$, and $H_0=70\;{\rm km\;s^{-1}\;Mpc^{-1}}$. 
All distances are proper unless otherwise stated. 

\section{Experimental Design}
\label{sec:methodology}

\subsection{The QPQ8 Sample}

The primary goal of this paper is to conduct a detailed absorption line analysis
of a statistical sample of CGM absorbers at proper impact parameters of 20 kpc to 300 kpc from 
$z\sim2$ quasars. The quasar pairs analyzed here are a subset of the sample studied previously for 
\ion{H}{1} Ly$\alpha$ absorption and metal line absorption (\ion{C}{2} and \ion{C}{4}) in QPQ6 and 
QPQ7 respectively. Imposed on this parent sample are selection criteria motivated by our detailed 
analysis of the quasar pair SDSSJ1204+0221 in QPQ3. We first required that the background quasar 
was observed with an echellette or echelle instrument, yielding a spectral resolution FWHM 
$\approx60\;{\rm km\;s}^{-1}$ for echellette and $\approx8\;{\rm km\;s^{-1}}$ for echelle. We 
further restricted the sample to those pairs where the average S/N at \ion{H}{1} Ly$\alpha$ 
exceeds 9.5 per 
resolution element. 
Table~\ref{tab:specs} provides a summary for the basic specs of different data sets, including 
spectral resolution, wavelength coverage, and S/N.
Spectra of such quality roughly resolve the \ion{H}{1} Lyman
series and yield metallicity and relative chemical abundance estimates
to a precision of 0.3~dex, and would allow for the construction of
photoionization models.

\begin{figure}
\includegraphics[width=3.5in]{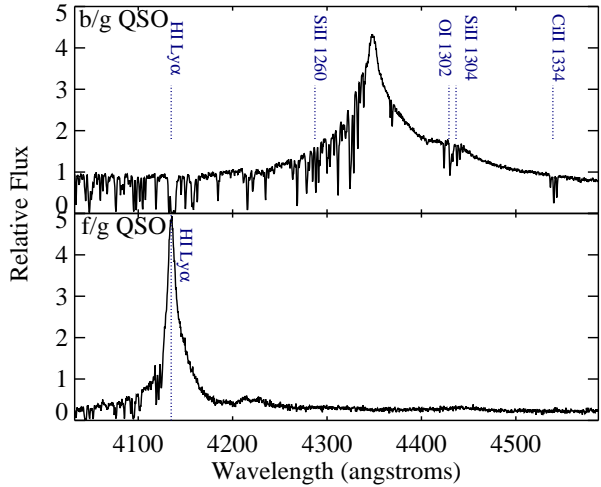} 
\caption{We show J0853-0011 as an example of a background-foreground quasar pair. Our line of
sight to the background quasar is transverse to the foreground quasar at an impact parameter
$R_\perp$, and intercepts its gaseous halo. The ionizing flux of the foreground quasar may or may
not have an opening angle $<4\pi$, and hence it may or may not suppress the cool CGM in the
transverse direction. In the background quasar spectrum we see strong Ly$\alpha$ and metal ion
absorption coincident with the foreground quasar's redshift.
}
\label{fig:qsopair_ex}
\end{figure}

Together, these criteria imply a cut on the apparent magnitude of the background quasar of 
approximately 19.5 mag. 
We limited the selection to close quasar pairs with projected physical distance 
$R_{\perp}<300$~kpc at the redshift of the foreground quasars to isolate the CGM. We imposed a cut 
on velocity difference between the redshifts of the two quasars $>3000\;{\rm km\;s}^{-1}$, to avoid 
ambiguity in distinguishing absorption intrinsic to the background quasar from absorption 
associated with the foreground quasar. Finally, we required that the \ion{C}{2}~1334 
transition at the foreground quasar's redshift lie redward of the background quasar's Ly$\alpha$ 
forest, i.e. $(1+z_{\rm fg})1334.5323\;{\rm\AA}>(1+z_{\rm  bg})1215.6701\;{\rm\AA}$. By placing 
this metal transition outside of the Ly$\alpha$ forest, we ensure access to a suite of rest frame 
far UV diagnostics free from confusion with intergalactic \lya\ absorption. This requirement 
corresponds to a relative velocity separation $\lesssim20000\text{--}30000\;{\rm km\;s}^{-1}$. 
Figure~\ref{fig:qsopair_ex} presents the spectra of J0853-0011 as an example of such 
background-foreground quasar pair. 

The final QPQ8 sample comprises \nqpq\ pairs. The observation journals, details related to data 
reduction and calibration of the 1D spectra are provided in QPQ6 and QPQ7. The spectra within the 
\lya\ forest were previously continuum-normalized with an automated principle component analysis 
algorithm \citep{Lee+12}. To enable the search for weak absorption lines in our higher S/N QPQ8 
sample, we here manually refitted the continuua. We generated a high order spline function that 
traces the obvious undulations and emission features of the background quasar. This analysis made 
use of the routine X\_CONTINUUM, distributed as part of the XIDL software package\footnote{
http://www.ucolick.org/\~{}xavier/xidl}.

Table~\ref{tab:sample} lists the QPQ8 sample and summarizes key properties of the pairs and 
spectral data set. In cases where multiple spectra taken by different instruments covered the same 
transition, we gave preference to the higher resolution spectra provided sufficient S/N.  
In terms of the spectral and photometric properties of the foreground quasar, the single 
sightline studied in QPQ3, J1204+0221, is unremarkable. J1204+0221FG has a systemic redshift of 
$z=2.4358$, while the median of the QPQ8 sample is $z=2.4$. J1204+0221FG has a bolometric 
luminosity of $10^{46.2}\;{\rm erg\;s^{-1}}$. Thus, J1204+0221 is representative of the larger 
statistical sample, and it is fair for us to rely on QPQ3 to form the selection criteria for QPQ8. 

\begin{deluxetable*}{cccccccccccc}
\tablewidth{0pc}
\tablecaption{QPQ8 Sample Summary\label{tab:sample}}
\tabletypesize{\scriptsize}
\setlength{\tabcolsep}{0in}
\tablehead{\colhead{Name} &
\colhead{f/g Quasar} & 
\colhead{$z_{\rm fg}$} &
\colhead{$\log\;L_{912}^a$} &
\colhead{$\log\;L_{\rm bol}^b$} &
\colhead{$g_{\rm UV}^c$} & 
\colhead{b/g Quasar} &
\colhead{$z_{\rm bg}$} &  
\colhead{$R_{\rm\perp}$} & 
\colhead{$\theta^d$} \\
 & & & & & & & & (kpc) & 
}
\startdata
J0225+0048 & J022517.68+004821.9 & 2.7265 & 30.33 & 46.34 & 535 &
J022519.50+004823.7 & 2.820 & 
226 & 27.4 \\
J0341+0000 & J034138.15+000002.9 & 2.1233 & 29.92 & 46.07 & 274 & 
J034139.19-000012.7 & 2.243 & 
190 & 22.1 \\
J0409-0411 & J040955.87-041126.9 & 1.7155 & 30.34 & 46.42 & 516 &
J040954.21-041137.1 & 2.000 & 
235 & 26.9 \\
J0853-0011 & J085358.36-001108.0 & 2.4014 & 29.82 & 45.86 & 645 &
J085357.49-001106.1 & 2.577 & 
112 & 13.2 \\
J0932+0925 & J093226.34+092526.1 & 2.4170 & 30.27 & 46.31 & 402 & 
J093225.60+092500.2 & 2.602 & 
238 & 28.1 \\
J1026+4614 & J102618.80+461445.2 & 3.3401 & 30.79 & 46.79 & 1119 & 
J102616.11+461420.8 & 3.421 & 
288 & 37.1 \\
J1038+5027 & J103857.37+502707.9 & 3.1322 & 30.90 & 46.90 & 2069 & 
J103900.01+502652.8 & 3.237 & 
233 & 29.4 \\
J1144+0959 & J114435.53+095921.6 & 2.9731 & 30.63 & 46.55 & 1639 & 
J114436.65+095904.9 & 3.160 & 
189 & 23.5 \\
J1145+0322 & J114546.54+032236.7 & 1.7652 & 29.93 & 46.05 & 559 &
J114546.21+032251.9 & 2.011 & 
139 & 15.9 \\
J1204+0221 & J120417.46+022104.7 & 2.4358 & 30.17 & 46.19 & 1424 &
J120416.68+022110.9 & 2.532 & 
112 & 13.2 \\
J1420+1603 & J142054.42+160333.3 & 2.0197 & 30.58 & 46.54 & 4298 & 
J142054.92+160342.9 & 2.057 & 
104 & 12.0 \\
J1427-0121 & J142758.88-012130.3 & 2.2736 & 30.63 & 46.63 & 17964 &
J142758.73-012136.1 & 2.354 & 
53 & 6.2 \\
J1553+1921 & J155325.60+192140.9 & 2.0098 & 29.70 & 45.81 & 3056 &
J155325.88+192137.6 & 2.098 & 
44 & 5.1 \\
J1627+4605 & J162738.63+460538.3 & 3.8137 & 30.66 & 46.69 & 1222 &
J162737.24+460609.3 & 4.110 & 
253 & 34.1 \\
\enddata
\tablenotetext{a}{The specific luminosity of the foreground quasar at the Lyman limit 
912~\AA, in unit of $\log\;{\rm erg\;s^{-1}\;Hz^{-1}}$}
\tablenotetext{b}{The bolometric luminosity of the foreground quasar, in unit of 
$\log\;{\rm erg\;s^{-1}\;Hz^{-1}}$.}
\tablenotetext{c}{The enhancement in flux relative to the extragalactic UV background, 
assuming the foreground quasar emits isotropically and a distance equal to the impact parameter 
$R_\perp$.}
\tablenotetext{d}{Angular separation between foreground and background quasar, in arcseconds.}
\end{deluxetable*}

\begin{deluxetable*}{cccccccc}
\tablewidth{0pc}
\tablecaption{QPQ8 Data Set Specs\label{tab:specs}}
\tabletypesize{\scriptsize}
\setlength{\tabcolsep}{0in}
\tablehead{\colhead{Name} &
\colhead{b/g Quasar Instrument} &
\colhead{Resolution in FWHM} &
\colhead{Wavelength Coverage} &
\colhead{S/N per \AA\ at Ly$\alpha$ at $z_{\rm fg}$} \\
 & & (${\rm km\;s^{-1}}$) & (\AA) 
}
\startdata
J0225+0048 & ESI, GMOS & 60, 125 & 3993-10556 & 76 \\
0341+0000  & MagE & 50 & 3044-10254 & 33 \\
0409-0411  & MagE & 62 & 3044-9459 & 15 \\
0853-0011  & MagE & 62 & 3042-10285 & 121 \\
0932+0925  & MagE & 51 & 3041-10284 & 74 \\
J1026+4614 & ESI & 49 & 3994-10197 & 184 \\
J1038+5027 & ESI & 48 & 3994-10197 & 70 \\
1144+0959  & MIKE & 9 & 3307-9167 & 205 \\
1145+0322  & MagE & 51 & 3042-10285 & 18 \\
1204+0221  & HIRES & 8 & 3448-6422 & 162 \\
J1420+1603 & MagE & 51 & 3042-10285 & 74 \\
J1427-0121 & MIKE, MagE & 8, 50 & 3309-9169 & 157 \\
1553+1921  & MagE & 51 & 3042-10285 & 33 \\
J1627+4605 & ESI & 45 & 3989-10198 & 98 \\
\enddata
\end{deluxetable*}
\begin{deluxetable*}{lccccccc}
\tablewidth{0pc}
\tablecaption{Journal of Near Infrared Observations\label{tab:nearIR}}
\tabletypesize{\scriptsize}
\setlength{\tabcolsep}{0in}
\tablehead{\colhead{Quasar} & \colhead{Observatory} 
& \colhead{Instrument} 
& \colhead{Date in UT} & \colhead{Exposure Time} 
& \colhead{Line} 
& \colhead{$z_{\rm em}$} & \colhead{$\sigma(z_{\rm em})$} 
\\ 
 &&&&(s)&&&(km s$^{-1}$) 
} 
 \startdata 
J0341+0000FG & Keck & MOSFIRE & 2014 Oct 1 & 960& H$\beta$& 2.1233 & 272$^a$\\ 
J0409-0411FG & Keck & MOSFIRE & 2014 Oct 1 & 960& H$\alpha$& 1.7155 & 272$^a$\\ 
J0853-0011FG & Keck & NIRSPEC & 2010 Jan 29 & 4800& [OIII]& 2.4014 & 44\\ 
J0932+0925FG & VLT & XSHOOTER & 2011 Apr 4 & 3600& [OIII]& 2.4170 & 44\\ 
J1038+5027FG & Gemini & NIRI & 2006 May 9 & 4800& [OIII]& 3.1323 & 44\\ 
J1144+0959FG & Keck & NIRSPEC & 2009 Jan 7 & 3000& [OIII]& 2.9731 & 44\\ 
J1204+0221FG & Gemini & GNIRS & 2006 Mar 27 & 5440& [OIII]& 2.4358 & 44\\ 
J1420+1603FG & VLT & XSHOOTER & 2011 Apr 28 & 2400& H$\alpha$& 2.0197 & 272$^a$\\ 
J1427-0121FG & Gemini & GNIRS & 2006 Mar 12 & 7200& [OIII]& 2.2736 & 44\\ 
J1553+1921FG & VLT & XSHOOTER & 2007 Jul 17 & 2400& [OIII]& 2.0098 & 44\\ 
J1627+4605FG & Gemini & NIRI & 2007 May 29 & 14400& [OIII]& 3.8137 & 44\\ 
\enddata 
\tablenotetext{a}{We quantified the uncertainties for H$\alpha$ emission redshift and H$\beta$ emission redshift to be $300\;{\rm km\;s^{-1}}$ and $392\;{\rm km\;s^{-1}}$ respectively. Hence when \ion{Mg}{2} is detected, we adopted the \ion{Mg}{2} emission redshift instead of the near IR redshift, for its smaller uncertainty of $272\;{\rm km\;s^{-1}}$.} 
\end{deluxetable*} 

\subsection{Quasar Redshifts}

\begin{figure*}
\includegraphics[width=6.5in]{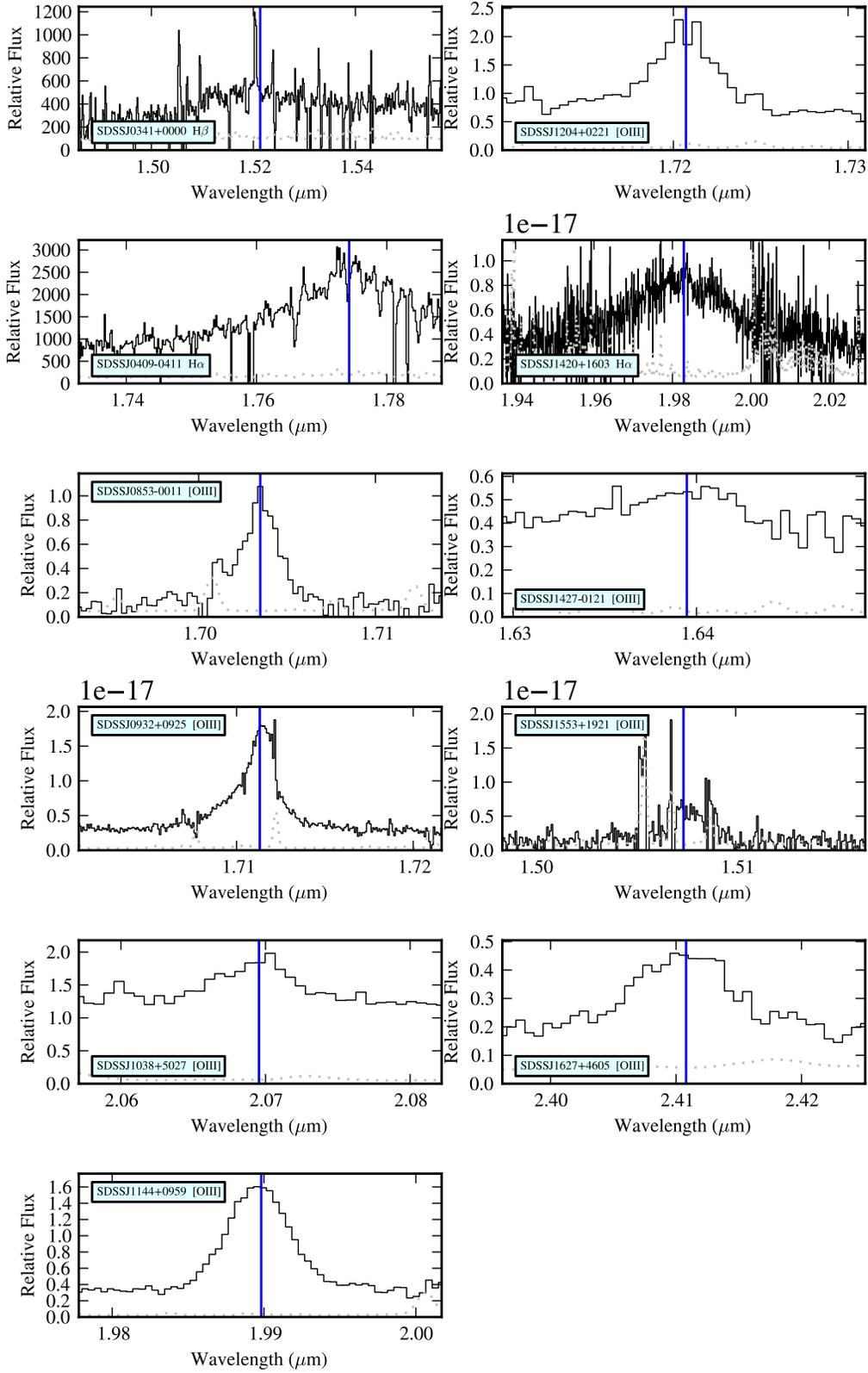} 
\caption{Near IR spectra of the foreground quasars for precisely determining the systemic
redshift. The gray dotted curve is the 1$\sigma$ uncertainty. We applied our custom centroiding  
algorithm on  the [\ion{O}{3}] $\lambda 5007$ emission line whenever it is present. When 
[\ion{O}{3}] is not detectable we used ${\rm H\alpha}$ or ${\rm H\beta}$. 
If no near IR data exists for the foreground quasar or no emission lines are found, we adopted the
QPQ6 systemic redshift which is obtained by fitting the full optical spectra.
}
\label{fig:nearIR}
\end{figure*}

The quasar emission redshifts $z_{\rm bg}$ were first taken directly from QPQ6. In QPQ6, the 
background quasar redshifts were taken from SDSS, while for the foreground quasars a custom 
line-centering algorithm was adopted to centroid one or more rest frame far UV emission lines 
\ion{Mg}{2}, [\ion{C}{3}], \ion{Si}{4} and \ion{C}{4}. 
We iterated a flux-weighted, line-centering scheme until the centroid converged.
Typical uncertainties range from $270\;{\rm km\;s^{-1}}$ to $790\;{\rm km\;s^{-1}}$. This 
precision is sufficient to define the QPQ8 sample.
To establish a robust association of absorption to the foreground quasar, however, we desire the 
most precise assessment of its redshift $z_{\rm fg}$. Ideally, the uncertainties should be less 
than the peculiar motions of gas within the massive halos hosting quasars. This requires more 
precise measurements for $z_{\rm fg}$ than the QPQ6 results. 

Our approach is to measure \zfg\ from rest frame optical narrow forbidden emission lines such as 
[\ion{O}{3}]~5007, or the \ion{H}{1} Balmer series. For $z\gtrsim2$ quasars, these lines are 
shifted into the near infrared. These lines have smaller systematic uncertainties of 
$400\;{\rm km\;s^{-1}}$ or lower. [\ion{O}{3}] has an average blueshift of $27\;{\rm km\;s^{-1}}$ 
and a dispersion of $44\;{\rm km\;s^{-1}}$ about this value \citep{Boroson05,Richards+02}. 
To account for this average shift due to the blue wing of the [\ion{O}{3}] line, we added 
$27\;{\rm km\;s^{-1}}$ to the vacuum rest wavelength of 5008.24~\AA\ when computing the redshift of 
the line. 
We have quantified the H$\alpha$ and H$\beta$ precision for an independent SDSS dataset to be 
$300\;{\rm km\;s^{-1}}$ and $392\;{\rm km\;s^{-1}}$ respectively about the systemic. We note that, 
in order to cover H$\alpha$ in the optical, these SDSS quasars need to be at low redshift and 
hence are often not as luminous as the quasars in our QPQ sample. For fainter quasars, H$\alpha$ 
is more peaked because of the narrow line region, and hence the redshifts will be more accurate 
than for a luminous sample. We are not using these redshift uncertainties in a very quantitative 
manner, however. 
We observed \nobsnir\ of the \nqpq\ 
foreground quasars using GNIRS \citep{GNIRS} and NIRI \citep{NIRI} on the Gemini North telescope, 
NIRSPEC \citep{NIRSPEC} on the Keck II telescope, and/or X-SHOOTER \citep{XSHOOTER11} on the 
Very Large Telescope. Table~\ref{tab:nearIR} provides a journal of the near IR observations. When 
\ion{Mg}{2}~2800 is detected, the \ion{Mg}{2} emission redshift is preferred over the H$\alpha$ or 
H$\beta $ emission redshifts for its smaller uncertainty of $272\;{\rm km\;s^{-1}}$. We have taken 
into account the median redshift of $97\;{\rm km\;s^{-1}}$ of \ion{Mg}{2} from \ion{O}{3} 
\citep{Richards+02}. 

The XSHOOTER spectra were reduced with a custom software package developed and 
kindly provided by George Becker, which includes nod sky subtraction on the slit and telluric 
corrections based on the European Southern Observatory SkyCalc sky model calculator 
\citep{Noll+12,Jones+13,Moehler+14}. 
Flat fielding of the detector was performed using dome flat exposures, and wavelength calibration 
of the near IR arm used night sky emission features. Sky subtraction implements a two dimensional 
b-spline algorithm and extraction was performed optimally. Significant residuals do persist at 
lines of the brightest sky emission. The remaining data was processed with algorithms in the 
LowRedux\footnote{http://www.ucolick.org/$\sim$xavier/LowRedux/index.html} package developed 
primarily by one of us, JFH. The processes are similar to those for XSHOOTER. The principal 
difference is sky subtraction where the LowRedux algorithms first perform image subtraction of 
dithered (AB) exposures before fitting a b-spline to sky residuals. For all spectra, fluxing was 
performed with a telluric standard observed close in time and position on the sky to the 
scientific target.

In the following analysis, we omit the near IR observations for (1) J1145+0322FG, whose 
H$\beta$~4862 and [\ion{O}{3}]~5007 lines fall outside the transmitting infrared atmospheric 
windows. The emission line analyzed for redshift was \ion{Mg}{2}, as described in QPQ6; (2) J
0225+0048FG, whose H$\beta$~4862, [\ion{O}{3}]~5007 and H$\alpha$ are all redshifted to 
wavelengths of low atmospheric transmission. The emission lines analyzed for redshift were 
\ion{Si}{4}, \ion{C}{4} and [\ion{C}{3}]. Although it has \ion{Mg}{2} at observable wavelengths, 
it falls outside of our spectral coverage; (3) J1026+4614FG, whose H$\beta$~4862 and 
[\ion{O}{3}]~5007 emission lines are weak and yield redshift estimates that are inconsistent. 
The emission lines analyzed for redshift were \ion{Si}{4} and [\ion{C}{3}]. The quasar too has 
\ion{Mg}{2} at observable wavelengths but outside of coverage. 
For these sources, we adopted the \zfg\ measurements from QPQ6.
Figure~\ref{fig:nearIR} presents the near IR spectra of the \nznir\ foreground quasars with 
a near IR redshift measurement. We have analyzed these data with a custom algorithm that centroids 
the emission lines and generates a best estimate for \zfg. The results and adopted uncertainties 
are listed in Table~\ref{tab:nearIR}.

\section{Analysis} 
\label{sec:analysis} 

In this section, we present column density measurements for the gas associated to the foreground 
quasars in the QPQ8 sample.  We begin with an analysis of associated metal line absorption, 
proceed to the \ion{H}{1} analysis, and then describe ionization modeling of the systems. Details 
for the individual systems are provided in the Appendix. Here we describe the methodology and 
present representative examples. 

\subsection{Metals Absorption in Multiple Ionization States}
\label{sec:ions}

We performed a search for associated metal line absorption within 1000~\kms\ of \zfg. This search 
window allows for large peculiar motions in the gas which may be common in the environments of 
these massive galaxies \citep[e.g. QPQ3,][]{Johnson+15}. 
At the median $z_{\rm fg}=2.4$ of our sample, this line-of-sight velocity window corresponds to 
8~Mpc physical. 
We emphasize that strong metal lines systems of equivalent width $>0.3\;{\rm \AA}$ are rare in the 
intervening IGM along quasar sightlines and are dominated by \ion{C}{4} absorption.
According to the calculations in QPQ7, the random incidence of \ion{C}{4} absorbers with 
equivalent width $>0.3\;{\rm \AA}$ is 2.1 per unit redshift.
QPQ6 measured a drop in \ion{H}{1} equivalent width and QPQ7 measured a drop in \ion{C}{2} 
and \ion{C}{4} equivalent width with impact parameter, especially at $>200$~kpc. They suggest the 
cool CGM gas is mostly contained in proximity to central galaxy and argues against a large 
contribution from Mpc scales.       
The positive detection of any such metals within 1000~\kms\ of \zfg\ is therefore 
unlikely to arise from gas at cosmological separations (see QPQ7 for further details). 
In \cite{Johnson+15}, it was suggested that the CGM of neighboring galaxies of other host halos 
could contribute a significant portion of the \ion{Mg}{2} absorption they observed around quasars. 
However, they plotted covering fractions due to all absorbers that fall within 
$\pm1500\;{\rm km\;s^{-1}}$, which include contributions from galaxies that are actually within 
300 kpc and hence are in the same host halo. 

\begin{figure}
\includegraphics[width=3.5in]{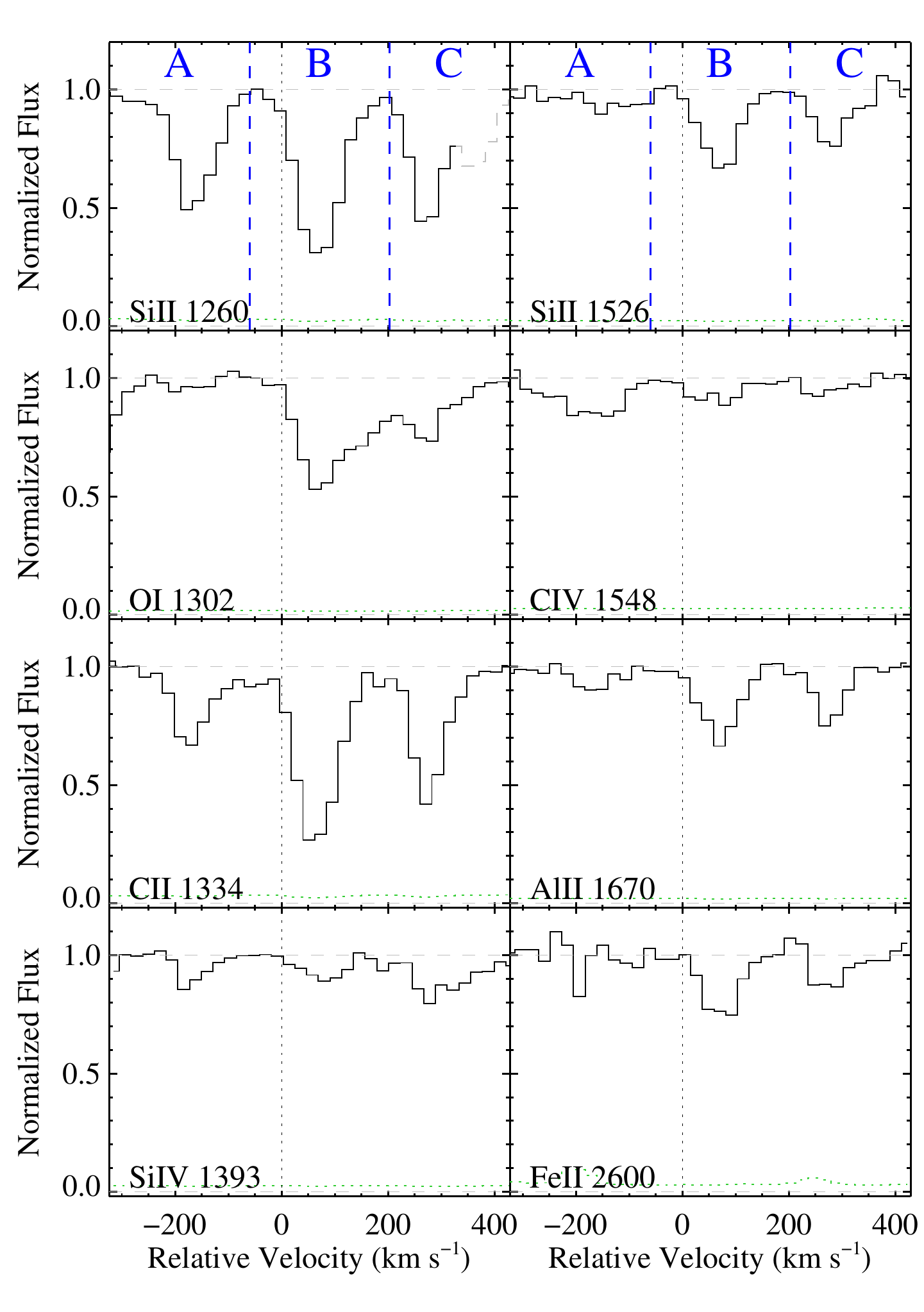} 
\caption{A subset of metal line transitions from the absorption associated to J0853-0011FG. The 
velocities are relative to the measured $z_{\rm fg}=2.4014$. Absorption well separated in distinct 
velocity intervals are designated as subsystems A, B and C as denoted by the vertical dashed lines 
in the top subpanels. Absorption that is presumed unrelated to the foreground quasar, e.g. 
Ly$\alpha$ features from unrelated redshifts, are presented as dashed, gray lines. The gas shows
multiple ionization states that roughly trace each other and span $\approx650\;{\rm km\;s^{-1}}$. 
The green histogram shows the $1\sigma$ noise in the normalized flux.
}
\label{fig:ex_velp}
\end{figure}

An absorption line system bearing heavy elements was detected in 12 out of \nqpq\ cases, an 
incidence that greatly exceeds random expectation. Furthermore, 8 of these systems exhibit low ion 
transitions (e.g. \ion{C}{2}~1334) which occurs even more rarely in random sightlines. 
For complex profiles, metal ion absorption well separated into distinct velocity intervals  
are grouped into subsystems A, B, C, etc, and are separately analyzed in what follows.
Figure~\ref{fig:ex_velp} shows the velocity profiles for the absorption associated to J0853-0011. 
The gas spans approximately 650~\kms, which we divided into three subsystems. Although multiple 
ionization states, e.g. C$^+$ and C$^{3+}$, tend to have optical depth ratios that vary across 
velocity, they roughly trace one another in velocity structure.

For each subsystem associated to each foreground quasar, we measured the ionic column densities 
from commonly detected metal line transitions in CGM gas \citep[e.g.][]{Werk+13}. For these 
measurements, we used the apparent optical depth method \citep[AODM;][]{SavageSembach91} which 
yields precise values for high spectral resolution observations. 
For non-detections, we report $3\sigma$ upper limits on column densities, obtained by integrating  
over a velocity window that encompasses most of the apparent optical depths of a subsystem. 

Line saturation, however, may affect the echellette spectra ($R_{\rm FWHM} \approx 6,000$) and we 
conservatively report lower limits for cases where the minimum normalized flux is below 
$\approx 0.4$. 
Metal absorption components are typically narrow $<10\;{\rm km\;s^{-1}}$, which is evident in 
the three echelle spectra included in this study. In the Appendix we show that, where the minimum 
normalized flux is greater than 0.4, echellette quality spectra are sufficient for accurate column 
density measurements. Hence with our criterion on the normalized flux, using echellette quality 
spectra does not introduce systematic biases.

We verified that the velocity intervals of the subsystems are chosen in a way that there is 
little apparent variation in the ionic ratios. 
The scientific results that involve assessing ionization do not sensitively depend on how many 
subsystems have been chosen, and we quote weighted average values of ionization parameter, chemical 
abundances, etc for each quasar-pair absorption system in what follows. For one absorption system, 
J1427-0121, for which an echelle spectrum was obtained, we fitted Voigt profiles to the 
unsaturated metal absorption components as discussed in the Appendix. As expected for resolved 
lines, the total ion column densities recovered for each subsystem agree with measurements from 
integration the apparent optical depths. In cases where there is no substantial ionization 
structure (as required for our subsystems), breaking down the absorption profile into individual 
components has limited scientific value. Uncertainties associated with ionization modeling and 
line saturation exceed the benefit gained from component-by-component fitting. Hence 
this study does not include a component-by-component analysis. 

A brief description of 
each system, figures for the velocity plots, and tables of all line measurements are presented in 
the Appendix. Table~\ref{tab:clm_summ} summarizes the integrated measurements for the QPQ8 sample.

\subsection{Lyman Series Voigt Profile Modeling}
\label{sec:HI}

\begin{figure}
\includegraphics[width=3.5in]{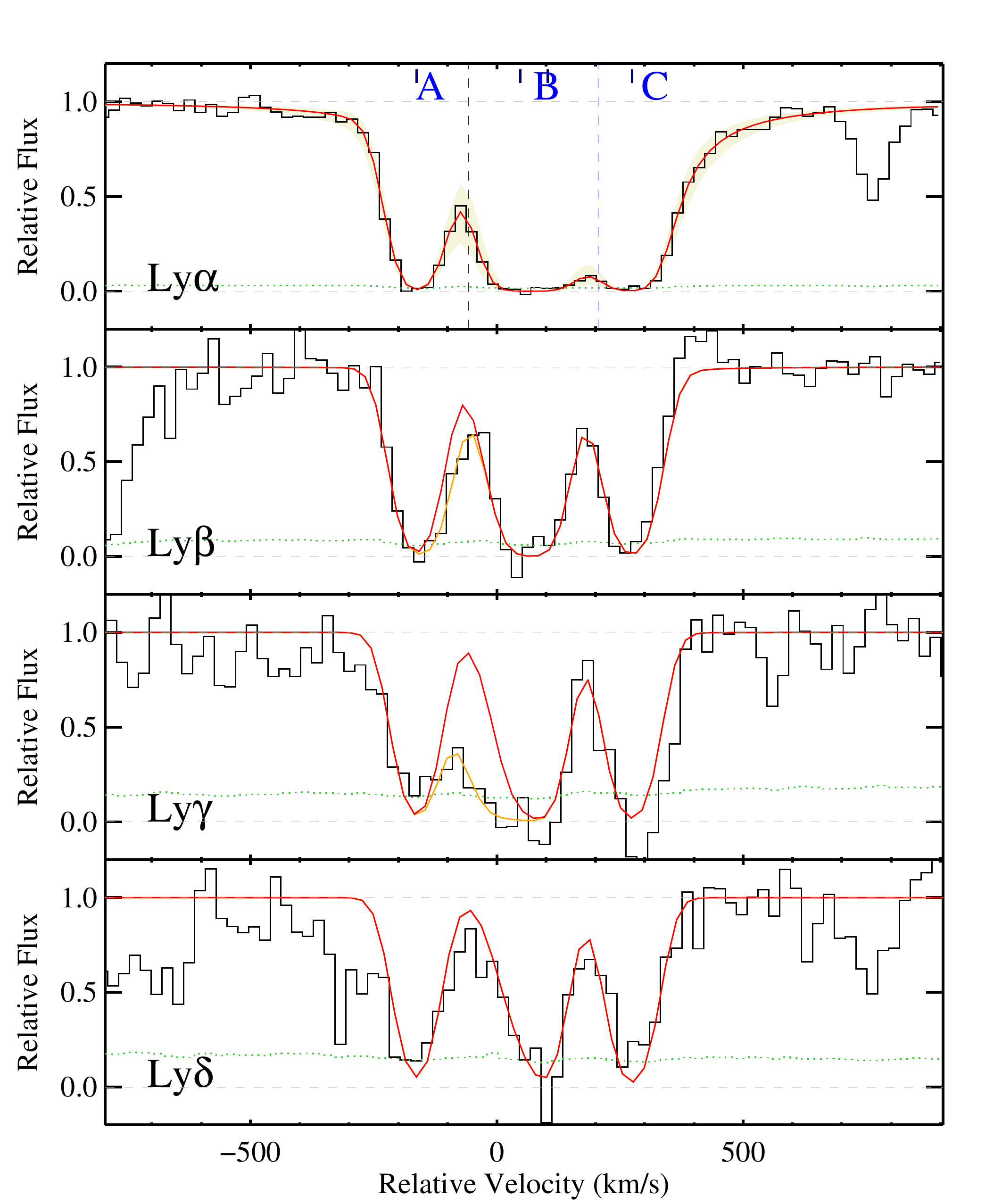} 
\caption{Example fit to \ion{H}{1} Lyman series absorption profiles for J0853-0011. The black
histograms show the Lyman series identified in the J0853-0011BG spectrum at velocities consistent
with J0853-0011FG. The green dotted curves show the $1\sigma$ noise in the normalized flux. The
relative velocity at $0\;{\rm km\;s}^{-1}$ corresponds to the redshift of J0853-0011FG. We
performed $\chi^2$ minimization Voigt profile modeling to assess the \ion{H}{1} column density.
Specifically we introduced \ion{H}{1} components centered at relative velocities traced by the
peak optical depths of the associated set of metal ions. Note in subsystem B there are two
components. The navy ticks in the top subpanel mark the centroid redshifts for components traced
by low ions. Figures of fits to Lyman series absorption for other QPQ8 systems, presented in the
Appendix, will also show ticks that mark the centroids for components traced by high ions and
ticks that mark \ion{H}{1} components not associated with metal ions.
Additional \ion{H}{1} components introduced to model Ly$\alpha$ forest blending are omitted in the
tickmarks. The red curve is the summation of all \ion{H}{1} components associated with
J0853-0011FG, and the beige shades mark the estimated $\pm1\sigma$ errors in \ion{H}{1} column
densities. The orange curve is the summation of all \ion{H}{1} components associated with
J0853-0011FG and Ly$\alpha$ forest contamination.
}
\label{fig:ex_HI}
\end{figure}

For each system we measured the \ion{H}{1} column density of the gas by modeling its Lyman 
series absorption. We have echellette or echelle resolution coverage of \lya\ for every system.
We also included higher order Lyman series lines where the S/N estimated from 
the quasar continuum exceeds 5 per pixel. 
When the echellette or echelle data do not cover blue 
enough cover for these transitions, we also used lower dispersion data when available. 
For many of the systems the Lyman series lines are all 
saturated yet do not exhibit damping wings, i.e. the lines fall on the saturated portion of the 
curve of growth. A precise measurement of \nhi\ is therefore difficult and is degenerate with 
assumptions made about the component structure and Doppler $b$-values one adopts in the modeling.
Our general approach is to first set conservative 
bounds on \nhi\ based on a wide range of allowed Doppler parameters from $b=15\;{\rm km\;s^{-1}}$ 
to $b=60\;{\rm km\;s^{-1}}$. Such estimates are dependent, however, on the number of 
assumed components and their relative velocities.  

We then performed Voigt profile modeling of the data by $\chi^2$ minimization using the Absorption 
LIne Software \citep[ALIS; ][]{Cooke+14}. ALIS uses the MPFIT package \citep{Markwardt09} which
employs a Levenberg--Marquardt least squares minimization algorithm, where the difference between 
the data and the model is weighted by the error in the data. 
We reduced the degrees of freedom by fixing the relative velocities of unblended \ion{H}{1} 
components or \ion{H}{1} components associated with metal absorption, preferably low ions, that 
show well matching velocity structures. The Doppler $b$ parameter is allowed to vary freely 
between the values of $15\;{\rm km\;s^{-1}}$ to $60\;{\rm km\;s^{-1}}$, which are typical of the 
high-$z$ Ly$\alpha$ forest \citep{KirkmanTytler97}. We adopted approximately the central 
$N_{\rm HI}$ values in the range of solutions allowed by ALIS that have nearly equivalent 
$\chi^2$. 
Since possible damping wings of a Voigt profile may as well be explained by a weak component that 
is not introduced to the model, ALIS best-fit solutions tend to give biased high values. For this 
reason, our adopted $N_{\rm HI}$ and the ALIS best-fit $N_{\rm HI}$ values need not agree. 

As an example, Figure~\ref{fig:ex_HI} shows the Lyman series data and our fit for the J0853-0011 
absorption system. For each subsystem we identified \ion{H}{1} components at the velocities 
corresponding to the peak optical depths of its associated set of metal lines. The velocity 
centroids of \ion{H}{1} components traced by low ions, e.g. \ion{C}{2} and \ion{O}{1}, are marked 
by navy ticks. In figures of Lyman series fits for other QPQ8 systems, presented in the Appendix, 
we also mark the centroids of \ion{H}{1} components traced by high ions, primarily \ion{C}{4}, and 
centroids of \ion{H}{1} components that have no associated metal ions. The red curve is 
the summation of all \ion{H}{1} components associated with the 
foreground quasar, while the orange curve is the summation of all \ion{H}{1} components 
associated with the foreground quasar and additional \ion{H}{1} components introduced to model 
Ly$\alpha$ forest contamination.

From trial-and-error of introducing additional components in the modeling, we found the possible 
presence of unresolved components introduces a systematic error of several tenths dex in the 
$N_{\rm HI}$ values. We also performed the same Lyman series modeling using a MagE spectrum of 
J1427-0121 and compared to the modeling results using the MIKE spectrum. We found that the 
best-fit $N_{\rm HI}$ values can be recovered within $1\sigma$ error, and hence echellette 
resolution spectra does not introduce significantly larger errors compared to echelle spectra. 
Moreover we estimated a systematic error of $\pm0.2$~dex in the 
$N_{\rm HI}$ values due to quasar continuum placement. For \ion{H}{1} components that we did not 
fix the velocities, due to line blending ALIS reported unphysically large uncertainties in 
$N_{\rm HI}$ values when the uncertainties in the relative velocities of the components are 
large. 
The Voigt model parameters as well as the total 
$N_{\rm HI}$ of each absorption system and subsystems are tabulated in Table~\ref{tab:NHI} in the 
Appendix. A detailed description of the model for each quasar pair is also presented in the 
Appendix.

\subsection{Ionization Modeling: the Ionization Parameter $U$}
\label{sec:cldy}

As summarized in the previous subsections and presented in Figures~\ref{fig:ex_velp} and 
\ref{fig:ex_HI}, the CGM of galaxies hosting quasars exhibit multiple ionization states (e.g. 
O$^0$, Si$^+$, S$^{2+}$, Si$^{3+}$), and the total \ion{H}{1} column densities are generally less 
than $10^{19} \cm{-2}$.
Both of these observations imply a predominantly ionized 
gas. Even if the quasar's ionizing flux is not directly impinging on the gas (as we have 
previously argued in the QPQ series), the extragalactic ultraviolet background (EUVB) may 
photoionize the medium, resulting in neutral fractions $x_{\rm HI} \ll 1$ 
(e.g. QPQ3). Therefore, we generated photoionization models for each of the systems 
exhibiting metal line absorption, to estimate $x_{\rm HI}$ and other physical properties of the 
gas. In the process, if we can make assumptions about the gas volume density or estimate it 
through fine structure excitation lines, we may 
compare the intensity of the ionizing radiation field that produces the observed ionic ratios with 
the predicted flux of the quasar at the impact parameter of the sightline. This way we may test 
the hypothesis of whether the nearby foreground quasars are shining on the gas. 

There are two primary processes that produce an ionized gas: collisional ionization and 
photoionization. 
If we assume the cooling function takes the form in \cite{SutherlandDopita93}, 
the cooling time $t_{\rm cool}\lesssim10^4$~yr is short for any reasonable density. 
A model where 
collisional ionization is the primary mechanism producing the observed ionic ratios would require 
a heat source to maintain the gas temperature in equilibrium. 
We therefore assumed photoionization is the dominant mechanism for setting the ionization structure 
of the gas, and also the dominant source of heat. 
Furthermore we note that the analysis in QPQ3 shows collisional ionization equilibrium solutions 
with $T\sim10^4$~K give very similar results to the photoionization models presented here.  

We calculated the ionization state of plane parallel slabs with version 10.00 of the Cloudy 
software package last described by \cite{Ferland+13}. The inputs to Cloudy are the $N_{\rm HI}$ 
of a subsystem, as modeled in the previous subsection, the total volume density (neutral plus 
ionized) $n_{\rm H}$ which we fix at a constant $0.1\;{\rm cm}^{-3}$, and an initial assumed 
metallicity [M/H] $=-0.5$~dex. We then varied the ionization parameter $U\equiv\Phi/n_{\rm H}c$ 
where $\Phi$ is the flux of ionizing photons having $h\nu\ge1$~Ryd, and iterated on [M/H] and $U$ 
until the results converge. The results are 
largely insensitive to the choice of volume density as it is nearly homologous with $U$, but they 
do vary with metallicity because this affects the cooling rate of the gas. It is convenient to fix 
$n_{\rm H}$ and vary $\Phi$, so that there is only one degree of freedom in the output which is 
the ionization parameter. For these calculations, 
we assumed the spectral shape of the extragalactic UV background field follows that computed by 
\cite{HaardtMadau12} and varied the amplitude. 
For radiation fields of $z\sim2$ quasars at a few 
Ryd, the EUVB is very similar to a power law spectrum 
$f_\nu\propto\nu^{-1.57}$. As the input $U$ parameter changes, the Cloudy algorithm varies the 
number of gas slabs to maintain a constant $N_{\rm HI}$ at the input value. For optically thick 
systems the results are sensitive to the assumed $N_{\rm HI}$. Larger $N_{\rm HI}$ values imply 
more self shielding of the inner regions which in turn demand a more intense radiation field to 
explain the observed ionization states.

For each system, we considered a series of ionic ratios and a wide range of ionization parameters. 
Our analysis focused on multiple ionization states of individual elements such as C$^+$/C$^{3+}$ 
and Si$^+$/Si$^{3+}$, to avoid uncertainties related to intrinsic abundance variations. We also 
considered ratios of low to high ion species of different elements for constraining $U$ assuming 
solar relative abundances. We adopted a correspondingly higher uncertainty for such constraints. 

\begin{figure}
\includegraphics[width=3.5in]{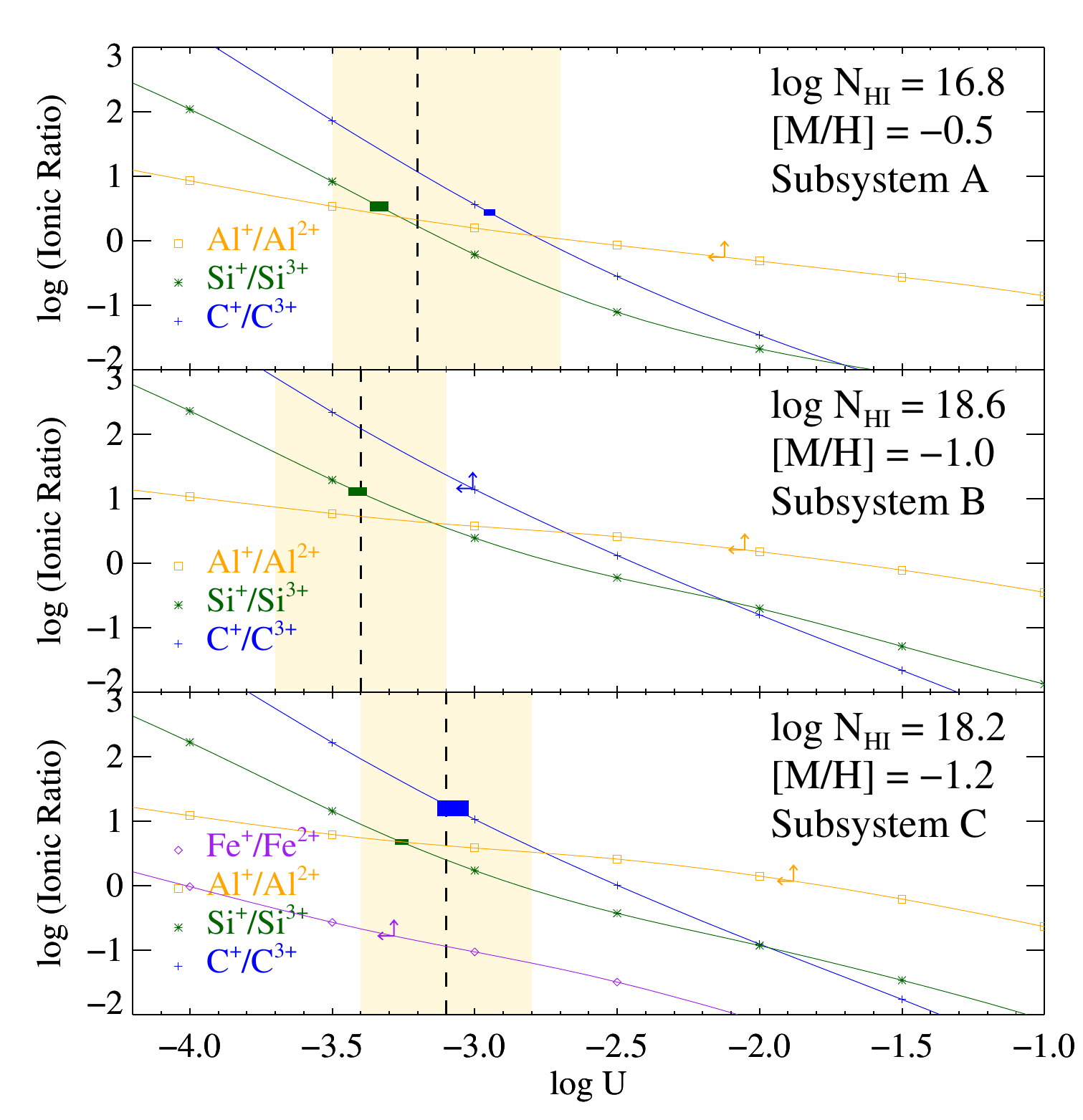} 
\caption{Cloudy example of constraining the ionization parameter $U=\Phi/n_{\rm H}c$ for
J0853-0011, where $\Phi$ is the ionizing photon flux. This figure presents a comparison of a
series of Cloudy models with constraints from the ionic ratios of the three subsystems A, B, and 
C. The observed ionic ratios and the corresponding $\log U$ are indicated by solid boxes, whose 
heights and breaths represent the 1-$\sigma$ uncertainties, or indicated by arrows for lower or 
upper limits. 
}
\label{fig:ex_cldy}
\end{figure}

Figure~\ref{fig:ex_cldy} presents the comparison of a series of Cloudy models with constraints 
from the ionic ratios of the three subsystems in J0853-0011. The observational constraints on the
ionic ratios and the corresponding $\log U$ values are indicated by solid boxes, or arrows for 
lower or upper limits. The ionic ratios for these three subsystems can be described by a 
photoionization model with $\log U \approx -3.3$. Occasionally, the observational constraints are 
not fully consistent with a single $U$ value, for example subsystem C of J0853-0011 shown in 
Figure~\ref{fig:ex_cldy}. 
Although this inconsistency may suggest the low and high ions arise in a multiphase or 
non-equilibrium medium, there are significant systematic uncertainties inherent to photoionization 
modeling including the assumed spectral shape for the radiation field, cloud geometry and the 
atomic data. In such cases, we preferred constraints from ionic ratios that are more 
sensitive to $U$. For example, for subsystems of J0853-0011, observational constraints from 
C$^+$/C$^{3+}$ and Si$^+$/Si$^{3+}$ are preferred to Fe$^+$/Fe$^{2+}$, and we adopted a $U$ value 
that is between that constrained by C$^+$/C$^{3+}$ and that by Si$^+$/Si$^{3+}$. 
We then proceeded conservatively by allowing for a substantial error on $U$. 
The uncertainties in 
the adopted $\log U$ values in this study are set to be at least 0.3~dex. For J0853-0011, the 
adopted uncertainties in $\log U$ encompass constraints on U from all ionic ratios. 
The estimated $U$ value will give a corresponding neutral fraction $x_{\rm HI}$ and a total 
hydrogen column density $N_{\rm H}\equiv N_{\rm HI}/x_{\rm HI}$. The error in the neutral fraction 
is roughly linear with the uncertainty in $U$ for $\log U>-4$.

\begin{deluxetable*}{lccccccccccccc}
\tablewidth{0pc}
\tablecaption{Total Ionic Column Densities \label{tab:clm_summ}}
\tabletypesize{\scriptsize}
\setlength{\tabcolsep}{0in}
\tablehead{\colhead{QSO Pair} & \colhead{$z_{\rm fg}$} 
& \colhead{$\log\;N({\rm C}^{+})$} 
& \colhead{$\log\;N({\rm C}^{+3})$} 
& \colhead{$\log\;N({\rm O}^0)$} 
& \colhead{$\log\;N({\rm Si}^{+})$} 
& \colhead{$\log\;N({\rm Si}^{+3})$} 
& \colhead{$\log\;N({\rm Fe}^{+})$} 
} 
 \startdata 
J0225+0048 & 2.7265 &$<13.86$ &$14.12 \pm 0.02$ &&$<12.76$ &$13.39 \pm 0.05$ &&\\ 
J0341+0000 & 2.1233 &$<14.00$ &$<13.28$ &&$<13.42$ &&$<13.62$ &\\ 
J0409-0411 & 1.7155 &$<13.97$ &$<13.31$ &&$<13.48$ &&$14.12 \pm 0.08$ &\\ 
J0853-0011 & 2.4014 &$>14.11$ &$13.25 \pm 0.03$ &$14.48 \pm 0.01$ &$13.36 \pm 0.01$ &$12.86 \pm 0.03$ &$13.01 \pm 0.03$ &\\ 
J0932+0925 & 2.4170 &$13.72 \pm 0.11$ &$13.93 \pm 0.02$ &&$<13.55$ &$13.14 \pm 0.04$ &$<13.19$ &\\ 
J1026+4614 & 3.3401 &$<13.14$ &$13.43 \pm 0.02$ &&$<12.12$ &$12.74 \pm 0.04$ &&\\ 
J1038+5027 & 3.1323 &$<13.38$ &$14.08 \pm 0.02$ &&$<13.39$ &$13.00 \pm 0.10$ &&\\ 
J1144+0959 & 2.9731 &$>13.26$ &$>13.84$ &$13.16 \pm 0.13$ &$12.84 \pm 0.03$ &$>13.06$ &$13.25 \pm 0.05$ &\\ 
J1145+0322 & 1.7652 &$>14.71$ &$>14.59$ &$14.08 \pm 0.14$ &$13.97 \pm 0.05$ &$13.74 \pm 0.03$ &$13.53 \pm 0.09$ &\\ 
J1204+0221 & 2.4358 &$>14.75$ &$13.80 \pm 0.01$ &$>14.46$ &$>14.27$ &$13.06 \pm 0.01$ &$13.59 \pm 0.04$ &\\ 
J1420+1603 & 2.0197 &$>14.22$ &$>14.12$ &$>14.61$ &$13.25 \pm 0.03$ &$13.27 \pm 0.03$ &$13.59 \pm 0.01$ &\\ 
J1427-0121 & 2.2736 &$14.01 \pm 0.02$ &$14.40 \pm 0.02$ &$14.12 \pm 0.03$ &$12.75 \pm 0.01$ &$12.91 \pm 0.03$ &$<12.58$ &\\ 
J1553+1921 & 2.0098 &$>14.92$ &$14.57 \pm 0.03$ &$>15.05$ &$15.26 \pm 0.14$ &$13.82 \pm 0.10$ &$14.02 \pm 0.03$ &\\ 
J1627+4605 & 3.8137 &$<13.17$ &$13.63 \pm 0.04$ &$<13.53$ &$<13.25$ &$<12.66$ &&\\ 
\enddata 
\tablecomments{Total logarithmic column densities for the absorption associated to each QPQ8 pair. 
One should adopt a systematic uncertainty of 0.05~dex related to continuum placement. 
}\end{deluxetable*} 

A discussion of the constraints and ionization modeling results for the individual absorption 
systems is presented in the Appendix. Figure~\ref{fig:UvsNHI} presents a scatter plot for all of 
the $U$ values derived from the dataset against the estimated \ion{H}{1} column density. Despite 
the large uncertainties in the measurements, there is a statistically significant anti-correlation 
between $U$ and \nhi. A Spearman's rank correlation test rules out the null hypothesis at 
$99.99\%$ confidence. This may be explained by either an increasing volume density $n_{\rm H}$ 
with increasing column density \nhi\ or a fixed volume density and a varying radiation field. In 
the latter case, when $\Phi$ decreases, $U$ decreases and $N_{\rm HI}$ increases. This would be 
the scenario if there is gas with similar density at different distances or if there are 
illuminated and obscured systems. 

\begin{figure}
\includegraphics[width=3.5in]{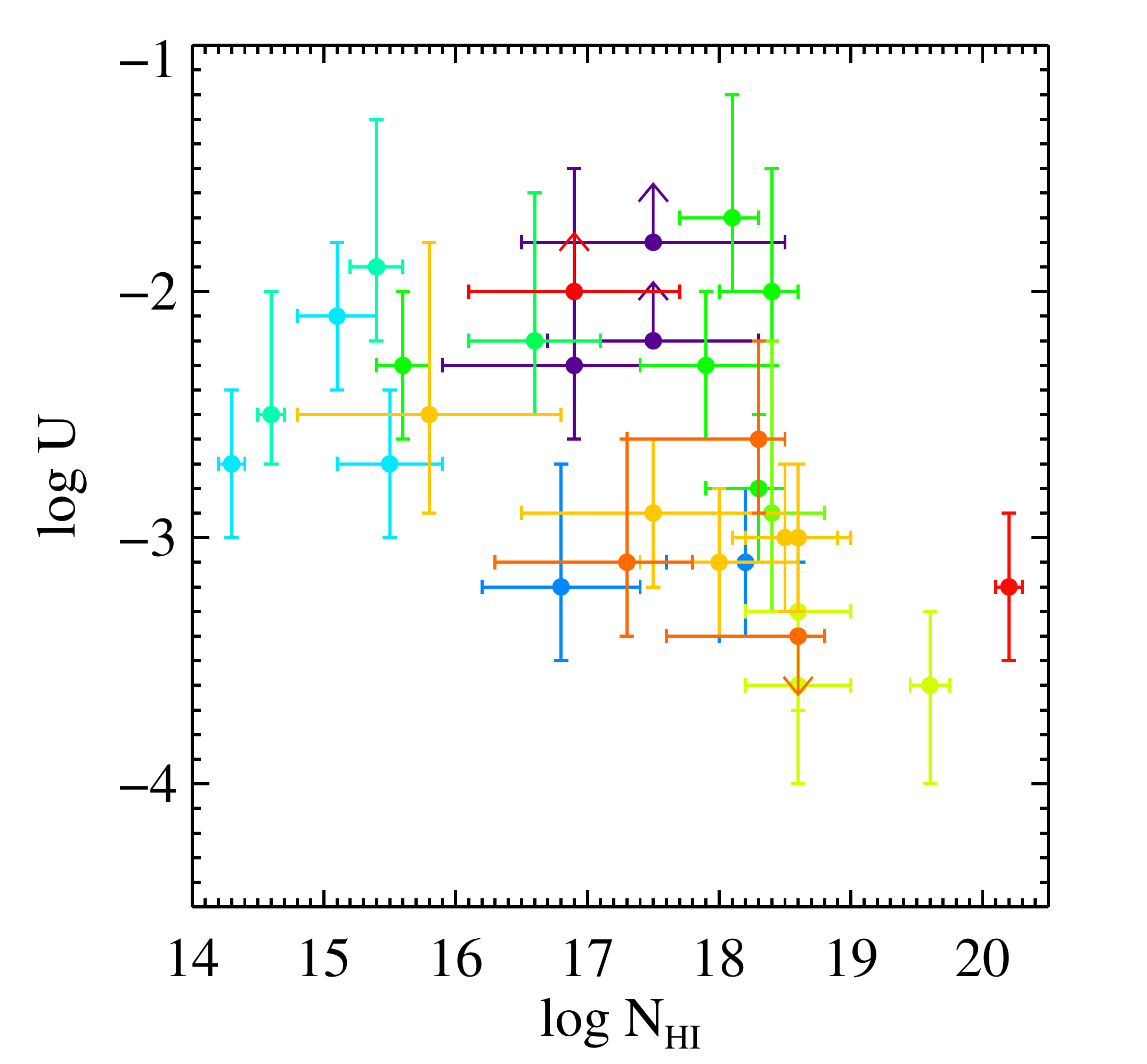} 
\caption{A scatter plot for all of the individual subsystems' $U$ values against the estimated 
$N_{\rm HI}$. For clarity, subsystems associated with different quasars are coded in different 
colors. An anti-correlation is significant at $>99.99\%$ confidence. 
}
\label{fig:UvsNHI}
\end{figure}

\begin{figure}
\includegraphics[width=3.5in]{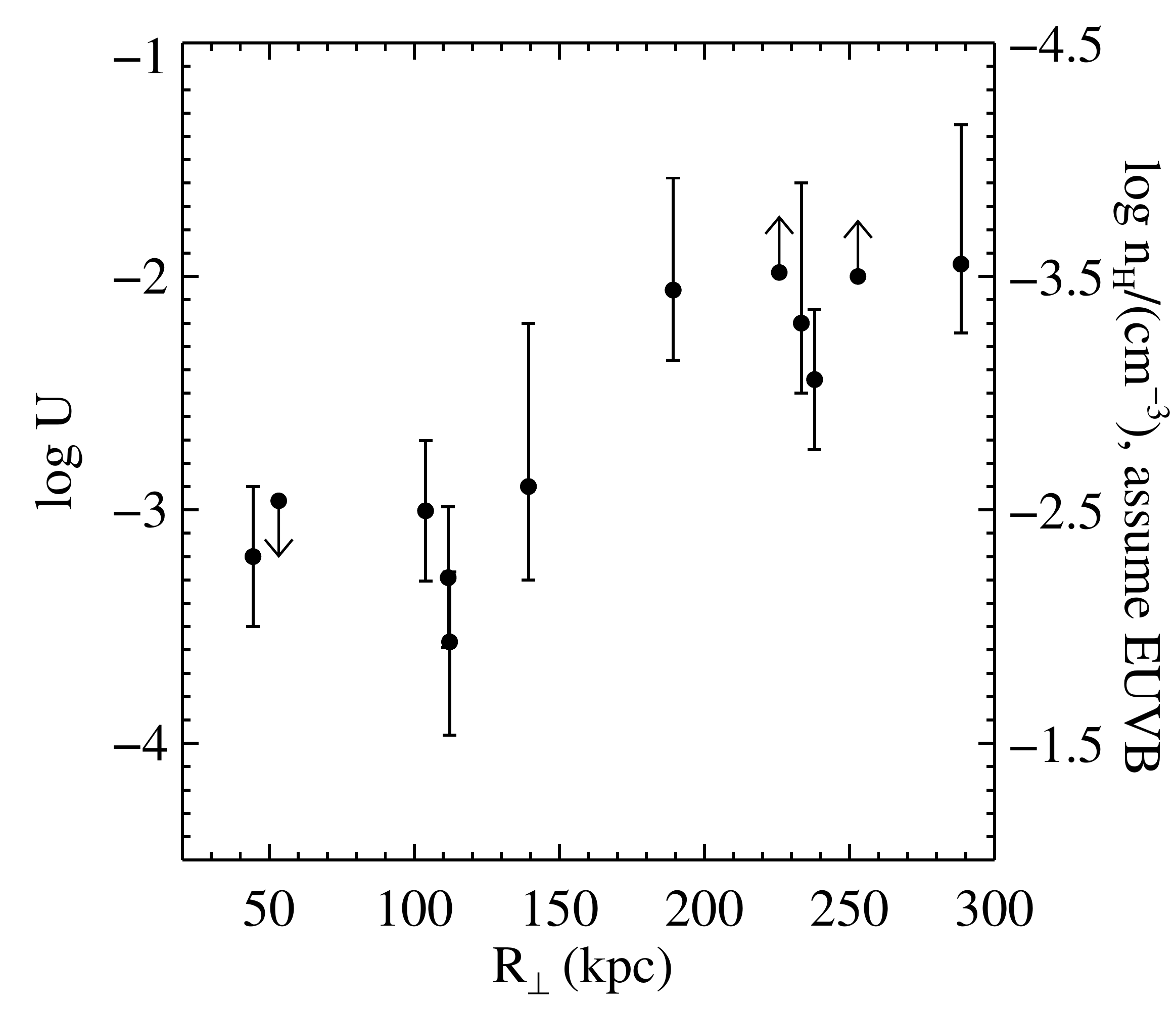} 
\caption{$<U>$ as a function of $R_\perp$. We estimated a characteristic average $<U>$ for each
quasar associated absorption system by weighting the $U$ values of the subsystems by their 
$N_{\rm HI}$. The correlation between $<U>$ and $R_\perp$ is significant at $99.8\%$ confidence. We 
also derived $n_{\rm EUVB}$, the density required to give $<U>$ if the gas is only illuminated by 
the extragalactic UV background and rescaled to match the mean opacity of the Ly$\alpha$ forest. 
These may be regarded as lower limits to the gas density.
}
\label{fig:UvsRperp}
\end{figure}

Granted the focus of this manuscript is on the cool CGM, we estimated a characteristic $U$ value 
for each pair by weighting the $U$ values of each subsystem by its corresponding \nhi\ value: 
\begin{equation}
\mavgU\equiv\frac{\smm(U\mnhi)}{\smm(\mnhi)}
\end{equation}
These values are listed in Table~\ref{tab:Usumm} together
 with the \nhi-weighted neutral fractions \xavg, and the individual constraints on $U$ for each 
subsystem. We also derived $n_{\rm EUVB}$, the density one will require to give the \avgU\ value 
if the gas is only illuminated by a EUVB with normalization rescaled to match the 
\cite{FaucherGiguere+08} mean opacity of the \lya\ forest. 
We list the $n_{\rm EUVB}$ results in Table~\ref{tab:Usumm}.
In this scenario, one would require gas densities of $10^{-2}\;{\rm cm^{-3}}$ or lower. These may 
be regarded as lower limits to the gas density. 
The results of $n_{\rm EUVB}$ together with \avgU\ as a function of the impact parameter $R_\perp$ 
are plotted in Figure~\ref{fig:UvsRperp}. There is a statistically significant positive 
correlation between \avgU\ and $R_\perp$. A Spearman's rank correlation test rules out the null 
hypothesis at $99.8\%$ confidence. This positive dependence of \avgU\ on $R_\perp$ runs contrary 
to expectation should the quasar dominate the ionizing flux received by the absorbers, 
unless the density profile is much steeper than that of $n_{\rm EUVB}$. 
Moreover, we found no statistically significant correlation between $U$ and quasar bolometric 
luminosity or the UV enhancement factor, again contrary to expectation if the quasar radiation 
dominates. The dependence of \avgU\ on $R_\perp$ further implies that $n_{\rm H}$ decreases with 
increasing $R_\perp$. This may also be explained as an increasing volume density with increasing 
neutral column density. Table~\ref{tab:Usumm} also lists $n_{\rm QSO}$, the density required to 
yield \avgU\ assuming the gas is located at a distance from the quasar equal to \rphys\ and that 
the quasar emits isotropically. We note that in QPQ2 we argued that the anisotropic clustering 
(i.e. transverse compared to line-of-sight) of optically thick absorbers around quasars suggests 
that most $N_{\rm HI} \gtrsim 10^{17}\,{\rm cm^{-2}}$ absorbers detected in background sightlines 
are not illuminated by the foreground quasar, and we came to similar conclusions based on the 
absence of fluorescent recombination emission from these absorbers in QPQ4.
The $n_{\rm QSO}$ values may be regarded as a 
rough upper bound to $n_{\rm H}$, unless there is an extra local source of radiation, where the 
$n_{\rm H}$ values would need to be correspondingly higher as the $U$ values are elevated. We 
discuss these values further in Section~\ref{sec:volumedensity}.

\begin{deluxetable*}{llllllllll}
\tablewidth{0pc}
\tablecaption{Summary of physical conditions \label{tab:Usumm}}
\tabletypesize{\scriptsize}
\setlength{\tabcolsep}{0in}
\tablehead{\colhead{Name} & \colhead{Subsystem} & \colhead{$\log U^a$} & \colhead{log \avgU$^b$} & \colhead{log \xavg$^c$} & \colhead{$\log n_{\rm EUVB}^d/{\rm cm^{-3}}$} & \colhead{$\log n_{\rm QSO}^e/{\rm cm^{-3}}$} & \colhead{$\log n_{\rm e}^f/{\rm cm^{-3}}$} & \colhead{Linear Size$^g$} \\ 
 & & & & & & & & (pc)}
\startdata
J0225+0048 &A &$>-1.8$ &$>-2.0$ &$-3.4$ &$-3.6$ &$-0.8$ &  & \\
 &B &$-2.3^{+0.8}_{-0.3}$ & & & & &  & \\
 &C &$>-2.2$ & & & & &  & \\
J0853-0011 &A &$-3.2^{+0.5}_{-0.3}$ &$-3.3$ &$-1.2$ &$-2.2$ &+$ 0.6$ &  & \\
 &B &$-3.4^{+0.3}_{-0.3}$ & & & & &  & \\
 &C &$-3.1^{+0.3}_{-0.3}$ & & & & & $< 0.9$ &$>  3.8$ \\
J0932+0925 &A &$-2.7^{+0.3}_{-0.3}$ &$-2.4$ &$-2.8$ &$-3.1$ &$-0.5$ & $< 1.8$ &$>0.01$ \\
 &B &$-2.1^{+0.3}_{-0.3}$ & & & & &  & \\
 &C &$-2.7^{+0.3}_{-0.3}$ & & & & &  & \\
J1026+4614 &A &$-1.9^{+0.6}_{-0.3}$ &$-1.9$ &$-3.3$ &$-3.7$ &$-0.6$ &  & \\
 &B &$-2.5^{+0.5}_{-0.2}$ & & & & &  & \\
J1038+5027 &A &$-2.2^{+0.6}_{-0.3}$ &$-2.2$ &$-3.3$ &$-3.4$ &$-0.1$ &  & \\
J1144+0959 &A &$-1.5^{+0.3}_{-0.3}$ &$-2.1$ &$-2.4$ &$-3.5$ &$-0.3$ &  & \\
 &B &$-1.7^{+0.5}_{-0.3}$ & & & & &  & \\
 &C &$-2.8^{+0.3}_{-0.3}$ & & & & &  & \\
 &D &$-2.3^{+0.3}_{-0.3}$ & & & & & $< 1.3$ &$> 11.2$ \\
 &E &$-2.3^{+0.3}_{-0.3}$ & & & & &  & \\
 &F &$-2.0^{+0.5}_{-0.3}$ & & & & & $< 0.4$ &$>239.3$ \\
J1145+0322 &A &$-2.9^{+0.7}_{-0.4}$ &$-2.9$ &$-1.8$ &$-2.6$ &+$ 0.1$ & $< 0.6$ &$> 11.8$ \\
J1204+0221 &A &$-3.3^{+0.3}_{-0.4}$ &$-3.6$ &$-0.3$ &$-1.9$ &+$ 1.2$ &  & \\
 &B &$-3.6^{+0.3}_{-0.4}$ & & & & &  & \\
 &C &$-3.6^{+0.3}_{-0.4}$ & & & & & $< 0.6$ &$>  2.4$ \\
J1420+1603 &A &$-2.5^{+0.7}_{-0.4}$ &$-3.0$ &$-1.6$ &$-2.5$ &+$ 1.1$ &  & \\
 &B &$-3.0^{+0.3}_{-0.3}$ & & & & &  & \\
 &C &$-3.1^{+0.3}_{-0.3}$ & & & & &  & \\
 &D &$-3.0^{+0.3}_{-0.3}$ & & & & &  & \\
 &E &$-3.0^{+0.3}_{-0.3}$ & & & & &  & \\
 &F &$-2.9^{+0.3}_{-0.3}$ & & & & &  2.2 &0.17 \\
J1427-0121 &A &$-3.1^{+0.5}_{-0.3}$ &$<-3.0$ &$-1.3$ &$-2.5$ &+$ 1.7$ &  & \\
 &B &$-2.6^{+0.4}_{-0.3}$ & & & & & $< 0.7$ &$> 23.5$ \\
 &C &$<-3.4$ & & & & &  1.2 &0.97 \\
J1553+1921 &A &$-3.2^{+0.3}_{-0.3}$ &$-3.2$ &$-0.3$ &$-2.3$ &+$ 1.2$ & $<2.5^h$ &$>  7.7$ \\
J1627+4605 &A &$>-2.0$ &$>-2.0$ &$-3.5$ &$-3.7$ &$-0.6$ &  & \\
\enddata
\tablenotetext{a}{The ionization parameter $U$ for each subsystem comes from Cloudy ionization modeling.}
\tablenotetext{b}{The average ionization parameter $<U>$ for the absorption system associated to the foreground quasar is calculated as the $U$ values of its subsystems weighted by their corresponding $N_{\rm HI}$ values.}
\tablenotetext{c}{For each subsystem the estimated $U$ will give a corresponding neutral fraction $x_{\rm HI}$. The average neutral fraction $<x_{\rm HI}>$ is calculated as the neutral fractions of the subsystems  weighted by their $N_{\rm HI}$.}
\tablenotetext{d}{The density required to give the $<U>$ value if the gas is only illuminated by the EUVB and rescaled to match the mean opacity of the Ly$\alpha$ forest.}
\tablenotetext{e}{The density required to yield $<U>$ assuming the gas is located at a distance from the quasar equal to the impact parameter and that the quasar emits isotropically.}
\tablenotetext{f}{Electron volume density calculated from fine structure excited state to ground state ratios, under the assumption of collisional equilibrium.}
\tablenotetext{g}{When the electron volume density is available, we estimated the linear size of the absorbing cloud by $\ell=N_{\rm H}/n_{\rm H}$, where $n_{\rm H}=n_e/(1-x_{\rm HI})$, not compensating for ionized helium contribution to electron density.}
\tablenotetext{h}{This is a damped Lyman $\alpha$ system and $n_{\rm HI}$ is reported instead of $n_{\rm e}$.}
\end{deluxetable*}


\section{Results}
\label{sec:results}

\subsection{Kinematics}
\label{sec:kin}

\begin{figure*}
\includegraphics[width=7in]{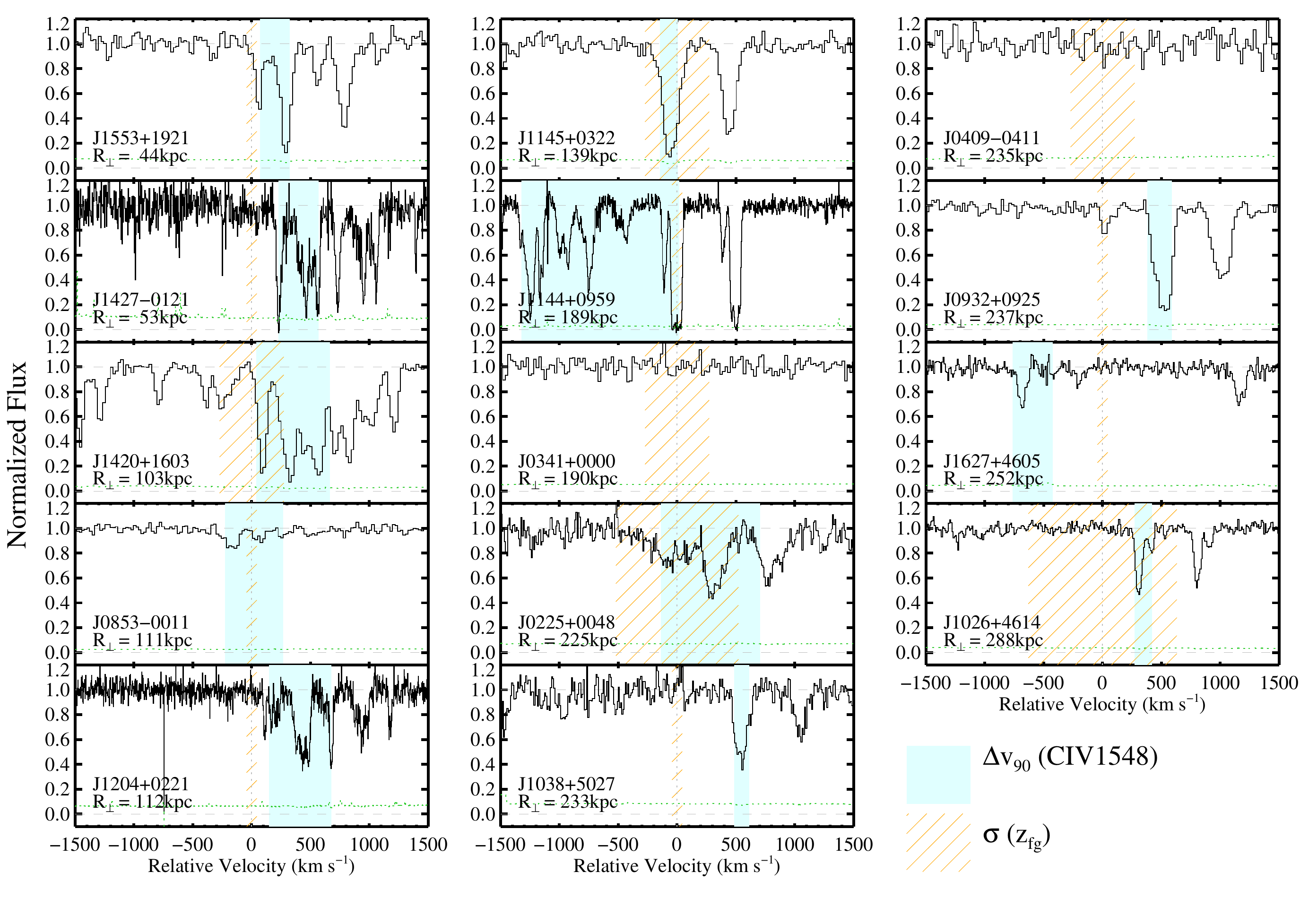} 
\caption{\ion{C}{4}~1548 absorption profiles associated with the foreground quasars. The cyan 
shade encompasses 90\% of the optical depth of the \ion{C}{4}~1548 transition. The orange shade 
marks the 1$\sigma$ uncertainty in $z_{\rm fg}$. When detected, the \ion{C}{2}~1334 profiles show
similar velocity widths. The gas exhibits velocity widths ranging from
$\approx50\;{\rm km\;s^{-1}}$ to nearly $1500\;{\rm km\;s^{-1}}$. The centroids tend to occur 
within $500\;{\rm km\;s^{-1}}$. Significant absorption rarely occurs at negative velocities. The 
green histogram shows the $1\sigma$ noise in the normalized flux.
}
\label{fig:civ_kin}
\end{figure*}

Combining the high spectral resolution of the QPQ8 dataset with our near IR estimates for \zfg, we 
may precisely characterize the kinematics of the absorbing gas. Such measurements resolve the 
dynamics of the cool gas in the massive halos hosting quasars and constrain physical scenarios for 
the origin of this medium. In particular, one may search for evidence of non-gravitational 
motions, i.e. galactic scale outflows powered by AGN, that are regularly invoked as a critical 
process in galaxy formation theory \citep[e.g.][]{FaucherGiguereQuataert12}.

In the following, we examine three statistical measures to characterize the kinematics: (i) the 
optical depth weighted velocity offset of the gas relative to \zfg:
\begin{equation}
\mdvstat\equiv\frac{\smm_i\tau_i\delta v_i}{\smm_i\tau_i}
\end{equation}
where $f_i$ the normalized flux and  $\tau_i = -\ln(f_i)$ is the optical depth per pixel. In cases 
where the absorption saturates, we adopt a value equal to one-half of the standard deviation in 
those pixels $\sigma(f_i)/2$.  This tends to limit $\tau_i$ to less than 4 per pixel; (ii) the 
velocity interval that encompasses $90\%$ of the total optical depth \delv\ 
\citep[e.g.][]{ProchaskaWolfe97}; (iii) the root-mean-square of the gas $\sigma_v$, measured from 
the optical depth weighted dispersion of the profiles.
Figure~\ref{fig:civ_kin} provides a qualitative picture of the kinematic characteristics. 
Plotted are the \ion{C}{4} doublets for the QPQ8 sightlines with $v=0\;\mkms$ corresponding to 
\zfg\ for the \ion{C}{4}~1548 transition. In each, we highlight the \delv\ interval relative to 
\zfg. For the few absorption systems where \delv\ exceeds the velocity separation of the 
$\lambda\lambda$~1548,~1550 doublet $\approx 500\;\mkms$, we estimated the kinematic measurements 
for \ion{C}{4}~1548 as described in the Appendix. A few results are evident. First, the gas 
exhibits a dynamic range in the \delv\ widths ranging from $\approx50\;\mkms$ to nearly 
$1500\;\mkms$. 
Second, the centroids of the absorption profiles tend to occur within $500\;\mkms$ 
of \zfg. Third, significant absorption rarely occurs at negative velocities.

\begin{figure}
\includegraphics[width=3in]{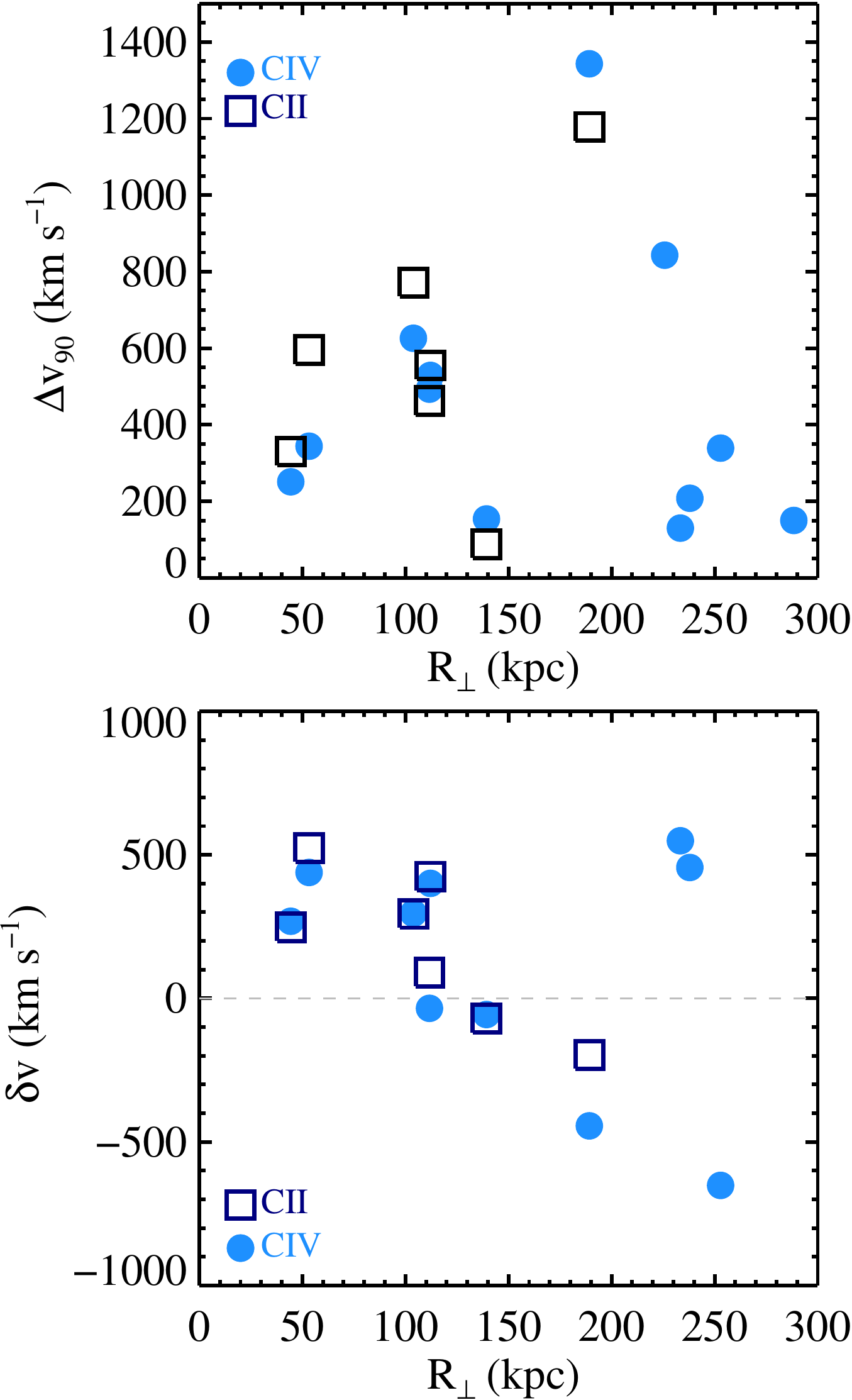} 
\caption{(Top) Measurements of the \delv\ statistic for low ion absorption in open squares and
high ion absorption in filled circles. The absorptions within $\mrphys \le 200$~kpc show very
large \delv\ values. This distribution exceeds all previous measurements for gas surrounding
galaxies \citep[e.g.][]{ProchaskaWolfe97,Neeleman+13,Werk+13}. (Bottom) Velocity offsets \dvstat\
between the optical depth weighted centroid of the absorption profiles and the systemic redshift
of the foreground quasar. The \dvstat\ values typically match or exceed the \delv\ value, 
indicating that a majority of gas occurs on one side of systemic.  In this small sample, there is
an apparent bias to positive velocities.
}
\label{fig:both_kin}
\end{figure}

These results are further described in Figure~\ref{fig:both_kin} which present the \delv\ and 
\dvstat\ measurements for \ion{C}{2}~1334 and \ion{C}{4}~1548 against impact parameter. These two 
ions generally exhibit similar kinematic characteristics, consistent with the high ionization 
fractions estimated for the gas. The \delv\ widths for the QPQ8 sample have median values of 
$555\;\mkms$ for \ion{C}{2} and $342\;\mkms$ for \ion{C}{4}, and exhibit no strong correlation 
with \rphys. These motions greatly exceed the values previously measured for gas tracing galaxies 
or CGM in absorption. This includes the damped \lya\ systems, whose median \delv\ for low ion 
absorption $\approx80\;\mkms$ and that for \ion{C}{4} $\approx170\;\mkms$ 
\citep{ProchaskaWolfe97,Neeleman+13}, 
and also the CGM of $L^*$ galaxies in the low-$z$ universe whose median \delv\ $\approx100\;\mkms$ 
\citep{Werk+13}.
Velocity widths exceeding several hundred 
\kms\ have only been routinely observed `down-the-barrel' to star forming galaxies and AGN 
themselves, where one probes gas within the galaxy \citep[e.g.][]{Steidel+10,Hamann98}. Presently, 
the QPQ8 sample exhibits the largest velocity widths probed in absorption on CGM scales at any 
epoch. 
Existing studies that found large velocity spreads along transverse sightlines around galaxies 
are limited to single sightlines with most of the gas within a few hundred ${\rm km\;s^{-1}}$ 
\citep[e.g.][]{Tripp+11}, gas tracing a higher ionization state \citep[e.g.][]{Churchill+12}, 
average velocity spread smaller than that 
measured in QPQ8 \citep[e.g.][]{Gauthier13,Muzahid+15,Zahedy+16}, 
or the velocity spreads are not well quantified \citep[e.g.][]{Johnson+15}. 
We further emphasize that this result follows from the systematically large equivalent 
widths observed in the full QPQ sample (e.g.\ QPQ5, QPQ7). Therefore despite the small sample size 
of QPQ8 we consider the distribution of large \delv\ values to be a statistically strong result.

The lower panel of Figure~\ref{fig:both_kin} shows the velocity offsets \dvstat\ for the sample, 
restricting to absorption systems with precise \ion{Mg}{2} or near IR measurements for \zfg. The 
\dvstat\ values range from $\approx -500\;\mkms$ to $\approx +500\;\mkms$ with an RMS of 261 \kms\ 
and 408 \kms\ for \ion{C}{2} and \ion{C}{4} respectively. The \dvstat\ values frequently match or 
exceed the velocity width, implying that the absorption often occurs to only positive or negative 
velocities relative to \zfg. Furthermore the \dvstat\ values of our sample appear to preferentially 
exhibit positive values.
For \ion{C}{4}, only 2 of the 10 profiles have $\mdvstat\ll0\;\mkms$ and the median measurement is 
$+282\;\mkms$.
Of course, the skewness of the \dvstat\ distribution may be dominated by sample variance and 
therefore must be confirmed with a larger sample of pairs (see QPQ9). The result 
does follow, however, a reported asymmetry in \ion{H}{1} in an independent quasar pair sample 
\citep{KirkmanTytler08}. 

We further illustrate the kinematic characteristics of the gas by constructing the average optical 
depth profiles for the QPQ8 sample. We interpolated the apparent optical depth of each transition 
onto a fixed velocity grid with 25 \kms\ pixels and normalized each to have a peak $\tau_i=1$. We 
then performed a straight average of all the profiles. For systems where the \ion{C}{4} doublet 
self blends, we estimated the optical depth of \ion{C}{4}~1548 in the overlap spectral region from 
the unblended portions of the doublet. The details of the treatment are in the Appendix. 
Figure~\ref{fig:avgtau} shows the average optical depth profiles. These further emphasize the 
large velocity widths and the tendency toward positive velocities.

\begin{figure}
\includegraphics[width=3in]{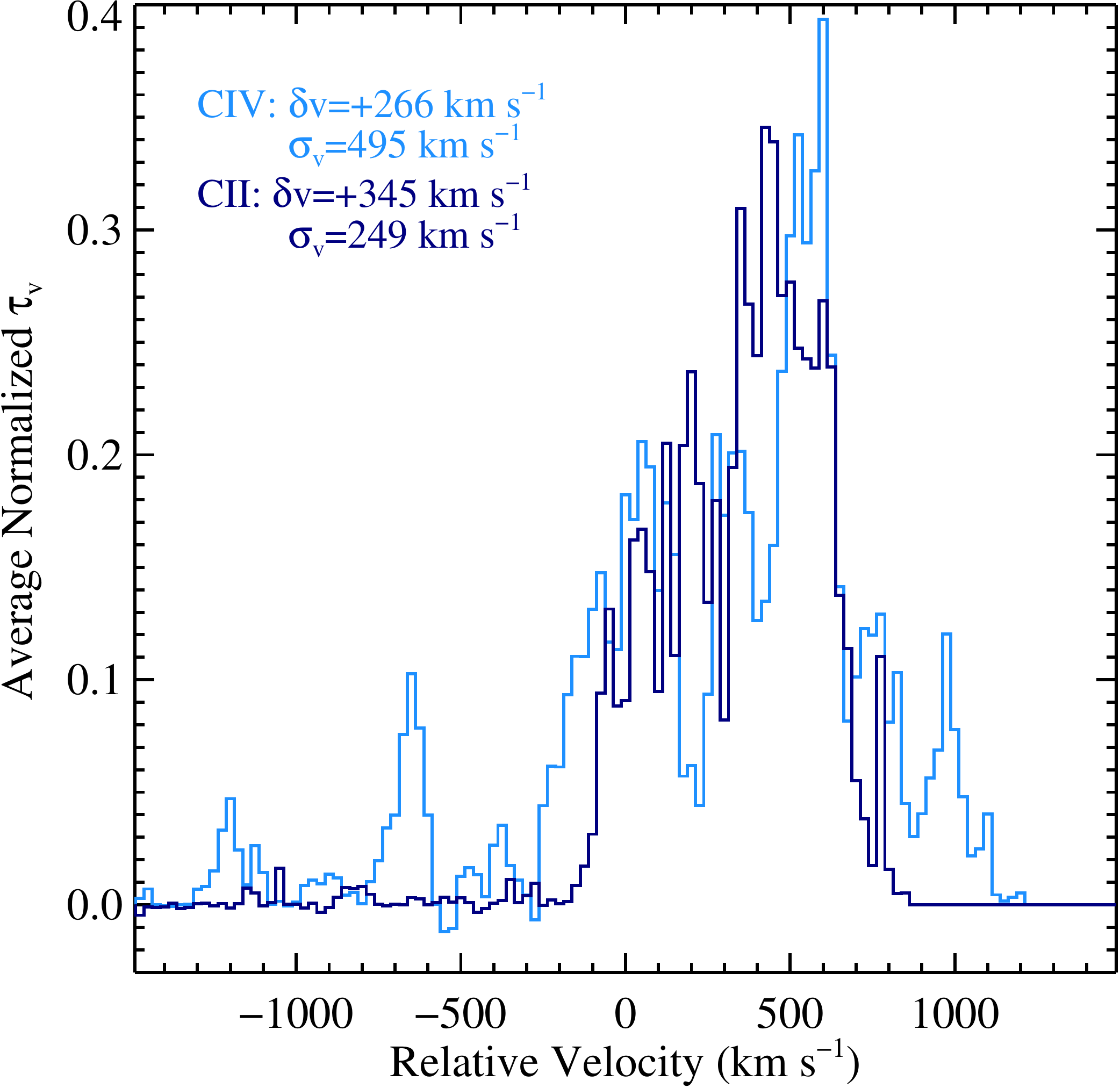} 
\caption{Average optical depth profiles of the QPQ8 sample, generated from the ensemble with each
normalized to peak optical depth of 1. These profiles stress the large velocity widths of the gas
and also a peculiar bias toward positive velocity relative to systemic.
}
\label{fig:avgtau}
\end{figure}

Taking the ensemble of QPQ8 data as a statistical representation of the CGM surrounding 
quasars, we may estimate the RMS of the CGM surrounding quasars $\sigma_v$ from the average 
profile. Measuring this dispersion relative to the profile centroid, 
instead of $v=0\;\mkms$, we recovered $\sigma_v=249\;\mkms$ for \ion{C}{2} and 
$\sigma_v=495\;\mkms$ for \ion{C}{4}. 
A large $\sigma_v$ value for \ion{C}{4} is representative of the sample. Even if we take out the 
two sightlines with the largest velocity dispersion, i.e. J0225+0048 and J1144+0959, we still get 
$\sigma_v=388\;\mkms$. 

In QPQ3, we argued the extreme kinematics of J1204+0221 ($\mdelv \approx 650\;\mkms$) could be 
consistent with the one dimensional velocity dispersion of a massive dark matter halo. We assumed 
a dark matter halo mass of $10^{13.3} \msol$ and estimated a line-of-sight velocity dispersion of 
$\sigma_{\rm 1D} \approx 431\;\mkms$ according to \cite{Tormen+97} and 
assuming an NFW halo with concentration parameter $c=4$. More recent analysis of quasar clustering 
prefers a lower typical mass ($10^{12.5} \msol$), giving $\sigma_{\rm 1D} = 212\;\mkms$.
At this mass scale, we consider it improbable that the velocity widths
we have observed are entirely virialized motions, and thus likely require
non-gravitational kinematics i.e. outflows. Note that satellite
galaxies within the halo should follow the potential and have a
similar line-of-sight velocity dispersion. If galaxies clustered to
the quasar host on larger scales also contribute to the observed
velocity widths, the widths will partly represent Hubble flow of
non-collapsed material. This scenario is rather unlikely, however, and
the probability intercepting a random optically thick absorber is only
$\approx4\%$, 
and clustering would only increase that to 24\% according to QPQ6 (also see the discussion in 
\cite{Johnson+15}).
We eagerly await advances in hydrodynamic simulations of massive $z\sim2$ galaxies to explore 
these scenarios. The greatest challenge may for such feedback to manifest itself as cool, 
optically thick gas on CGM scales \citep{FaucherGiguereQuataert12}. Although 
\cite{FaucherGiguere+16} is able to reproduce the high covering fraction of optically thick gas, 
their velocity fields are not extreme enough. 

Recently, \cite{Johnson+15} have studied the CGM surrounding $z\sim1$ quasars using QPQ techniques, 
and reported 40\% of the \ion{Mg}{2} absorption occurs at radial velocities exceeding the expected 
average virial velocity $300\;{\rm km\;s}^{-1}$, which they interpret as quasar driven outflows. 
For the QPQ8 sample, we found 4 out of 7 \ion{C}{2} systems and 8 out of 10 \ion{C}{4} systems 
exhibit \dvstat\ exceeding the one-dimensional virial velocity (see Figure~\ref{fig:both_kin}). 
To allow a one-to-one comparison between our results and \cite{Johnson+15}, we quote 2 out of 7 
\ion{C}{2} systems and 6 out of 10 \ion{C}{4} systems exhibit \dvstat\ exceeding 
$300\;{\rm km\;s}^{-1}$. This is consistent with \cite{Johnson+15}, especially considering 
the higher random probability for \ion{C}{4} absorbers. 
 
However, we caution that the mere presence of a few large kinematic offsets cannot be used to make 
the case against gravitational motions, unless the shift is of order $\sim1000\;{\rm km\;s}^{-1}$, 
such as the J1144+0959 absorption system of QPQ8. Clustering only constrains the average mass of 
the dark matter halos occupied by quasars, but not the distribution function. The halo mass 
occupation distribution could be very broad, or have significant tails, which could both give rise 
to occasional large velocity shifts, which would not require non-gravitational kinematics. 
(On a side note, despite the large velocity shift of the J1144+0959 absorption system from 
$z_{\rm fg}$, in the Appendix we demonstrated that it is likely the gas is physically 
associated to the quasar. For random incidence, the probability of finding one such strong 
\ion{C}{4} absorber is only 3\%, with an even smaller probability for \ion{C}{2} absorber. 
Clustering would at most quadrupole this estimate.)

Moreover, the [\ion{O}{3}] emission occasionally exhibits a significant blue-shifted tail relative 
to the systemic defined by [\ion{O}{2}] (see Figure~\ref{fig:nearIR}). 
Systematic shifts from redshift errors on the order of $200\;{\rm km\;s}^{-1}$ are occasionally 
expected \citep{Boroson05}. A second caveat is that we used a 
mixture of [\ion{O}{3}] and \ion{Mg}{2} systemic redshifts, the latter being less accurate. 


Now consider the putative bias to positive velocities that is apparent in Figure~\ref{fig:avgtau}. 
Because the sightlines penetrate the entire halo, there is an inherent symmetry to the experiment. 
Both random velocity fields (e.g. random sampling of a virialized ensemble of CGM gas) and 
coherent velocity fields (e.g. outflow) will therefore yield symmetric absorption about 
$v=0\;\mkms$. If the apparent asymmetry in Figure~\ref{fig:avgtau} is confirmed with a larger 
sample, one may require a non-dynamical process to provide the asymmetry. One possibility is 
an asymmetric radiation field that preferentially ionizes the gas moving towards the observer, 
i.e. where the quasar is known to shine. The quasar is obscured in the direction pointing away 
from us, and the gas observed in absorption lies preferentially behind the quasar and is shadowed. 
One may either interpret the redshift as an organized velocity field of flow away 
from the galaxy, or interpret the velocity as a proxy for distance along the line-of-sight, as it 
would be if the material is in the Hubble flow and hence this asymmetry arises from a transverse 
proximity effect. 
There is no known case for a quasar shining only towards us, however.
Another explanation is that, 
given the finite lifetime of quasar episodes, light from the
background quasar may arrive at absorbers behind the foreground quasar
before the ionizing radiation from the foreground quasar arrives. On
the other hand, light from the background quasar will need to travel
larger distances to reach the absorbers in front of the foreground
quasar and hence allow more time for the foreground quasar's radiation
to reach and ionize them \cite[see Figure 8 of][]{KirkmanTytler08}. 
We note, however, \cite{ShenHo14} and \cite{Shen16} found a correlation between the fraction 
of total [\ion{O}{3}] flux that is in the blueshifted wing and quasar luminosity. The putative 
redshift of absorbing components relative to the systemic is subject to redshift determination 
errors. We defer a more elaborated discussion to future papers which focus on the kinematics. 
We are assembling a much larger sample of pairs with precise redshift
measurements to better resolve this kinematic signature, using metal
absorption lines to study on gas on CGM scales (Prochaska et al. in
prep.; QPQ9) and \ion{H}{1} to probe gas on larger scales in the
Ly$\alpha$ forest (Hennawi et al. in prep.).

\subsection{Chemical Abundances}
\label{sec:abundance}

The frequent detection of metal line absorption in the CGM surrounding quasars, even at low 
spectral resolution (QPQ5, QPQ7), implies a significant enrichment of the gas in heavy elements. 
Quantitative estimates for the metallicity, however, require an assessment of the ionization state 
because the observed atoms in a given ionization state may only comprise a small fraction of the total hydrogen and metal 
column densities. Furthermore, high spectral resolution is necessary to precisely estimate the 
column densities and to assess line saturation. With the majority of the QPQ8 sample, we satisfy 
both of these requirements and may constrain the chemical abundances for a set of elements.  

Consider first an estimate for the metallicity [M/H] of each pair in the QPQ8 sample, i.e. an 
assessment of the heavy element enrichment for the gas comprising the CGM. In what follows, we 
focus on the cool gas by emphasizing the subsystems that dominate the \ion{H}{1} absorption in 
each pair. 
While the more highly ionized subsystems may have a large total column $N_{\rm H}$ 
\citep[e.g.][]{Crighton+13}, these components are subject to greater uncertainties in the 
ionization modeling and may better track a more highly ionized phase that is physically distinct 
from the cool CGM. Therefore, we report average abundances by weighting the results of each 
subsystem by our \nhi\ estimate instead of $N_{\rm H}$. 
We have calculated [M/H] values weighted by $N_{\rm H}$ instead of $N_{\rm NHI}$, and found very 
similar values for each absorption system.

Although we have constrained ionization models for the majority of our sample, we emphasize that 
the data exhibit positive detections of the \ion{O}{1}~1302 transition for 7 of the \nqpq\ 
pairs. In these cases, we may estimate [M/H] directly from the measured O$^0$/H$^0$ ratio, i.e. 
[O/H]~$=\log\N{O^0}-\log\mnhi-\epsilon_{\rm O}+12$, where $\epsilon_{\rm O}$ is the solar 
abundance of oxygen on logarithmic scale, with the expectation that ionization corrections are 
small. This assertion follows from the very similar ionization potentials of these atoms and the 
charge exchange reactions that couple their ionization for a range of physical conditions. When 
$\mnhi<10^{19}\cm{-2}$ or $\log U>-2$, however, the O$^0$/H$^0$ ratio may underestimate [O/H], 
especially in the presence of a hard radiation field (e.g. QPQ3). In this respect, [O/H] estimated 
without any ionization correction provides a conservative lower
bound to [M/H]. We are further 
motivated to focus first on oxygen because this element frequently dominates by mass and number 
among the heavy elements. 

Roughly half of the systems showing \ion{O}{1} detections have saturated profiles yielding only 
lower limits to [O/H]. For these cases we set an additional upper bound to [M/H] from our analysis 
of [Si/H]. 
We set the upper bound to [M/H] from the Si$^+$/H$^0$ measurements incremented by the 
$2\sigma$ uncertainty in the estimated ionization corrections. For systems where \ion{Si}{2} 
absorption occurs in multiple subsystems, we adopt the \nhi\ weighted [Si/H] value to emphasize 
the cool gas expected to dominate the \ion{H}{1} absorption. We found [O/H] is often larger than 
[Si/H], and for cases where the [Si/H] value exceeds the lower limit from [O/H] we adopted the 
former. In one other case where the [O/H] estimate requires a large ionization correction, 
J1144+0959 (see below and the Appendix), [Si/H] is adopted for the metallicity estimate.

\begin{figure}
\includegraphics[width=3in]{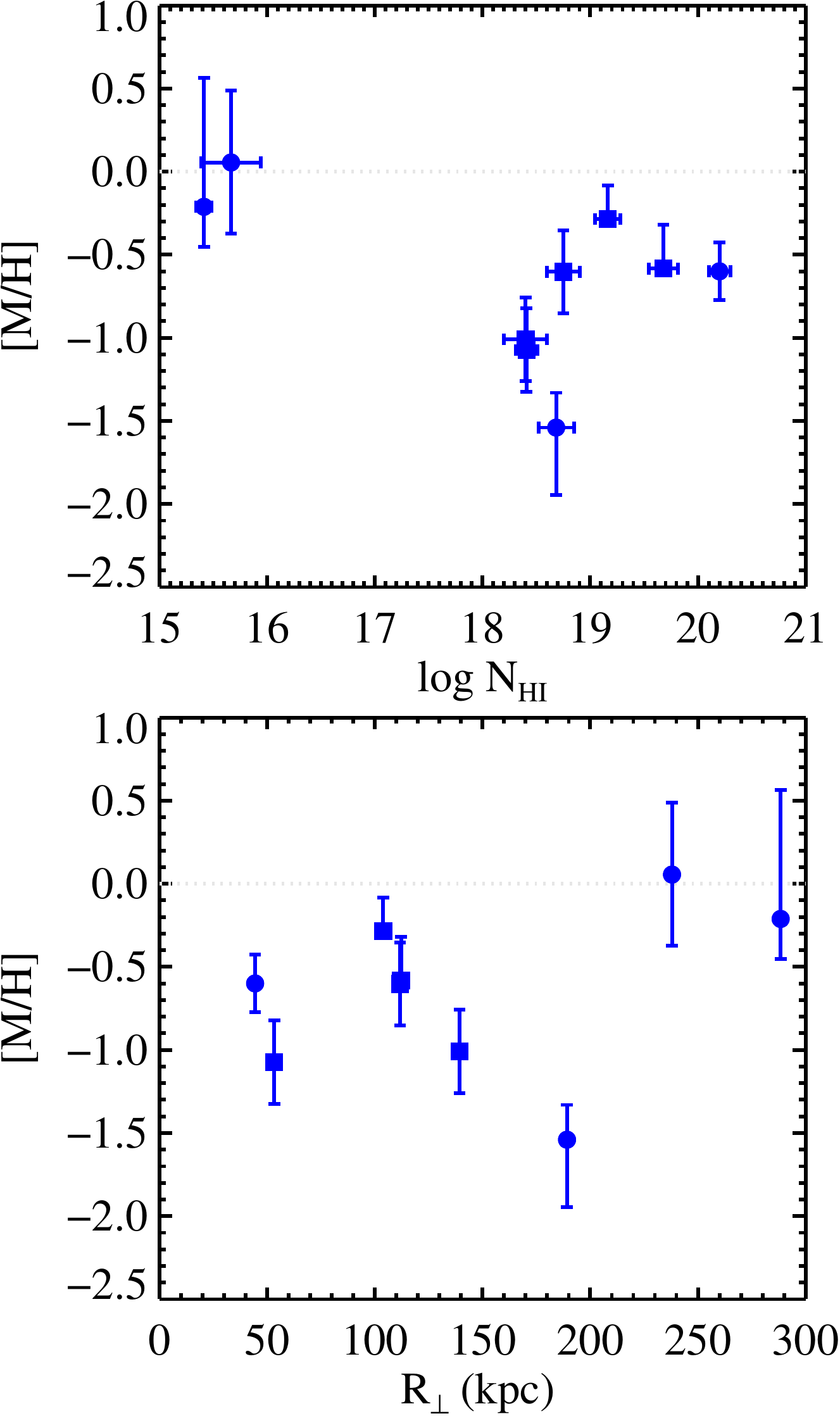} 
\caption{All metallicity estimates for the cool gas of the full sample, estimated from the
\ion{O}{1}, \ion{Si}{2}, \ion{Si}{3} or \ion{Si}{4} columns with ionization corrections from
Cloudy. Estimations from \ion{O}{1} are plotted with squares and estimations from Si ions are 
plotted with circles. All measurements exceed 1/10 solar and the median [M/H]$=-0.60$~dex. The 
measurements exhibits no correlation with $N_{\rm HI}$ or $R_\perp$. Significant enrichment exists 
even beyond the estimated virial radius of the host halos at $\approx\;160$~kpc. Note in the [M/H] 
versus $R_\perp$ subplot, there are two almost overlapping points at $R_\perp=112$~kpc.
}
\label{fig:MH}
\end{figure}

The resulting distribution of [M/H] for these 7 systems, all of which have $\mnhi>10^{18}\cm{-2}$, 
is presented in Figure~\ref{fig:MH}. All of the measurements exceed 1/10 solar and the median 
metallicity is $-0.60$~dex. This is a conservative value because three of the measurements are 
formally lower limits and because ionization corrections to O$^0$/H$^0$ would only increase [M/H]. 
Furthermore, the \nhi\ values estimated for these systems are more tightly bounded on the
upper end by the absence of damping wings, and lower \nhi\ values would yield even higher [M/H]. 
We conclude that the cool CGM surrounding massive $z \sim 2$ galaxies hosting quasars has a median 
[M/H] of at least 1/3 solar.

\begin{figure}
\includegraphics[width=3.5in]{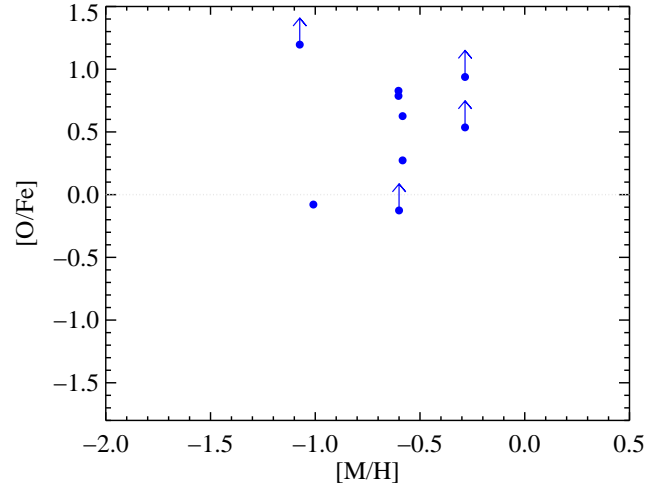} 
\caption{[O/Fe] estimates after adopting the ionization corrections to O$^0$/Fe$^+$. Nearly all
measurements indicate supersolar O/Fe ratio, implying a significant fraction of the CGM
surrounding quasars must have enhanced $\alpha$/Fe abundance. Our findings suggest a star
formation history similar to elliptical galaxies and a starburst that lasted less than 1~Gyr.
}
\label{fig:OFe}
\end{figure}

\begin{deluxetable*}{lllllllllllllllllll}
\tablewidth{0pc}
\tablecaption{Chemical Abundances \label{tab:XH}}
\tabletypesize{\scriptsize}
\tablehead{\colhead{Quasar Pair} & \colhead{$\log N_{\rm HI}/{\rm cm^{-2}}$} &  \colhead{[M/H]} & 
\colhead{[O/H]$^a$} & 
\colhead{[Si/H]$^a$} & 
\colhead{[Fe/H]$^a$} & 
\colhead{[C/H]$^a$} & 
 \colhead{Comment}}
\startdata
J0225+0048 & 18.9 & ${-1.5^{+1.0}_{-1.0}}^g$ &  &${-1.5^{+1.0}_{-1.0}}^k$ & &${-2.8^{+9.8}_{-1.1}}^k$ &\nhi\ very uncertain. \\ 
J0341+0000 & 16.6 &  & &&&&No metals detected.  \\ 
J0409-0411 & 14.2 &  & &&&&No metals detected.  \\ 
J0853-0011 & 18.8 & ${-0.6^{+0.2}_{-0.2}}^b$ & ${-0.6^{+0.2}_{-0.2}}^h$ &${-1.0^{+0.3}_{-0.6}}^i$ &${-1.4^{+0.2}_{-0.6}}^i$ &${>-1.4}^i$ & \\ 
J0932+0925 & 15.7 & ${ 0.1^{+0.4}_{-0.4}}^f$ &  &${ 0.1^{+0.4}_{-0.4}}^j$ & &${-0.0^{+0.7}_{-0.8}}^i$ & \\ 
J1026+4614 & 15.4 & ${-0.2^{+0.8}_{-0.2}}^g$ &  &${-0.2^{+0.8}_{-0.2}}^k$ & &${-0.5^{+0.5}_{-0.4}}^k$ & \\ 
J1038+5027 & 16.9 & ${-1.5^{+0.8}_{-0.5}}^g$ &  &${-1.5^{+0.8}_{-0.5}}^k$ & &${-1.6^{+0.9}_{-0.7}}^k$ &$U$ value very uncertain. \\ 
J1144+0959 & 18.7 & ${-1.5^{+0.2}_{-0.4}}^e$ & ${-1.9^{+0.2}_{-0.2}}^h$ &${-1.5^{+0.2}_{-0.4}}^i$ &${-0.3^{+0.9}_{-0.6}}^i$ &${>-2.1}^i$ & \\ 
J1145+0322 & 18.4 & ${-1.0^{+0.2}_{-0.2}}^b$ & ${-1.0^{+0.2}_{-0.2}}^h$ &${-1.1^{+0.4}_{-0.4}}^i$ &${-0.8^{+0.6}_{-0.4}}^i$ &${>-1.2}^i$ & \\ 
J1204+0221 & 19.7 & ${-0.6^{+0.3}}^c$ & ${>-0.6}^h$ &${>-0.7}^i$ &${-1.3^{+0.2}_{-0.2}}^i$ &${>-1.0}^i$ & \\ 
J1420+1603 & 19.2 & ${-0.3^{+0.2}}^c$ & ${>-0.3}^h$ &${-1.0^{+0.4}_{-0.4}}^i$ &${-0.9^{+0.4}_{-0.4}}^i$ &${>-0.7}^i$ & \\ 
J1427-0121 & 18.4 & ${-1.1^{+0.2}_{-0.2}}^b$ & ${-1.1^{+0.2}_{-0.2}}^h$ &${-2.1^{+0.2}_{-1.3}}^i$ & &${-1.7^{+0.2}_{-1.4}}^i$ & \\ 
J1553+1921 & 20.2 & ${-0.6^{+0.2}_{-0.2}}^e$ & ${>-1.8}^h$ &${-0.6^{+0.2}_{-0.2}}^i$ &${-1.7^{+0.1}_{-0.1}}^i$ &${> 0.2}^i$ & \\ 
J1627+4605 & 16.9 &  & &&&&Only \ion{C}{4} detected. \\ 
\enddata
\tablenotetext{a}{Errors in abundances are propagated from errors in HI and metal ion column densities and errors in ionization parameter values.}
\tablenotetext{b}{Adopted [O/H] from OI/HI values.}
\tablenotetext{c}{Adopted [O/H] from OI/HI as lower limit, capped by [Si/H] measurement. See the main text for details.}
\tablenotetext{d}{Adopted [O/H] lower limit. No QPQ8 systems match this case.}
\tablenotetext{e}{Adopted [Si/H] from SiII/HI values with ionization corrections, including cases where [Si/H] $>$ [O/H] lower limit. See the text.}
\tablenotetext{f}{Adopted [Si/H] from SiIII/HI values with ionization corrections.}
\tablenotetext{g}{Adopted [Si/H] from SiIV/HI values with ionization corrections.}
\tablenotetext{h}{Using OI/HI as a proxy.}
\tablenotetext{i}{Measured from the first ionization state of the element, with ionization corrections applied.}
\tablenotetext{j}{Measured from the second ionization state of the element, with ionization corrections applied.}
\tablenotetext{k}{Measured from the third ionization state of the element, with ionization corrections applied.}
\end{deluxetable*}

There are four highly ionized systems at $\mrphys>160$~kpc that offer estimates on the enrichment 
of the extended CGM from analysis of Si$^{2+}$ (J0932+0925) or Si$^{3+}$ (J0225+0048, J1026+4614, 
J1038+5027). In two of these cases, the estimates have uncertainties exceeding 1.0 dex because of 
poor constraints on the ionization state and/or \nhi. These are not considered further. One other 
highly ionized system at $\mrphys>160$~kpc (J1627+4605) exhibits only C$^{3+}$ and is also not 
considered further. Lastly, two systems exhibit no positive detection of heavy elements 
(J0341+0000, J0409-0411). While this could reflect a very metal poor gas, the observed \ion{H}{1} 
absorption is also very weak ($\mnhi<10^{15}\cm{-2}$) and the absence of heavy elements may simply 
reflect the absence of a substantial cool CGM along those sightlines.

Figure~\ref{fig:MH} shows all of these [M/H] estimates for the cool gas of the full QPQ8 sample, 
and Table~\ref{tab:XH} lists their values.   
The measurements exhibit no strong correlation with \ion{H}{1} column density or impact 
parameter. Indeed, there is evidence for significant enrichment of the gas even beyond the 
estimated virial radius of the halos hosting quasars. In QPQ7, we conservatively 
estimated a gas minimum metallicity of 1/10 solar for the CGM and inferred metal masses 
$>10^7\msol$. In QPQ5, we estimated a maximum metallicity of 1/2 solar and inferred an upper 
bound on metal masses $<10^9\msol$. 
The results presented here indicate metallicities that are several 
times higher. With the Si measurements scaled to O assuming solar relative abundance, we 
derive a median metal column density of $10^{16.7} \cm{-2}$ for the 7 sightlines within 200~kpc. 
This corresponds to a total oxygen mass in the cool CGM of 
$M_{\rm O}^{\rm CGM}\approx4.9\times10^8\msol\,(\mrphys/160\,{\rm kpc})^2$. 
Furthermore, assuming that oxygen accounts for 44\%\ of the metal mass \citep{Asplund+09}, we 
refined our mass estimate to be $M_{\rm metal}^{\rm CGM} 
=1.1\times10^9\msol\,(\mrphys/160\,{\rm kpc})^2$. 
Theoretical calculations of core-collapse supernovae nucleosynthesis yields 
\citep{Sukhbold+16,Nomoto+06} predict an oxygen yield 
mass fraction of 0.008, scaled to a Kroupa initial mass function. Thus, the total stellar mass 
that must have formed to synthesize all the oxygen in the cool CGM, not accounting for oxygen 
locked up in stellar content, in the ISM or in the warm-hot/hot phase CGM, has to be 
$M_* >6.1\times10^{10}\msol\,(\mrphys/160\,{\rm kpc})^2$. To form this stellar mass by $z=2.4$, the 
median $z_{\rm fg}$ of the QPQ8 sample, the average star formation rate required is 
$>34\msol\,{\rm yr}^{-1}$ if the galaxies first formed at $z\sim6$. From halo abundance matching 
techniques \citep{Behroozi+13}, the typical host galaxy stellar mass of the QPQ8 sample is 
$M_{\rm gal}=(6\pm3)\times10^{10}\msol$. \cite{Whitaker+14} gives the star formation rate of the 
star forming sequence at $z\sim2\text{--}3$ at this galaxy mass to be $100\msol\,{\rm yr}^{-1}$. 
Our estimate of the minimum average star formation rate required, and hence the total oxygen mass 
in the cool CGM, is consistent with the instantaneous total star formation rate inferred from 
these scaling relations. 

We have also examined the abundance ratios of a subset of the elements, restricting this analysis 
to low ions to minimize the effects of ionization. We focus on O/Fe which is a chemical signature 
of the supernovae that have enriched the gas. High values are indicative of massive stellar 
nucleosynthesis, i.e. core-collapse supernovae, whereas low values imply substantial enrichment by 
Type~Ia supernovae \citep[e.g.][]{Tinsley79}.

Figure~\ref{fig:OFe} presents [O/Fe] estimates after adopting the ionization corrections from our 
favored models against the \nhi\ measurements. Nearly all of the measurements indicate a 
supersolar O/Fe ratio, implying a significant fraction of the CGM surrounding quasars must have an 
enhanced $\alpha$/Fe abundance. The obvious exception is subsystem F from the J1144+0959 pair. 
Perhaps not coincidentally, this subsystem has the highest estimated $U$ parameter of those that 
exhibit O or Fe, and therefore has the largest ionization correction for \psol{O}{0}{Fe}{+}. The 
corrected [O/Fe] value is still subsolar, although we caution that the uncertainty is many tenths 
dex given the high $U$ value. Furthermore, the [O/Fe] values from QPQ8 show a large dispersion 
spanning two orders of magnitude. 

In summary, the cool gas surrounding $z\sim2$ massive galaxies hosting quasars to $\approx200$~kpc 
is highly enriched and $\alpha$-enhanced. This implies that the gas expelled from these galaxies 
(and their progenitors) was enriched primarily by core-collapse supernovae. Furthermore, if 
supernovae explosions are the principal factor in transporting metals to the CGM, we may speculate 
that core-collapse supernovae dominate in high$-z$ massive galaxies hosting quasars.

The high [$\alpha$/Fe] enhancement suggest a star formation history similar to elliptical galaxies 
\citep[e.g.][]{Matteucci94,ConroyGravesvanDokkum14}, which these quasar host galaxies are expected 
to evolve into. Currently there are three competing models that explain [$\alpha$/Fe] enhancement 
in the {\it stellar} content of elliptical galaxies (1) selective loss of Fe via galactic winds 
\citep{Trager+00b}; (2) a short starburst that ceased before Fe produced from Type~Ia supernovae 
became available for incorporation into new stars; and (3) a variable initial mass function 
flattened at the high mass end \citep{Thomas99}. Our finding that the CGM shows a high 
[$\alpha$/Fe] enhancement strongly disfavors the scenario where more Fe than $\alpha$ elements is selectively ejected through outflows. It is more likely that the supersolar $\alpha$/Fe ratios 
reflect an intrinsic nucleosynthetic enhancement. This conclusion is further strengthened by X-ray 
observations of the hot interstellar medium of early type galaxies 
\citep{Loewenstein+10,Loewenstein+12,Konami+14} and the hot intracluster medium 
\citep{Sato+07,Mushotzky+96} where enhanced $\alpha$/Fe ratios are similarly found 
\citep[but see][]{Simionescu+15}. Furthermore, 
if without a flatter initial mass function, the starburst driven galactic scale winds must occur 
before the Fe production from Type~Ia supernovae starts to become important, hence we expect the 
duration of starburst lasts less than 1 Gyr. In turn this implies a minimum star formation rate 
$>55\;{\rm M}_\odot\;{\rm yr}^{-1}$ based on the inferred total stellar mass formed for 
synthesizing all oxygen in the cool CGM.

\subsection{Surface Density Profiles}
\label{sec:surfacedensity}

\begin{figure}
\includegraphics[width=3.5in]{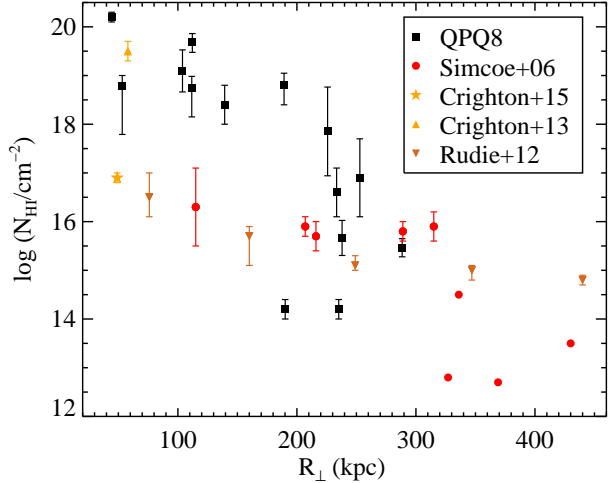} 
\caption{Scatter plots of $N_{\rm HI}$ measurements against projected quasar pair separation
calculated at $z_{\rm fg}$. The \ion{H}{1} exhibits a statistically significant decline in
surface density with $R_\perp$. Strong \ion{H}{1} absorption exists even beyond the estimated
virial radius $\approx\;160$~kpc. For comparison we also plot CGM $N_{\rm HI}$ measurements of
$z\sim2\text{--}3$ galaxies from the literature as a function of projected distance from the 
galaxies. At $R_\perp\leq200$~kpc, the $N_{\rm HI}$ values of QPQ8 predominate those of coeval 
galaxies.
}
\label{fig:NHIvsRperp}
\end{figure}

\begin{figure}
\includegraphics[width=3.5in]{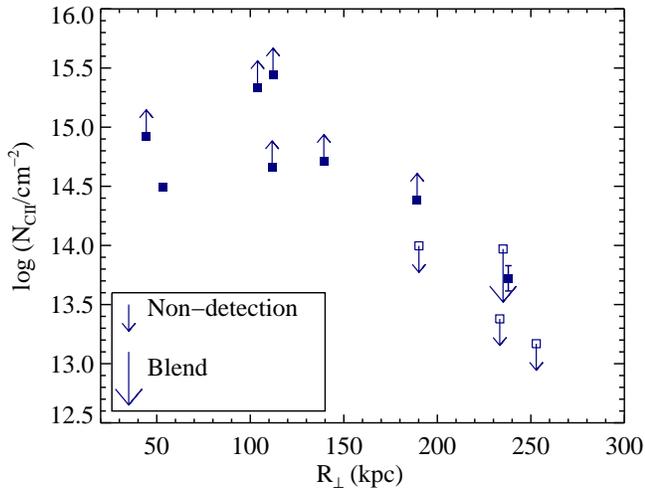} 
\caption{Scatter plots of $N_{\rm CII}$ measurements against $R_\perp$. Similar to \ion{H}{1}, the
 \ion{C}{2} ion exhibits a statistically significant decline in surface density with $R_\perp$. The
observed low ion absorption thus traces the CGM gas of the foreground quasar.
}
\label{fig:NCIIvsRperp}
\end{figure}

Figure~\ref{fig:NHIvsRperp} presents scatter plots of the \ion{H}{1} column density measurements 
against projected quasar pair separation $R_\perp$ calculated at $z_{\rm fg}$. 
Figure~\ref{fig:NCIIvsRperp} presents scatter plots of the \ion{C}{2} column density measurements 
against projected quasar pair separations $R_\perp$ calculated at $z_{\rm fg}$. Both ions exhibit 
a decline in surface density with increasing $R_\perp$. Statistically a Kendalls's tau test gives 
a $99.8\%$ probability that the null hypothesis of no correlation between $N_{\rm HI}$ and 
$R_\perp$ is ruled out. A generalized Kendall's tau test gives a $98.7\%$ probability that the null 
hypothesis of no correlation between $N_{\rm CII}$ and $R_\perp$ is ruled out, and we expect the 
correlation to be stronger because the $N_{\rm CII}$ absorptions detected at small $R_\perp$ are 
saturated and  $N_{\rm CII}$ is not detected at 3-$\sigma$ confidence at large $R_\perp$. The 
observed low ion absorption is thus tracing the CGM gas of the foreground quasar. We further note 
that strong \ion{H}{1} absorption exists even beyond the estimated virial radius $\approx160$~kpc.

For comparison, in Figure~\ref{fig:NHIvsRperp} we also plot the work of 
\cite{Simcoe+06,Rudie+12,Crighton+13,Crighton+15} of $N_{\rm HI}$ as a function of $R_\perp$ in 
the CGM of $z\sim2\text{--}3$ galaxies, in which $R_\perp$ refers to the transverse, physical 
distance. Their work uses background quasars as sightlines to probe the 
CGM of foreground galaxies. Their galaxies typically have  
$M_*\sim1\times10^{10}\;{\rm M_\odot}$, which characterize the typical stellar mass in star 
forming galaxies at $z\sim2\text{--}3$, while the typical host galaxy of the QPQ8 sample inferred 
from halo abundance matching has $M_*\sim6\times10^{10}\;{\rm M_\odot}$, i.e. much more massive. 
At 
$R_\perp\;\approx200\text{--}300$~kpc, the QPQ8 $N_{\rm HI}$ values smoothly join those measured 
for the more typical $z\sim2\text{--}3$ star forming galaxy population, which are still 
significantly enhanced compared to the IGM absorption at $\gtrsim300$~kpc. At 
$R_\perp\lesssim200$~kpc, we find the $N_{\rm HI}$ values of the QPQ8 sample
are significantly larger that that of the coeval 
galaxies, as inferred previously from equivalent width measurements (QPQ5, QPQ6). 

We calculated the \ion{H}{1} mass in the cool CGM by considering the $R_\perp$ values as annular 
bins. We calculated the mass in the $i$-th annulus by 
$M_i=N_{\rm HI}(R_i)m_{\rm H}\pi(R_i^2-R_{i-1}^2)$ and summed up the mass in all annuli. The total 
\ion{H}{1} within 160~kpc is then found to be 
$M^{\rm CGM}_{\rm HI} = 1.3\times10^{10}\;{\rm M_\odot}$. 
Even without any ionization correction, we infer the baryonic mass of the cool CGM approaches the 
stellar mass. 
In a highly clustered environment, the observed gas could be contributed by CGM of neighboring 
gaalxies. In this regard, the \ion{H}{1} mass estimate can be considered an upper limit. 

Concerning the \ion{C}{2} ion, the absorption within $R_\perp<200$~kpc is strong and saturated, 
with a median $N_{\rm CII}>14.7\;{\rm cm^{-2}}$, indicating a substantial metal mass. At 
$R_\perp>200$~kpc, we have one solid detection of \ion{C}{2} in the J0932+0925 system, two 
non-detections in the J1038+5027 system and the J1627+4605 system  
shown as 3-$\sigma$ upper limits in Figure~\ref{fig:NCIIvsRperp}, and one system J0409-0411 whose 
\ion{C}{2} is blended with unrelated intergalactic absorption, depicted by a large downward arrow. 
Our results quantitatively assert the conclusion of QPQ7 that the gas surrounding massive, 
$z\sim2$ galaxies hosting quasars represents the pinnacle of the cool CGM, in terms of the neutral 
hydrogen mass and the enrichment. 

We detect cool enriched gas transverse to the sightline to the foreground quasars to at least 
$\approx200$~kpc. Our results indicate the quasar plays a minor role in producing the cool CGM, and 
our argument is as follows. Observations of spatial clustering of quasars \citep{HaimanHui01,
MartiniWeinberg01,Martini04}, observations of the transverse proximity 
effect in \ion{He}{2} \citep{Jakobsen+03}, 
as well as galaxy merger simulations that include supermassive black holes
\citep{Hopkins+05}, all constrain the quasar lifetime to be $10^6\text{--}10^8$~years, with a 
preference for $10^7$~years.
Should the velocities represent outflows, given that the line-of-sight 
velocity dispersion is $249\;{\rm km\;s}^{-1}$ in \ion{C}{2} and $495\;{\rm km\;s}^{-1}$ in 
\ion{C}{4}, as discussed in Section~\ref{sec:kin}, even if the observed quasar episode has been 
active for $10^8$~years and accelerate material to $500\;{\rm km\;s}^{-1}$, this gas would only 
reach $50$~kpc. It is unlikely the observed quasar episode alone could transport all of this cool 
material from the interstellar medium of the galaxies. 
Furthermore, we found no statistically significant trend between \ion{C}{2} column density and 
quasar luminosity $L_{\rm Lbol}$ or the UV enhancement factor $g_{\rm UV}$, contrary to the claims 
in \cite{Johnson+15} where they investigated \ion{Mg}{2} in $z\sim1$ quasar CGM. 
We caution the readers, however, that the mean ${L_{\rm bol}}$ of QPQ8 is 
$10^{46.4}\;{\rm erg\;s^{-1}}$, which is nearly an order of magnitude higher than the mean 
${L_{\rm bol}}$ of the \cite{Johnson+15} sample at $10^{45.5}\;{\rm erg\;s^{-1}}$. 

\begin{figure}
\includegraphics[width=3.5in]{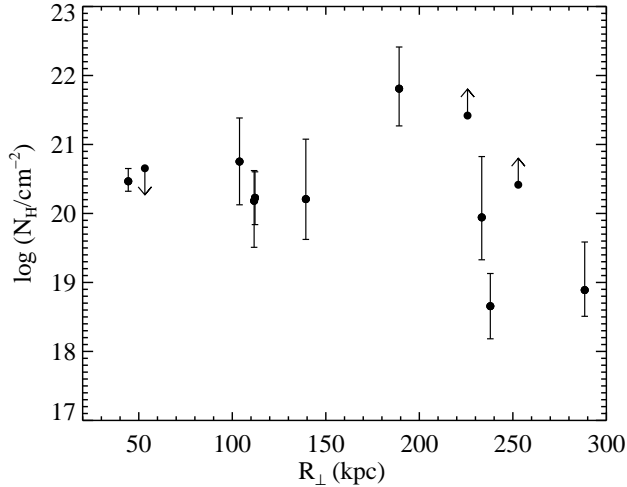} 
\caption{The total H column from ionization state modeling as a function of $R_\perp$. We see
little evolution in $N_{\rm H}$ up to $\approx200$~kpc, as both $N_{\rm HI}$ and the neutral
fraction $x_{\rm HI}$ anticorrelate with $R_\perp$. The median
$N_{\rm H}=10^{20.5}\;{\rm cm^{-2}}$.
}
\label{fig:NHvsRperp}
\end{figure}

As described in Section~\ref{sec:cldy}, we modeled the ionization state and calculated $N_{\rm H}$ 
for each absorption system. In Figure~\ref{fig:NHvsRperp} we plot $N_{\rm H}$ as a function 
of $R_\perp$. Consistent with expectation, we do not observe any significant 
variation of $N_{\rm H}$ with $R_\perp$
up to 
$\approx200$~kpc, as both $N_{\rm HI}$ and the neutral fraction $x_{\rm HI}$ anticorrelate with 
$R_\perp$. 

\begin{figure}
\includegraphics[width=3.5in]{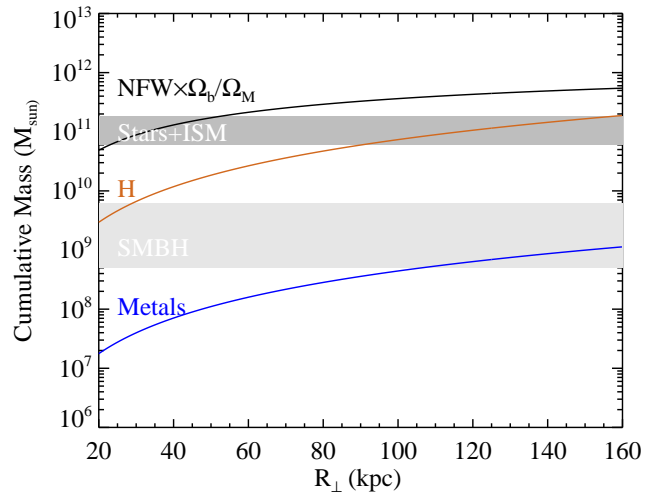} 
\caption{Using the median $N_{\rm H}$ and the median [M/H] within 200~kpc, we constructed
cumulative mass profiles of total H and metals in the cool CGM. For reference we also included the
expected baryonic mass projected profile of an NFW halo with $c=4$ and dark matter mass
$M_{\rm DM}=10^{12.5}\;{\rm M}_\odot\;(R_\perp/160\;{\rm kpc})^2$. We also plotted the typical
host galaxy mass, as well as the range of supermassive black hole mass of the QPQ8 sample. We
estimated the total mass of the cool CGM as
$M_{\rm cool}^{\rm CGM}\approx1.9\times10^{11}\;{\rm M}_\odot\;(R_\perp/160\;{\rm kpc})^2$. This
accounts for 1/3 of the total expected baryonic mass.
}
\label{fig:cum_HvsRperp}
\end{figure}

Using the median $N_{\rm H}=10^{20.5}\;{\rm cm^{-2}}$ and the median [M/H] = $-0.6$ 
(Figure~\ref{fig:MH}) within 200~kpc, we constructed cumulative mass profiles of total H and 
metals in the cool CGM, and the results are plotted in Figure~\ref{fig:cum_HvsRperp}. 
For reference, we also included the expected baryonic mass 
projected profile of an NFW halo with concentration parameter $c=4$ \citep{NavarroFrenkWhite97} 
and dark matter mass $M_{\rm DM}=10^{12.5}\;{\rm M}_\odot\;(R_\perp/160\;{\rm kpc})^2$. We assumed 
the cosmic baryon fraction 0.17. 
We 
also plotted the typical host galaxy mass of the QPQ8 sample using halo abundance matching 
techniques \citep{Behroozi+13} and assuming a 50\% gas fraction, as well as the range of 
supermassive black hole mass of the QPQ8 sample, calculated 
from the bolometric luminosity of the quasars and assuming an eddington ratio $f_{\rm Edd}=0.1$. 
We estimated the total mass of the cool CGM as 
$M_{\rm cool}^{\rm CGM}(R_\perp)\approx1.9\times10^{11}\;{\rm M_\odot}\;(R_\perp/160\;{\rm kpc})^2
$. 
{\rm Since the median $N_{\rm H}$ of the sample is not sensitive to one system with the highest 
$N_{\rm H}$, namely J1144+0959, our estimate of the cool CGM mass is representative of the sample. 
}
In QPQ7 we estimated an infall time $\tau_{\rm infall}\lesssim1$~Gyr. The corresponding cool gas 
inflow rate $M_{\rm cool}^{\rm CGM}/\tau_{\rm infall}$ exceeds 
$100\;{\rm M_\odot}\;{\rm yr^{-1}}$. This is comparable to the star formation rate of massive 
star forming galaxies at $z\sim2$, and is enough to fuel star formation for at least 1~Gyr. The 
QPQ8 results further strengthen the conclusion of QPQ3 and QPQ7 that quasars are unlikely to 
quench star formation at $z\sim2$. 

It can be seen in Figure~\ref{fig:cum_HvsRperp} that at the virial radius the cool gas fraction 
accounts for only $\approx1/3$ of the total expected baryonic mass. Together with the galaxy's 
stellar and gas mass, they account for $\approx56\%$ of the total baryonic budget. Modern X-ray 
observations of the hot intracluster medium at $z\gtrsim1$ \citep{Baldi+12,Andreon12} report 
metallicities $\approx1/3$ solar, which is consistent with the median [M/H] = $-0.6$ dex found for 
the cool CGM in this work. If we assume [M/H] = $-0.6$ is a good representation of the metallicity 
of the cool and the hot phase CGM, then likewise to the total H, at the virial radius the cool CGM 
only accounts for 1/3 of the total expected metal mass. A massive reservoir of warm/hot, enriched 
gas within the QPQ halos is thus required to complete the baryonic mass budget. Such warm/hot 
intracluster medium is predicted to be already fully in place at $z\sim2$ in massive halos 
\citep[e.g.][]{Fumagalli+14}. 

\subsection{Volume Density and Linear Size of The Absorbers}
\label{sec:volumedensity}

From our analysis, there are two standard methods for assessing the volume density $n_{\rm H}$ of 
the gas and, in turn, offering an assessment of the linear scale $\ell$ of the absorption system. 
One approach is to assume an intensity for the incident radiation field intensity $J_\nu$ and then 
convert the estimated ionization parameter $U$ into an estimate of \nhv. The uncertainties in this 
analysis are large: we have no direct constraint on $J_\nu$, the error in $U$ is substantial, and 
systematic uncertainties in the photoionization modeling influence this treatment. Therefore, we 
proceed in a conservative fashion.

For $J_\nu$, we consider two limits:
 (1) the gas is illuminated by only the EUVB, where we adopt the spectral energy distribution of 
the cosmic ultraviolet background normalized to the \cite{FaucherGiguere+08} estimate for the 
effective IGM \lya\ opacity;
 (2) the gas is fully illuminated by the foreground quasar which shines isotropically and
  the gas is at a distance $r$ equal to the impact parameter $R_\perp$.
The first limit sets a lower bound to $J_\nu$ and therefore the density while the latter sets
an upper limit to \nhv.  For the EUVB case, we would require $\mnhv \approx 10^{-3} \cm{-3}$. This
is a somewhat low value for overdense and optically thick gas, but is comparable to 
predictions from simulations \citep{RosdahlBlaizot12}. A Ly$\alpha$ fluorescence analysis 
\citep{ArrigoniBattaia+16} suggests $\mnhv \approx0.6\times10^{-2}\;{\rm cm^{-3}}$, which
is not that far from our result given  the uncertainties. 
Furthermore, the 
resultant length scale $\ell \equiv N_{\rm H}/\mnhv$ would exceed 100~kpc per cloud for the 
typical $N_{\rm H}^{\rm QPQ8}\approx10^{20.5}\;{\rm cm}^{-2}$. Although not strictly ruled out for 
the majority of our sample (see below), we suspect the densities are significantly higher and that 
$\ell$ is correspondingly smaller. If the quasar emits isotropically and $r \approx \mrphys$, then 
the implied densities are $\mnhv \approx 1-10 \cm{-3}$. Such values are characteristic of the 
diffuse interstellar medium of galaxies. In this extreme, we recover $\ell\approx1-10$~pc, which 
has been previously reported for some absorption systems \citep{Sargent+79,Simcoe+06} and yet may 
be considered extreme.
In another study, metal enriched gas clumps in the circumgalactic medium at 
$z\sim2.5$ are found to have sizes $100\text{--}500\;{\rm pc}$ \citep{Crighton+15}.


An independent assessment of the density may be obtained from analysis of the fine structure 
absorption of C$^+$ and Si$^+$ \citep[e.g.][QPQ3]{Prochaska99,SilvaViegas02}. Under the 
assumption that electron collisions dominate the excitation of these ions in our photoionized gas, 
the ratio of the excited state to the ground state yields a precise estimate for $n_{\rm e}$. We 
have previously assessed in QPQ3 whether the gas could be excited indirectly by UV pumping 
\citep[e.g.][]{ProchaskaChenBloom06}, and find the quasars are too faint. From the QPQ8 dataset, 
we report two positive detections and several upper limits of \ion{C}{2*}~1335 absorption. 
Following the methodology in QPQ3, we assumed an electron gas temperature of 20000~K and obtained 
$n_{\rm e}=\frac{106}{2(N({\rm C}^+_{J=1/2})/N({\rm C}^+_{J=3/2}))-1}$, where $J=1/2$ represents 
the ground state and $J=3/2$ is the first excited level.
In 
Table~\ref{tab:Usumm} we present the resultant $n_{\rm e}$ estimates and the corresponding linear 
size per cloud $\ell=N_{\rm H}/n_{\rm H}$, where $n_{\rm H}=n_e/(1-x_{\rm HI})$, not compensating 
for ionized helium contribution to electron density. For the positive 
detections, we find $n_e > 10 \cm{-3}$. For subsystem~F of J1420+1603FG, $n_e$ even exceeds the 
electron density expected for isotropic quasar illumination $n_{\rm QSO}$ 
(Section~\ref{sec:cldy}), suggestive of local sources of radiation. 

%

\subsection{Peculiarities of the Quasar Circumgalactic Medium} 
\label{sec:peculiarities} 

The environments of $z \sim 2$ quasars must be considered extreme
compared with co-eval Lyman break galaxies, at least in terms of
overdensity and possibly an elevated, local radiation field.
In this respect, the high metallicities, large CGM gas masses, and
complex kinematics may follow naturally.  As such, we are further
motivated to search for peculiar features in the absorption line data
that occur rarely, if at all, along random sightlines. We now
summarize a series of examples which on their own are remarkable and
together paint a highly unusual picture of the $z \sim 2$ universe.

\subsubsection{Evidence for Elevated Radiation Field}
\label{sec:peculiar_radiation}

\begin{figure}
\includegraphics[width=3.5in]{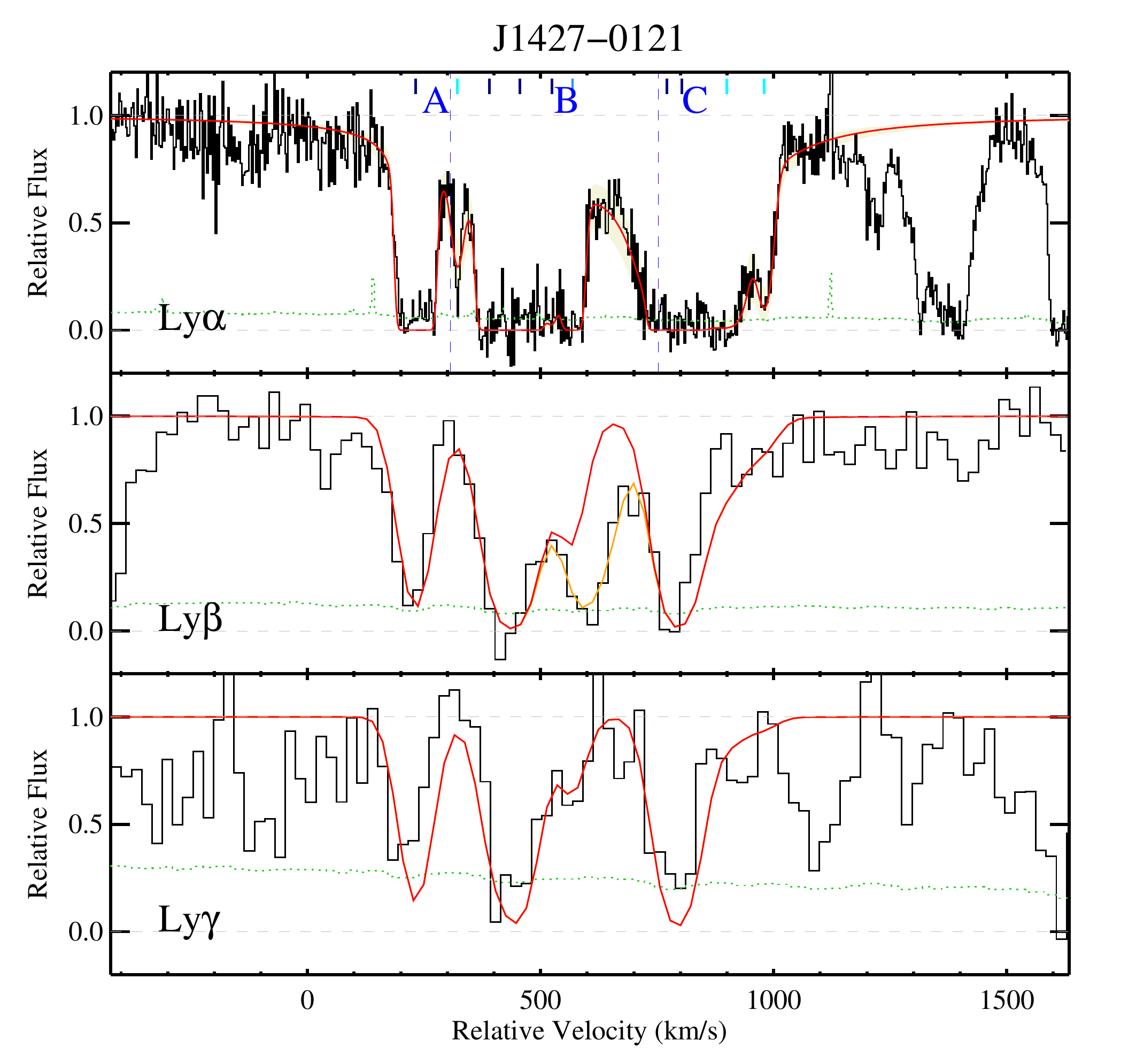} 
\caption{\ion{H}{1} Lyman series absorption profiles for J1427-0121. The black histograms show the 
Lyman series identified in the J1427-0121BG spectrum at velocities consistent with J1427-0121FG. 
The navy ticks mark the centroid redshifts for \ion{H}{1} components traced by low metal ions. The 
cyan ticks mark \ion{H}{1} components not associated with metal ions. The red curve is the Voigt 
profile modeling fit of summing up all \ion{H}{1} components associated with J1427-0121FG. The 
orange curve is the modeling fit including Ly$\alpha$ forest contamination. The positive flux at 
Ly$\alpha$ suggests $N_{\rm HI}<10^{14}\;{\rm cm^{-2}}$ whereas the positive detections of low 
ions and the high opacities at Ly$\beta$ and Ly$\gamma$ indicate 
$N_{\rm HI}\gg10^{14}\;{\rm cm^{-2}}$. We suspect the flux at Ly$\alpha$ is unresolved Ly$\alpha$ 
emission as predicted for gas illuminated by the ionizing flux of a nearby quasar. The green 
histogram presents the $1\sigma$ noise in the normalized flux.
}
\label{fig:J1427_lyseries}
\end{figure}

The most extreme case, in our opinion, is the putative \lya\ emission in subsystem~A of
J1427-0121FG. As shown in the Appendix, the positive flux at \lya\ suggests 
$\mnhi < 10^{14} \cm{-2}$ whereas the positive detections of low ions and the high opacities at 
\lyb\ and Ly$\gamma$ indicate $\mnhi \gg 10^{14} \cm{-2}$. Although metal bearing systems with 
$\mnhi\sim10^{14}\cm{-2}$ have been reported previously \citep[e.g.][]{BoksenbergSargent14}, these 
are very rare and generally dominated by \ion{C}{4} and \ion{O}{6} absorption 
\citep{SimcoeSargent+04}. We strongly suspect that the flux at \lya\ is unresolved \lya\ emission 
as predicted for gas illuminated by the ionizing flux of a nearby quasar 
\citep[e.g.][]{Cantalupo+14,QPQ4,Hennawi+15}. In Figure~5 of QPQ4, we showed the PSF subtracted 
Gemini spectrum of the J1427-0121 pair and there is no revealing residual Ly$\alpha$ emission to 
sensitivity limit 
${\rm SB}_{1\sigma}=0.14\times10^{-17}\;{\rm erg\;s^{-1}\;cm^{-2}\;arcsec^{-2}}$.  
Thus this emission is spatially unresolved, i.e. it is consistent with the point spread function 
of the foreground quasar. 
This phenomenon, however, is rare in our quasar pair sample 
(QPQ4) and we identify no other example in the 8~cases from QPQ8 where we have coverage of \lyb\ 
or higher order Lyman series lines. 

To estimate the Ly$\alpha$ flux more accurately we prefer a lower dispersion spectrum than our 
echelle/echellette resolution follow-up spectra. Based on the SDSS spectrum continuum, we 
estimated the Ly$\alpha$ flux 
$F_{\rm Ly\alpha}=1.5\times10^{-17}\;{\rm erg\;s^{-1}\;cm^{-2}}$. Since the size of the emitting 
cloud has to be smaller than typical seeing disk of $1''$, the Ly$\alpha$ surface brightness 
${\rm SB}_{\rm Ly\alpha}>1.5\times10^{-17}\;{\rm erg\;s^{-1}\;cm^{-2}\;arcsec^{-2}}$. 
If the absorber is optically thin, we may estimate the surface brightness of subsystem~A following 
the formalism of QPQ4. Specifically, we assumed a volume density that is the same as that of 
subsystem~C, which is calculated to be $n_{\rm H}=16\;{\rm cm^{-3}}$ from the 
$N_{\rm CII}/N_{\rm CII*}$ ratio. We adopted the $N_{\rm H}$ value from photoionization modeling. 
The expected 
${\rm SB}_{\rm Ly\alpha}\approx3.3\times10^{-17}\;{\rm erg\;s^{-1}\;cm^{-2}\;arcsec^{-2}}$.  
If the absorber is optically thick, the formalism of QPQ4 then gives 
${\rm SB}_{\rm Ly\alpha}\approx9.5\times10^{-17}\;{\rm erg\;s^{-1}\;cm^{-2}\;arcsec^{-2}}$. Since 
the measured Ly$\alpha$ flux corresponds to lower Ly$\alpha$ surface brightness, the 
characteristic angular size of the absorber should be smaller than the typical seeing disk $1''$, 
corresponding to $<8\;{\rm kpc}$. Furthermore, again following the formalism of QPQ4, should this 
absorber be optically thin, this level of Ly$\alpha$ fluorescence would imply 
$N_{\rm HI}=1.1\times10^{17}\;{\rm cm}^{-2}$, and hence the optically thin assumption breaks down. 
The absorber probably lies in the transition from the optically thin to the optically thick limit. 
Alternatively, Ly$\alpha$ emission can result from collisional excitation. In the absence 
of significant metal enrichment, collisionally excited Ly$\alpha$ emission is the primary coolant 
for $T\sim10^4$~K gas. Collisional excitation of neutral hydrogen requires a non-negligible neutral 
fraction, and the resulted surface brightness is exponentially sensitive to gas temperature via 
the collisional excitation rate coefficient $q_{\rm 1s\rightarrow2p}\;(T)$. Thus the electron 
density estimated will be uncertain by orders of magnitude. 
However, we expect recombination dominates cooling radiation. \cite{Hennawi+15} reports the 
Ly$\alpha$ emission from 
collisional excitation is at most 20\% of the Ly$\alpha$ emission from recombination, for the  
nebula illuminated by a quasar quartet discovered by them. 


In previous works (QPQ2, QPQ4, QPQ6), we have argued that quasars emit anisotropically owing to the 
high incidence of optically thick, cool gas in the transverse direction relative to that observed 
along the direct sightline. With the high fidelity of our QPQ8 sample, we may search for more 
subtle signatures that at least a portion of the transverse sightline is illuminated by a luminous 
and hard radiation field. One such signature would be the \ion{N}{5} doublet 
$\lambda\lambda$~1238,~1242, which may also trace a distinct warm-hot phase. With an ionization 
potential of 5.5~Ryd, production of the N$^{4+}$ ion requires a gas with $T \gtrsim 10^5$\,K for 
collisional ionization or a high intensity of extreme UV photons. Indeed, \ion{N}{5} is very 
rarely detected in random, intervening systems along quasar sightlines \citep{Fox+09} or even in 
the CGM of Lyman break galaxies \citep{Turner+14}, but is more 
frequently detected in gas associated to the quasar itself. For each pair in the QPQ8 sample, we 
have coverage of the \ion{N}{5} doublet but these lines are free of the \lya\ forest in only 
3~cases. 
Of these, we report one positive detection (J1026+4614). Our photoionization model for 
this gas yields $\log U\approx-1.9$~dex (see the Appendix). To achieve this $U$ value with the 
EUVB, one would require a very low gas density $n_{\rm H}\approx10^{-4}\cm{-3}$ and an absorber 
size $\ell=N_{\rm H}/n_{\rm H}\approx13$~kpc. Given the rarity of \ion{N}{5} detections along 
random quasar sightlines, we contend this gas is illuminated by the foreground quasar, 
although the flux need not be as bright as the observed along our sightline, or the flux may also 
exhibit temporal variability. 
The absence of \ion{N}{5} in the other 2 pairs with coverage outside the \lya\ forest and the 4  
cases with minimal IGM blending suggests that only small portions of the transverse sightlines are 
illuminated. This conclusion is tempered by the possible underabundance of N but large deviations 
from solar relative abundances tend to occur in gas with metallicities much lower than those 
estimated for this CGM \citep[e.g.][]{HenryEdmundsKoppen00}. 
Taking the median values of [M/H] = $-0.6$ and $N_{\rm H}=10^{20.5}\;{\rm cm}^{-2}$,
using Cloudy we estimated for the QPQ8 systems 
$N_{\rm NV}=10^{9.3}\text{--}10^{14.7}\;{\rm cm}^{-2}$, varying $\log U$ from $-4$ to $-2$. Hence 
only for high values of $U$ parameter will \ion{N}{5} become strong enough to detect in most of 
our echellette resolution data, whose detection limit is on average 
$N_{\rm HV}<10^{13.6}\;{\rm cm^{-2}}$. 
Another doublet that may trace a quasar illuminated gas or a 
distinct warm-hot phase is the \ion{O}{6} doublet $\lambda\lambda$~1031,~1037. However, in the 
QPQ8 data we are unable resolve this doublet from coincident absorption in the Ly$\alpha$ forest. 

The subsystem~F of J1144+0959FG exhibits significant low ion absorption, e.g. \ion{O}{1} and
\ion{Fe}{2}, and has an estimated $\mnhi \approx 10^{18} \cm{-2}$. These properties characterize
optically thick, partially ionized Lyman limit system. The very strong medium and high ions,
however, yield an ionization parameter $\log U \approx -2$,
giving a neutral fraction of $x_{\rm HI} \approx 10^{-3}$ and a total
$N_{\rm H} \approx 10^{21} \cm{-2}$. If the volume density is close to the rough upper bound 
$n_{\rm QSO}=10^{-0.3}\;{\rm cm^{-3}}$, together these suggest an elevated radiation field.

Given the rarity of such phenomena along random quasar sightlines, we may hypothesize that where 
one observes similar cases an AGN may be present or recently was shining 
\citep{OppenheimerSchaye13}. In this study, evidence for an elevated radiation field is found for 
only 2 out of 14 quasar environments studied. We do not believe that the quasar CGM is 
qualitatively peculiar relative to other massive galaxies, however to date QPQ is the only 
statistical sample to characterize the physical properties of the CGM of massive galaxies at 
$z\sim2$. For the typical bolometric luminosities of these quasars, observations imply star 
formation rates comparable to inactive star forming galaxies of similar masses 
\citep[e.g.][]{Rosario+13}. 

\cite{Farina+13} studied a mass-controlled sample and found that, if the mass of the galaxies is 
taken into account as an additional parameter that influences the extent of the gaseous halos, the 
distribution of \ion{Mg}{2} absorbers around quasars is consistent with that for normal galaxies. 

\subsubsection{Evidence for Elevated Volume Density}
\label{sec:peculiar_density}

In Section~\ref{sec:volumedensity}, we assessed the density of the gas through analysis of 
\ion{C}{2*}~1335 absorption and reported two positive detections. Among these, subsystem~F of 
J1420+1603FG has a \ion{C}{2*}~1335 optical depth exceeding 0.8 at its peak and an excited to 
ground state ratio of $\approx1.1$. To our knowledge, this exceeds any such measurement along an 
extragalactic sightline including GRBs \citep{ProchaskaChenBloom06}, whose gas is radiatively 
excited.
In turn, this requires an electron density $n_{\rm e} = 10^{2.2} \cm{-3}$
which defies conventional wisdom for diffuse CGM gas \citep{Werk+14}.
Similar inferences, however, have been drawn from the \lya\ nebulae surrounding $z \sim 2$ quasars 
\citep{Cantalupo+14,Hennawi+15,ArrigoniBattaia+15a,ArrigoniBattaia+15b}
and the extended narrow emission line regions and nebulae of other AGNs and radio galaxies 
\citep{Humphrey+07,Dey+05,Stockton+02,FuStockton07}.
Given that the inferred absorption pathlength is 
$\ell \lesssim 1$~pc, one must invoke a large population of such clouds to explain our 
intersecting even one given $(\ell/\mrphys)^2 < 10^{-10}$. Furthermore, small and dense clumps 
will be disrupted by hydrodynamic instabilities as they move through a hotter CGM phase, unless 
the hot plasma pressure is able to confine the compact clumps \citep{ArrigoniBattaia+15a,
Crighton+15}. 
Furthermore, such a small cloud is 
representative of the dense knots inside giant molecular clouds. However, the absence of 
Lyman-Werner bands implies $N_{\rm H_2}<10^{18}\;{\rm cm}^{-2}$.



\section{Summary and Future Outlook}
\label{sec:summary}

In the ``Quasars Probing Quasars'' series of papers, we have
introduced a novel technique to study the 
CGM surrounding quasar host galaxies, which provides clues to the physics of massive galaxy 
formation and the nature of quasar feedback. We mined the existing quasar catalogs for closely 
projected, physically unassociated quasar pairs and confirmed them with follow up spectroscopy. 
From a total of $\approx700$ confirmed pairs, we selected 14 pairs with projected separations 
$<300$~kpc and high dispersion, high S/N data, to study the CGM surrounding quasars at 
$z\sim$2--3. We analyzed the velocity fields of the absorbing gas, the \ion{H}{1} and metal ion 
column densities and the ionization state characterized by the ionization parameter $U$. These 
analyses constrain the physical state of the cool gas near the foreground quasars, 
including its kinematics, chemical abundance patterns, surface density profiles, volume density, 
size of the absorbers and the intensity of the impinging radiation field, as well as to test for 
the presence of a hotter phase of CGM. Our key findings are as follows. 

Model-Independent Constraints:

\begin{itemize}
\item
The low (e.g. \ion{C}{2}) and high (e.g. \ion{C}{4}) ions roughly trace each other in velocity 
structure. The velocity widths exceed all previous measurements of gas surrounding any galaxy 
populations, with a RMS $\sigma_v=495\;{\rm km\;s^{-1}}$ for \ion{C}{4} and 
$\sigma_v=249\;{\rm km\;s^{-1}}$ for \ion{C}{2} . The velocity centroids of the absorption 
profiles are frequently biased to positive (redshifted) velocities from the systemic redshift of 
the foreground quasars.  
\item 
The \ion{H}{1} and low metal ion (traced by \ion{C}{2}) surface density declines with $R_\perp$. 
\ion{H}{1} absorption is strong even beyond the estimated virial radius. The \ion{H}{1} column 
densities are significantly larger than those of co-eval galaxies.
\item 
In one case, subsystem~A of J1427-0121FG, we suspect the presence of unresolved Ly$\alpha$ 
emission, a prediction for gas illuminated by the foreground quasar. In another case, J1026+4614, 
we detected \ion{N}{5} absorption and we contend the gas is at least partially illuminated by the 
foreground quasar. The non-detection of \ion{N}{5} in all other absorption systems, however, 
suggests that only small portions of the transverse sightlines are illuminated and the flux needs 
not be as high as along the line of sight. 
\end{itemize}

Model-Dependent Constraints: 

\begin{itemize}
\item
The ionization parameter $U$ positively correlates with projected distance from the foreground 
quasar. This runs contrary to expectation should the foreground quasar dominate the ionizing 
radiation field. 
\item 
The CGM is significantly enriched even beyond the estimated virial radius of the host dark matter 
halos ($\approx$~160~kpc). Within $R_\perp\approx200$~kpc, the median metallicity 
is [M/H] = $-0.6$~dex. 
\item
The O/Fe ratio is supersolar in nearly all measurements. A significant fraction of the CGM must 
have an enhanced $\alpha$/Fe abundance, suggestive of a star formation history similar to massive 
ellipticals with a short starburst duration. 
\item
We did not find any evolution in the total H column up to $R_\perp \approx200$~kpc, consistent 
with the finding that both the \ion{H}{1} column and the neutral gas fraction decline with 
$R_\perp$. The median total H column is $N_{\rm H} \approx 10^{20.5}\;{\rm cm^{-2}}$. 
\item
Within the estimated virial radius, we found the total mass of the cool phase CGM is substantial: 
$M_{\rm cool}^{\rm CGM}\approx1.5\times10^{11}\;{\rm M}_\odot\;(R_\perp/160~{\rm kpc})^2$. This 
accounts for 1/3 of the dark halo baryonic budget.
\item 
For 2 absorption subsystems with positive detection of the \ion{C}{2}* fine structure line, we 
estimated the electron volume density and the corresponding linear size per cloud, and found 
$n_e>10\;{\rm cm^{-3}}$. In one case, subsystem F of J1420+1603FG, the \ion{C}{2}* to \ion{C}{2} 
column ratio exceeds any previous measurement along extragalactic sightlines. The implied 
$n_e=10^{2.2}\;{\rm cm^{-3}}$ even 
defies conventional wisdom that the CGM is primarily diffuse. 
\end{itemize}

Below we list several directions of future inquiry.

\begin{itemize}
\item 
Assemble a larger sample of quasar pairs with precise redshift measurements to better 
quantify any anisotropy in the velocity fields of metal ions on CGM scale (QPQ9),  and \ion{H}{1} 
on larger scales (Hennawi et al. in prep.). 
\item 
Model the transverse proximity effect out to large scales to
  determine the average opening angle or variability timescale of the
  quasar radiation, which would help in the interpretation of the CGM
  measurements, in particular how often these CGM absorbers are
  expected to be illuminated.
\item 
Cosmological galaxy formation simulations that include feedback from AGN and/or
  star-formation powered winds to determine whether they can reproduce the observed
  cool and often optically thick quasar CGM. 
\item 
Use projected submillimeter galaxy-quasar pairs to study the CGM
  surrounding submillimeter galaxies, whose clustering strength is
  comparable to quasars. This will help isolate the impact of quasar
  feedback on the CGM.
\item 
Narrow band \citep{Cantalupo+14} and integral field \citep{Martin+14} imaging of the CGM. 
\end{itemize}

\acknowledgements

We thank the anonymous referee for a careful review and help in consolidating the findings. 
JXP and MWL acknowledge support from the National Science Foundation (NSF) grant AST-1010004, 
AST-1109452, AST-1109447 and AST-1412981. We acknowledge the contributions of Sara Ellison, 
Crystal Martin, and George Djorgovski in obtaining the ESI spectra. We 
thank Ryan Cooke for his software ALIS and technical support. We thank 
Guillermo Barro for technical support for MOSFIRE data reduction. We thank George Becker for 
helpful advices on XSHOOTER data reduction. We thank Camille Leibler for comments on chemical 
enrichment patterns. We thank Justin Brown for comments on supernova nucleosynthesis yields. We 
thank John O'Meara for sharing his metal enriched IGM absorption data. We thank Emanuele P. Farina 
for examining the manuscript. MWL thanks Sprite Chu for professional computing support. 

Much of the data presented herein were obtained at the W. M. Keck Observatory, which is operated as 
a scientific partnership among the California Institute of Technology, the University of 
California, and the National Aeronautics and Space Administration. The Observatory was made 
possible by the generous financial support of the W. M. Keck Foundation. 

Some of the data herein were obtained at the Gemini Observatory, which is operated by the 
Association of Universities for Research in Astronomy, Inc., under a cooperative agreement with 
the NSF on behalf of the Gemini partnership: the NSF (United States), the Science and Technology 
Facilities Council (United Kingdom), the National Research Council (Canada), CONICYT (Chile), the 
Australian Research Council (Australia), Minist\'erio da Ci\^encia, Tecnologia e Inova\c c\~ao 
(Brazil) and Ministerio de Ciencia, Tecnolog\'ia e Innovaci\'on Productiva (Argentina).

The authors wish to recognize and acknowledge the very significant cultural role and reverence 
that the summit of Mauna Kea has always had within the indigenous Hawaiian community. We are most 
fortunate to have the opportunity to conduct observations from this mountain.

Some of the data were obtained with the 6.5 meter Magellan Telescopes located at Las Campanas 
Observatory, Chile. 

Some of the data were obtained with the European Southern Observatory Very Large Telescope under 
program ID 087.A-0610(A).

\appendix

\section{Treatment of \ion{C}{4} in Kinematic Analysis}

For \ion{C}{4}~1548, we required a special treatment for the absorption systems where the velocity 
width exceeds the doublet separation $c\Delta\lambda/\lambda\approx500\;\mkms$. In these cases the 
optical depth profiles overlap and we made the following modifications to estimate \delv\ and 
\dvstat. Specifically we examined the optical depth profile of the entire doublet and boosted the 
opacity by 1.5 in the region of overlap and by 2 at velocities where only absorption by 
\ion{C}{4}~1550 is present. These factors account for the 2:1 ratio in the oscillator strengths of 
the two transitions. We then calculated \delv\ and \dvstat\ from the optical depth profile of the 
full doublet and offset the derived quantities by the doublet separation.

\section{Robustness of Column Density Measurements}

\begin{figure}
\includegraphics[width=0.5\textwidth]{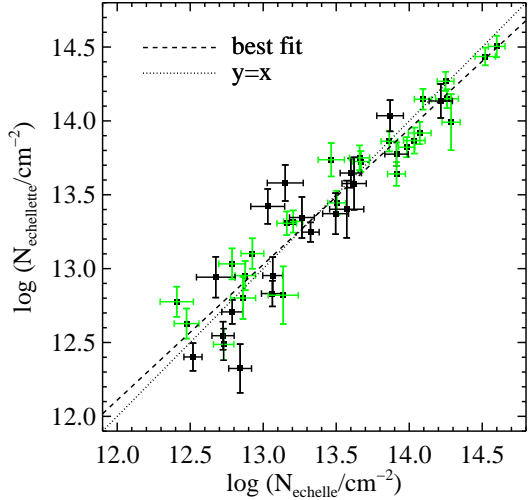} 
\caption{Column densities of ions of the subsystems of J1144+0959, J1204+0221, and J1427-0121, 
measured from echellette spectra versus echelle spectra, where the normalized flux is greater than 
0.4 in the echellette spectra. Error bars colored in green indicate multiple transitions of the 
same species is available for measurement. The plot shows that our criterion of normalized flux 
going below 0.4 for reporting lower limits well captures saturated components.}
\label{fig:echellette_vs_echelle}
\end{figure}

We examined the effect of saturated narrow features appearing unsaturated in low-resolution 
spectra, which would result in underestimated ion column densities. The paper conservatively 
reports lower limits for cases where the normalized flux goes below 0.4. In this paper we use 
echelle spectra to analyze the metal absorption subsystems of J1144+0959, J1204+0221, and 
J1427-0121. We have obtained echellette spectra of them using MagE. 
Figure~\ref{fig:echellette_vs_echelle} plots the column densities of the metal absorption 
subsystems measured from echellette spectra versus echelle spectra, where the normalized flux is 
greater than 0.4 in the echellette spectra. The data points include all metal ions such as C$^+$, 
C$^3+$, O$^0$, etc and do not differentiate them.  
For an ion species where multiple transitions are available for measurement, the column density 
reported is the mean of the measurements weighted by inverse variance. The data points in 
Figure~\ref{fig:echellette_vs_echelle} are color coded to indicate those with more than one 
transition analyzed. We do not see any systematic bias in measurement using strong versus weak 
lines. 

The data points show an increased scatter with decreasing column density. This is what we would 
expect intuitively, as the relative error in weaker absorption is larger. The linear best fit is 
very close to the $y=x$ line, and there is only one data point that deviates more than 3-$\sigma$ 
from the $y=x$ line. If we force the linear best fit to pass through the origin, which would be 
the case were S/N infinite, then the slope equals 1.00. Figure~\ref{fig:echellette_vs_echelle} 
shows both data points that lie above and below the $y=x$ line. The overall mean agreement 
demonstrates our criterion on the minimum normalized flux for reporting lower limits well captures 
saturated components. Moreover, there is also no evidence of systematic bias to higher column 
density measurements from echellette spectra that may originate from contamination from neighboring 
absorption features that is unresolved. 

We also fitted Voigt profiles to unsaturated metal absorption components to the echelle spectrum 
of J1427-0121, and summed the total column densities in each subsystem. Our aim was only to verify 
that Voigt profile fitting and AODM give the same column densities. For the $\chi^2$ 
minimization process, we set the Doppler $b$ values to be $5\;{\rm km\;s^{-1}}$ minimum as set by 
resolution limit, although the model preferred lower values for a few components. Conventionally, 
due to turbulence the minimum $b$ value is $8\;{\rm km\;s^{-1}}$, and contributions from turbulence 
dominates over thermal broadening. This is just as expected for an instrumental resolution high 
enough to resolve the lines. 

We have defined absorption subsystems as those separated in distinct velocity intervails, with no 
obvious ionization gradient within. The three echelle sightlines also allow us to examine 
any remaining ionization gradient within a subsystem. We considered the J1427-0121 sightline. We 
performed a component-by-component photoionization modeling for its subsystem C, and omitted its 
subsystems A and B for that analysis. There is residual Ly$\alpha$ emission at the 
absorption trough of subsystem A, which makes it difficult to deblend the two absorption components 
in \ion{H}{1} at velocities defined by the metal ions. The three components of subsystem B which 
contain low ions are all weak absorption, which makes it difficult to analyze them separately. 
Subsystem C contains two components and we found $\log U < -3.6$ and $\log U < -3.4$ respectively. 
In the ionization modeling treatment in this study, we adopted $\log U <-3.4^{+0.3}$ for subsystem 
C. We conclude that measuring column densities using AODM is robust to whtin the uncertainties 
allowed for ionization modeling-dependent analysis.

\section{Notes on Individual Pairs}

We present a detailed description, including figures and tables, for the absorption associated to 
each of the foreground quasars in the QPQ8 sample. See Section~\ref{sec:analysis} for a 
description of the techniques employed in the analysis. In Table~\ref{tab:NHI}, we present the 
\ion{H}{1} column density measurements of each absorption subsystem. 

\begin{figure*}
\includegraphics[width=7in]{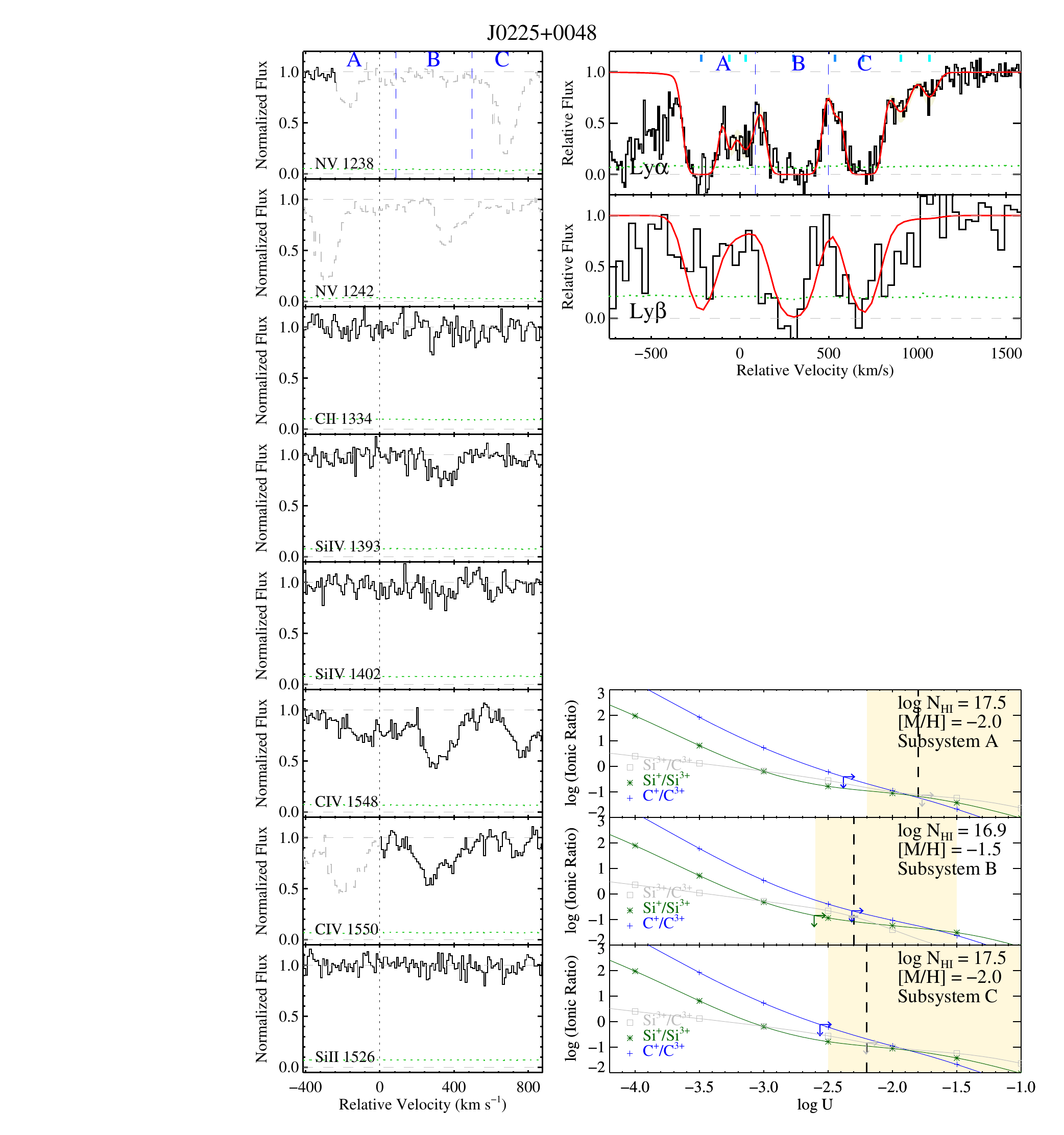} 
\caption{Combined figure for the J0225+0048 pair showing: 
(Left) Metal line transitions from the absorbers identified at velocities consistent with 
$\mzfg = 2.7265$. Absorption well separated in distinct velocity intervals are designated as 
subsystems A, B, C, etc. and marked by the vertical dashed lines in the upper panels. Absorptions 
that are unrelated to the foreground quasar, e.g. blends in the \lya\ forest, are presented as 
dashed, gray lines. The green histograms show the $1\sigma$ noise in the normalized flux. 
(Top right) Lyman series absorption profiles identified in the background quasar spectrum at 
velocities consistent with the foreground quasar of each projected pair. The green histograms show 
the $1\sigma$ noise in the normalized flux. The relative velocity $v=0{\rm \;km\;s^{-1}}$ 
corresponds to the redshift of the foreground quasar. For each system we performed $\chi^2$ 
minimization Voigt profile modeling. We introduced \ion{H}{1} components centered at relative 
velocities traced by the peak optical depths of the associated metal ion absorptions. The navy 
ticks mark the centroid redshifts for components traced by low ions, while the blue ticks mark the 
centroids for components traced by high ions. \ion{H}{1} components not associated with any metal 
ions are marked with cyan ticks. Additional \ion{H}{1} components introduced to model Ly$\alpha$ 
forest blending are omitted in the tick marks. The red curve is the convolved fit of all 
\ion{H}{1} components associated with the foreground quasar, and the beige shades mark the 
estimated $\pm1\sigma$ errors in \ion{H}{1} column densities.The orange curve is the convolved fit 
of all \ion{H}{1} components associated with the foreground quasar and the Ly$\alpha$ forest 
contaminations, if any. 
(Bottom right) Cloudy modeling of the ionization parameter $U$ for each of the 12 quasar 
associated absorption systems where metal ion column measurements are available. Solid curves show 
predicted ionic ratios as a function of $U$ for a series of ion pairs. Overplotted on the curves 
are observational constraints of the subsystems, indicated by solid boxes, whose edges are the 
1-$\sigma$ uncertainties, or indicated by arrows for lower and upper limits. The observations 
indicate a varying $U$. 
}
\label{fig:j0225}
\end{figure*}

{\it J0225+0048}---
Figure~\ref{fig:j0225} reveals absorption for this pair in roughly three distinct velocity 
intervals spanning a total of $\Delta v\approx1000\;{\rm km\;s}^{-1}$. We define three subsystems 
based on the strong features in the \ion{H}{1} Lyman series absorption. We present the ionic 
column densities measured with the apparent optical depth method in 
Table~\ref{tab:J0225+0048_colm}.
Absorption from high ions Si$^{3+}$ and C$^{3+}$ is strong in subsystems A and B and moderate in 
subsystem C. No low ion transitions are detected for any of the subsystems, indicating a highly 
ionized gas.
Regarding the \nhi\ analysis, we have limited constraints given that the echellette resolution data 
only covers the Ly$\alpha$ transition. There is no indication of damping wings and the lines are 
saturated, i.e. on the flat portion of the curve of growth. We also present the lower dispersion 
GMOS spectrum that covers Ly$\beta$. The gas shows absorption separated in roughly three distinct 
velocity intervals. If we assume minimum Doppler $b_{\rm A,B,C}$ values of $15\;{\rm km\;s}^{-1}$, 
the lack of strong damping wings in all three subsystems yields strict upper limits of 
$N_{\rm HI}^{\rm A}<10^{18.6}\;{\rm cm}^{-2}$, $N_{\rm HI}^{\rm B}<10^{18.7}\;{\rm cm}^{-2}$ and 
$N_{\rm HI}^{\rm C}<10^{18.5}\;{\rm cm}^{-2}$. If we assume maximum Doppler $b_{\rm A,B,C}$ values 
of $60\;{\rm km\;s}^{-1}$ and single components, the equivalent widths demand 
$N_{\rm HI}^{\rm A}>10^{15.6}\;{\rm cm}^{-2}$, $N_{\rm HI}^{\rm B}>10^{16.1}\;{\rm cm}^{-2}$ and 
$N_{\rm HI}^{\rm C}>10^{16.0}\;{\rm cm}^{-2}$. 
We fitted these data with the ALIS software package assuming 7 components (see Table~
\ref{tab:NHI}), with redshifts set by the metal line absorption. We find solutions with nearly 
equivalent $\chi^2$ that range from $\mnhi = 10^{16.4} \cm{-2}$ to $10^{18.9} \cm{-2}$.  
We adopted approximately the central \nhi\ values in this range and a correspondingly large 
uncertainty for each subsystem.
Because no low ion states are detected, the ionic ratios Si$^+$/Si$^{3+}$ and C$^+$/C$^{3+}$ 
impose lower limits on $U$. Our ionization modeling yields $\log U_{\rm A}>-2.4$, 
$\log U_{\rm B}>-2.3$, and $\log U_{\rm C}>-2.6$.
To further constrain $U$ we also considered the Si$^{3+}$/C$^{3+}$ ratios. They imply yet higher 
$U$ values, provided the relative abundances are roughly solar. These $U$ values suggest either 
the radiation field is much stronger than the EUVB and the foreground quasar is shining on the 
gas, or that $n_{\rm H}\ll0.1\;{\rm cm}^{-3}$. 

\begin{figure*}
\includegraphics[width=7in]{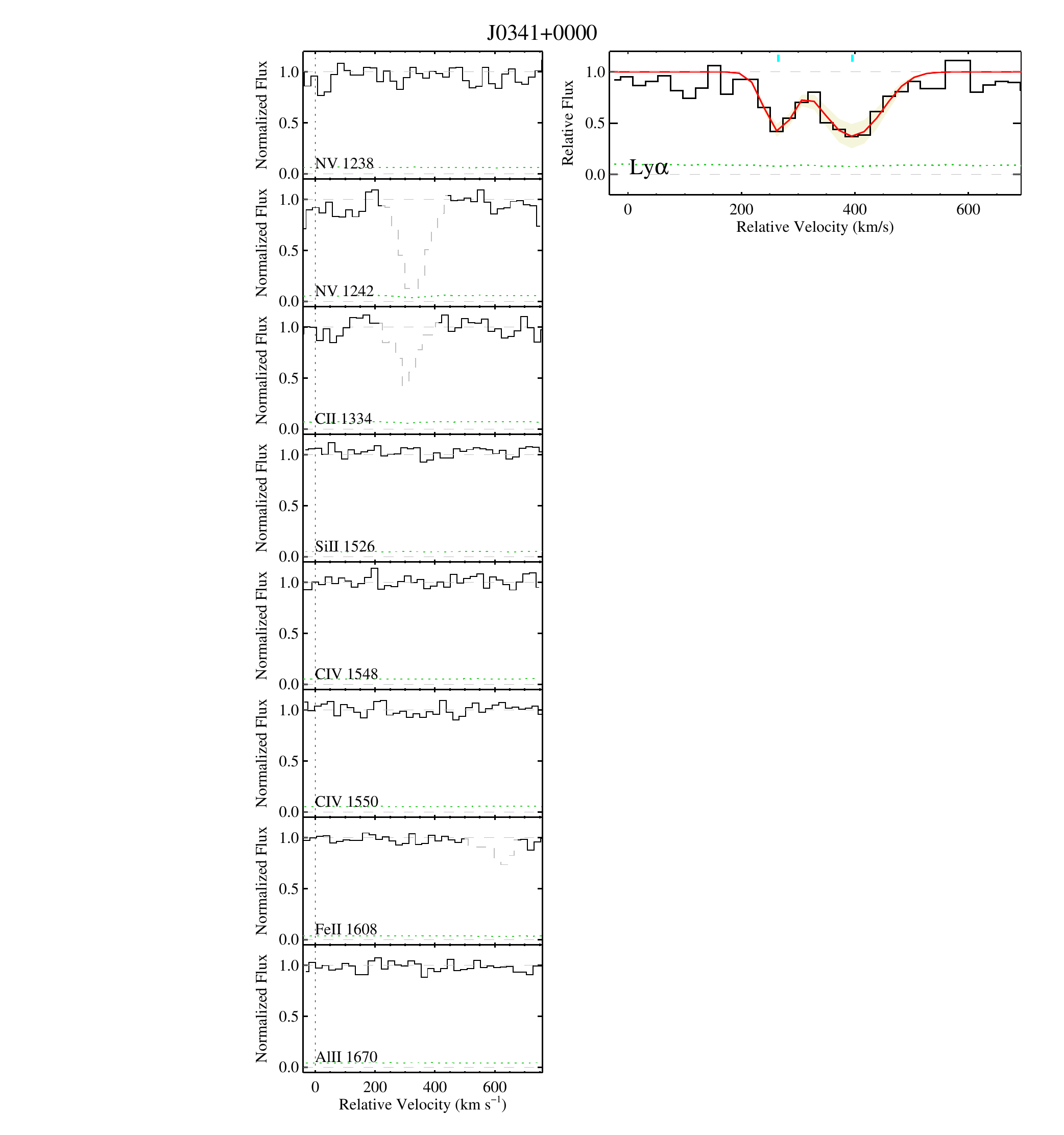} 
\caption{Similar to Figure~\ref{fig:j0225} but for J0341+0000 at $\mzfg = 2.1233$.
}
\label{fig:j0341}
\end{figure*}

{\it J0341+0000}---
Within $\pm1000\;{\rm km\;s^{-1}}$ of \zfg\ we identify no system with a rest equivalent width for 
\lya\ exceeding 0.3~\AA. We present AODM measurements of the upper limits to any associated metal 
ion columns in Table~\ref{tab:J0341+0000_colm}. Associating the strongest absorption line to the 
quasar at $z=2.1286$ (Figure~\ref{fig:j0341}), we measured 
$N_{\rm HI}=10^{14.2\pm0.2}\;{\rm cm}^{-2}$ from a fit to 
the data. Although there is uncertainty dominated by possible line saturation, we strictly 
constrain $N_{\rm HI}<10^{14.4}\;{\rm cm}^{-2}$ assuming $b>15\;{\rm km\;s}^{-1}$.

With the absence of metal line detections, we have no constraints on the ionization state of the 
gas. The low $N_{\rm HI}$ column, however, implies a highly ionized medium. Together the data 
suggests either an extreme ionization state, low metallicity, or little gas along the sightline.

\begin{figure*}
\includegraphics[width=7in]{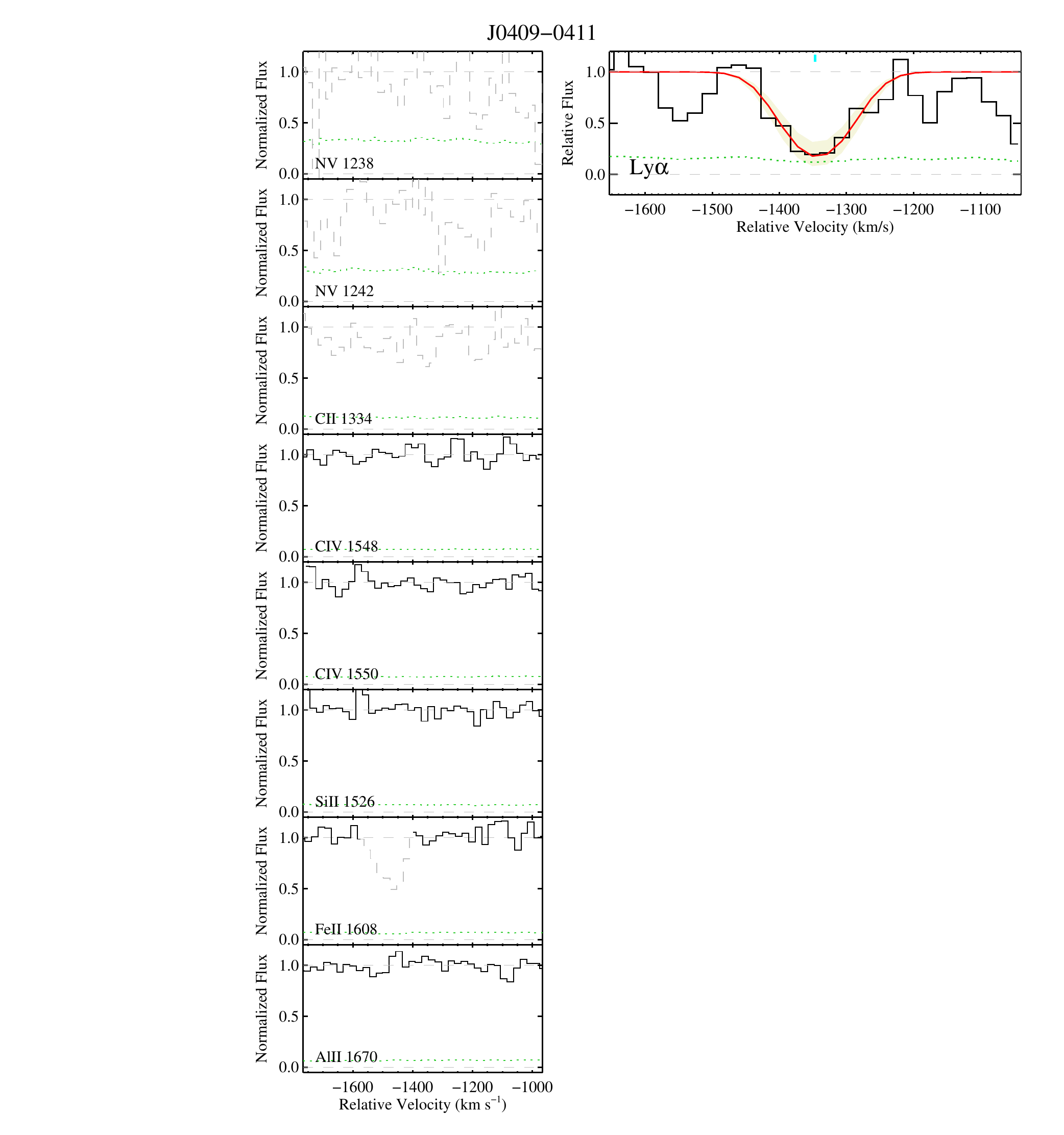} 
\caption{Similar to Figure~\ref{fig:j0225} but for J0409-0411 at $\mzfg = 1.7155$.
}
\label{fig:j0409}
\end{figure*}

{\it J0409-0411}---
Similar to J0341+0000, this system shows no \lya\ line with a peak optical depth greater than 2 
within $1000\;{\rm km\;s}^{-1}$ of \zfg\ (Figure~\ref{fig:j0409}). We present AODM measurements of 
the upper limits to any associated metal ion columns in Table~\ref{tab:J0409-0411_colm}. The 
strongest \lya\ line at $z=1.7027$ has an equivalent width $W_{{\rm Ly}\alpha}=0.45\pm0.04$~\AA. A 
line profile analysis gives $N_{\rm HI}=10^{14.2\pm0.2}\;{\rm cm}^{-2}$ for the complex with the 
uncertainty dominated by possible line saturation.

With the absence of metal line detections, we have no constraints on the ionization state of the 
gas. The low \nhi column, however, implies a highly ionized medium. 

\begin{figure*}
\includegraphics[width=7in]{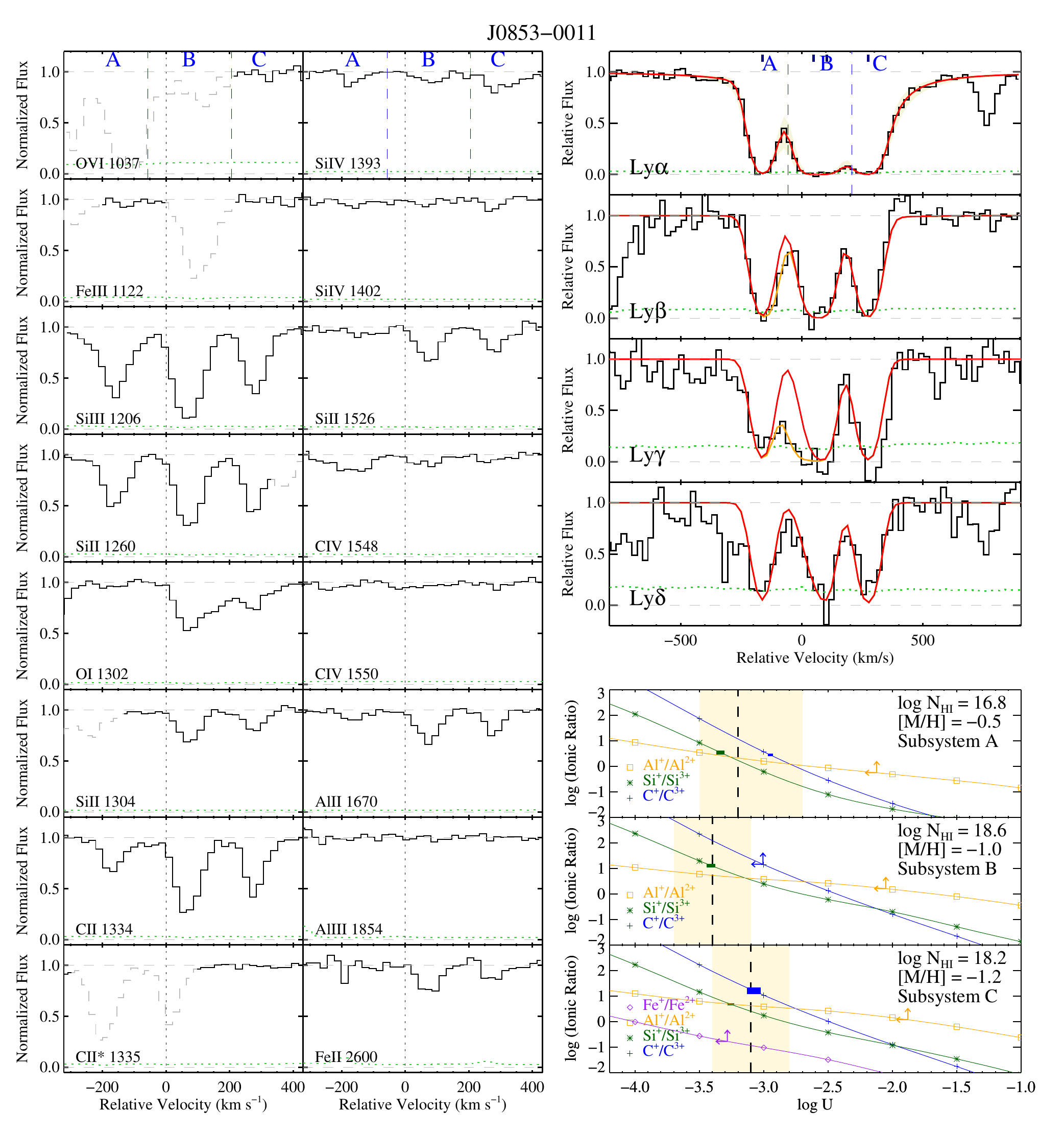} 
\caption{Similar to Figure~\ref{fig:j0225} but for J0853-0011 at $\mzfg = 2.4014$.
}
\label{fig:j0853}
\end{figure*}

{\it J0853-0011}---
The ions show absorption in roughly three distinct velocity intervals spanning a total of 
$\Delta v\approx650\;{\rm km\;s}^{-1}$ (Figure~\ref{fig:j0853}). We define three subsystems across 
the complex. We present the ionic column densities measured with AODM in 
Table~\ref{tab:J0853-0011_colm}. 
One notes moderate absorption in subsystem A and strong absorption in subsystems B and C from a 
series of low ions, including O$^0$, Si$^+$, C$^+$, Fe$^+$ and Al$^+$. Subsystems B and C are 
characteristic of Lyman limit systems where the large \ion{H}{1} opacity self shields gas from 
local or background UV sources \citep[e.g.][]{Prochaska+15}. Elements therefore occupy the 
first ionization state with an ionization potential greater than 1 Ryd. Furthermore, there is only 
weak \ion{Si}{4} and \ion{C}{4} absorption. Altogether the data suggests an optically thick gas. 

The Ly$\beta$, Ly$\gamma$ and Ly$\delta$ profiles show absorptions in three distinct 
velocity intervals corresponding to the three subsystems. 
The absence of strong damping wings in the the Ly$\alpha$ absorption profiles demand 
$N_{\rm HI}^{\rm A}<10^{18.0}\;{\rm cm}^{-2}$, $N_{\rm HI}^{\rm B}<10^{18.8}\;{\rm cm}^{-2}$ and 
$N_{\rm HI}^{\rm C}<10^{18.4}\;{\rm cm}^{-2}$. Assuming maximum $b_{\rm A,B,C}$-values of 
$60\;{\rm km\;s^{-1}}$ and absorption dominated by single components, as implied by the \ion{H}{1} 
and metal line profiles, the measured equivalent widths imply 
$N_{\rm HI}^{\rm A}>10^{14.8}\;{\rm cm}^{-2}$, $N_{\rm HI}^{\rm B}>10^{16.0}\;{\rm cm}^{-2}$ and 
$N_{\rm HI}^{\rm C}>10^{15.2}\;{\rm cm}^{-2}$. The asymmetric Ly$\alpha$ absorption profile of 
subsystem B implies modest blending with a weak \ion{H}{1} component. The Ly$\gamma$ profile is 
blended with an unrelated \ion{H}{1} component of $N_{\rm HI}\approx10^{14.8}\;{\rm cm}^{-2}$ at 
$z\approx1.7211$. We adopt $N_{\rm HI}^{\rm A}=10^{16.8\pm0.6}\;{\rm cm}^{-2}$, 
$N_{\rm HI}^{\rm B}=10^{18.6_{-0.6}^{+0.2}}\;{\rm cm}^{-2}$ and 
$N_{\rm HI}^{\rm C}=10^{18.2_{-0.6}^{+0.3}}\;{\rm cm}^{-2}$, where the error in $N_{\rm HI}$ is 
dominated by Ly$\alpha$ forest contamination and line saturation.

Multiple ionization states of the same element Fe$^+$/Fe$^{2+}$, Si$^+$/Si$^{3+}$, 
Al$^+$/Al$^{2+}$ and C$^+$/C$^{3+}$ provide observational constraints on $U$. The gas shows 
systematically stronger low ion absorption and correspondingly lower $U$ values, $\log U<-3$, than 
the majority of the QPQ8 sample.
We adopt $\log U_{\rm A}=-3.2$, $\log U_{\rm B}=-3.4$ and $\log U_{\rm C}=-3.1$ giving 
$x_{\rm HI,A}=0.01$, $x_{\rm HI,B}=0.08$ and $x_{\rm HI,C}=0.02$ and
$N_{\rm H}^{\rm A}=10^{18.9}\;{\rm cm}^{-2}$, $N_{\rm H}^{\rm B}=10^{19.7}\;{\rm cm}^{-2}$ and 
$N_{\rm H}^{\rm C}=10^{20.0}\;{\rm cm}^{-2}$. 

\begin{figure*}
\includegraphics[width=7in]{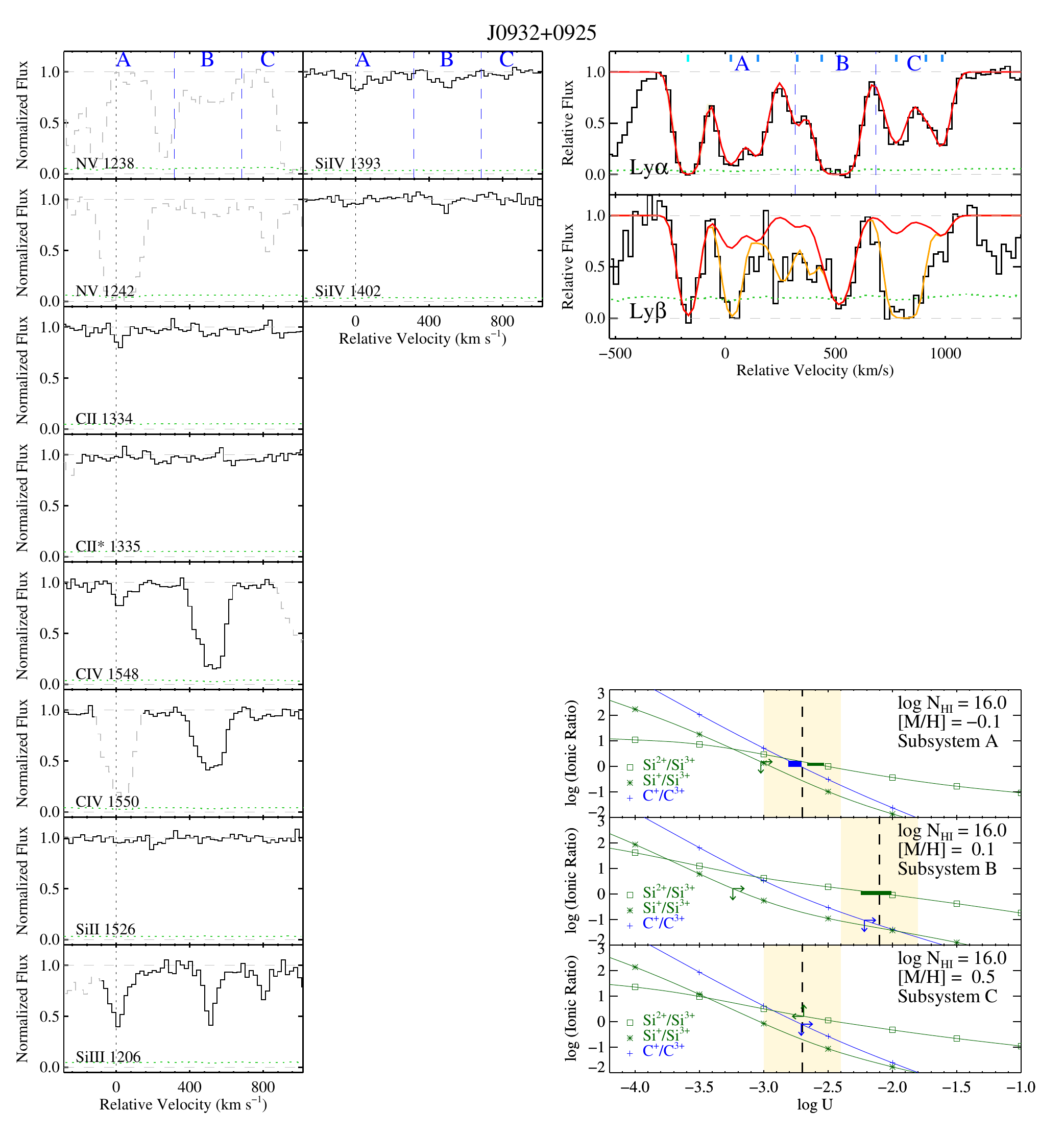} 
\caption{Similar to Figure~\ref{fig:j0225} but for J0932+0925 at $\mzfg = 2.4170$.
}
\label{fig:j0932}
\end{figure*}

{\it J0932+0925}---
The ions show absorption in roughly three distinct velocity intervals spanning a total of 
$\Delta v\approx1200\;{\rm km\;s}^{-1}$ (Figure~\ref{fig:j0932}) that demarcate three subsystems. 
We present the ionic column densities measured with AODM in Table~\ref{tab:J0932+0925_colm}. 
One notes strong \ion{C}{4} absorption in subsystem B and moderate \ion{C}{4} absorption in 
subsystems A and C. \ion{C}{2} absorption is weak in subsystem A and absent in subsystems B and C. 
The intermediate ion \ion{Si}{3}~1206 transition locates in a relatively clean region of the 
Ly$\alpha$ forest. Qualitatively the data indicates a highly ionized gas.

Only the Ly$\alpha$ and the Ly$\beta$ absorption profiles have sufficient S/N for analysis. The 
stronger absorption features of subsystems A and B are well constrained by both the Ly$\alpha$ and 
the Ly$\beta$ profiles. Subsystem C and the weaker components of subsystems and B which have their 
Ly$\beta$ profiles blended with Ly$\alpha$ forest lines, all have $N_{\rm HI}$ values on the 
linear part of the curve of growth.  For the strongest component in subsystem A, the measured 
equivalent width constrains $10^{14.5}\;\cm{-2} < N_{\rm HI}^{\rm A}<10^{17.5}\;{\rm cm}^{-2}$ 
assuming $b_A$ ranges between $15\text{--}60\;{\rm km\;s}^{-1}$. The strongest absorption feature 
of subsystem B is asymmetric in Ly$\alpha$ and Ly$\beta$, suggesting that it contains two 
\ion{H}{1} components, as modeled. 
Our best fit solution gives a total $N_{\rm HI}^{\rm B}=10^{15.1_{-0.3}^{+0.3}}\;{\rm cm}^{-2}$, 
where the errors are dominated by uncertainty in continuum placement and Ly$\alpha$ forest line 
blending. The \nhi\ values for subsystem C are well constrained by the unsaturated \lya profile.

This system has Si$^{2+}$ and Si$^{3+}$ detections which constrain the $U$ value for the three 
subsystems. These are also consistent with the constraints derived from C$^+$/C$^{3+}$.

\begin{figure*}
\includegraphics[width=7in]{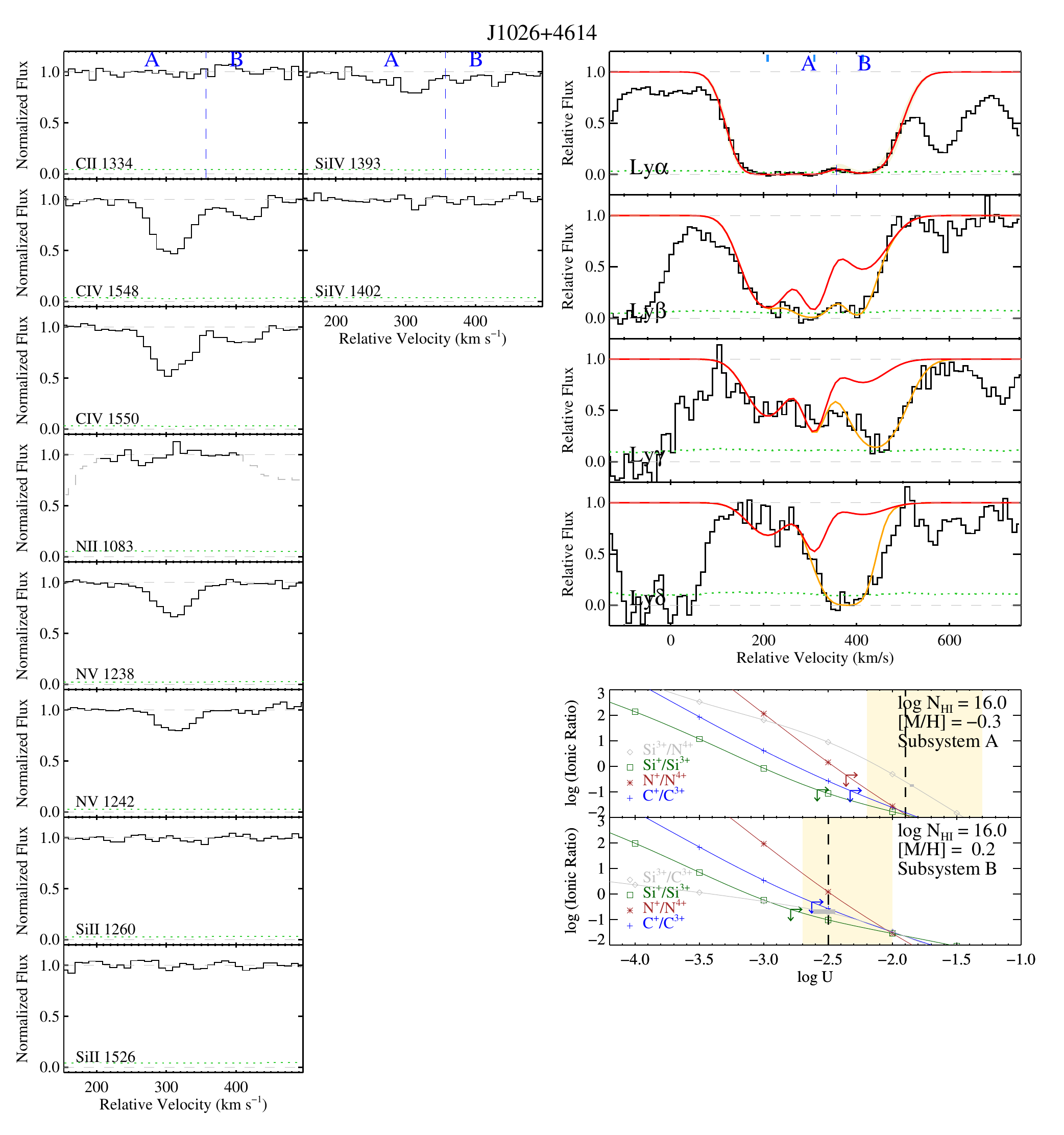} 
\caption{Similar to Figure~\ref{fig:j0225} but for J1026+4614 at $\mzfg = 3.3401$.
}
\label{fig:j1026}
\end{figure*}

{\it J1026+4614}---
The ions show absorption in two velocity intervals spanning a total of 
$\Delta v\approx240\;{\rm km\;s}^{-1}$ (Figure~\ref{fig:j1026}). We present the ionic column 
densities in Table~\ref{tab:J1026+4614_colm}. 
This is the only member of the QPQ8 sample with strong absorption from the \ion{N}{5} doublet, 
although we also note that it is also one of the few where those transitions lie outside the \lya\ 
forest. Together with the lack of low ions, the data suggests the ionization state of the gas is 
extreme. It is peculiar that in subsystem A the seemingly unsaturated \ion{C}{4} doublet 
$\lambda\lambda$~1548,~1550 have similar optical depth, i.e. inconsistent with the oscillator 
strength ratio of 2 to 1. We suspect hidden saturation as there is no evidence for partial 
covering in the other observed doublets, and treat the \ion{C}{4}~1548 measurement as a lower 
limit.
Figure~\ref{fig:j1026} presents the the Ly$\alpha$, Ly$\beta$, Ly$\gamma$ and Ly$\delta$ profiles 
of this absorption system. 

All of the components become unsaturated by Ly$\gamma$ and therefore yield precise measurements 
for the column densities. The Ly$\beta$ absorption profile is blended with two Ly$\alpha$ forest 
lines of respectively $N_{\rm HI}\approx10^{14.7}\;{\rm cm}^{-2}$ and 
$N_{\rm HI}\approx10^{15.3}\;{\rm cm}^{-2}$ located at respectively $z\approx2.6658$ and 
$z\approx2.6669$. The Ly$\gamma$ absorption profile is blended with a Ly$\alpha$ forest line of 
$N_{\rm HI}\approx10^{14.2}\;{\rm cm}^{-2}$ at $z\approx2.4773$. The Ly$\delta$ absorption profile 
is blended with a Ly$\alpha$ forest line of $N_{\rm HI}\approx10^{15.8}\;{\rm cm}^{-2}$ at 
$z\approx2.3951$.  Modeling these blends together with the Lyman series absorption, we recovered a 
best-fit solution with $N_{\rm HI}^{\rm A}=10^{15.4\pm0.2}\;{\rm cm}^{-2}$ and
$N_{\rm HI}^{\rm B}=10^{14.6\pm0.1}\;{\rm cm}^{-2}$. The errors are dominated by continuum 
placement at the higher Lyman series lines. 

The non-detection of lower ionization states in subsystem A constrain $\log U > -2.3$. If we 
assume relative solar abundances, the measured Si$^{3+}$/N$^{4+}$ ratio implies 
$\log U\approx-1.9$. We adopt a larger uncertainty towards higher $U$ values to account for 
non-solar abundances. Our model for subsystem B, which does not show \ion{N}{5} absorption, has a 
lower $U$ value consistent with the various constraints. We adopt a larger uncertainty to higher 
$U$ values to allow for an enhanced intrinsic Si/C abundance ratio.

\begin{figure*}
\includegraphics[width=7in]{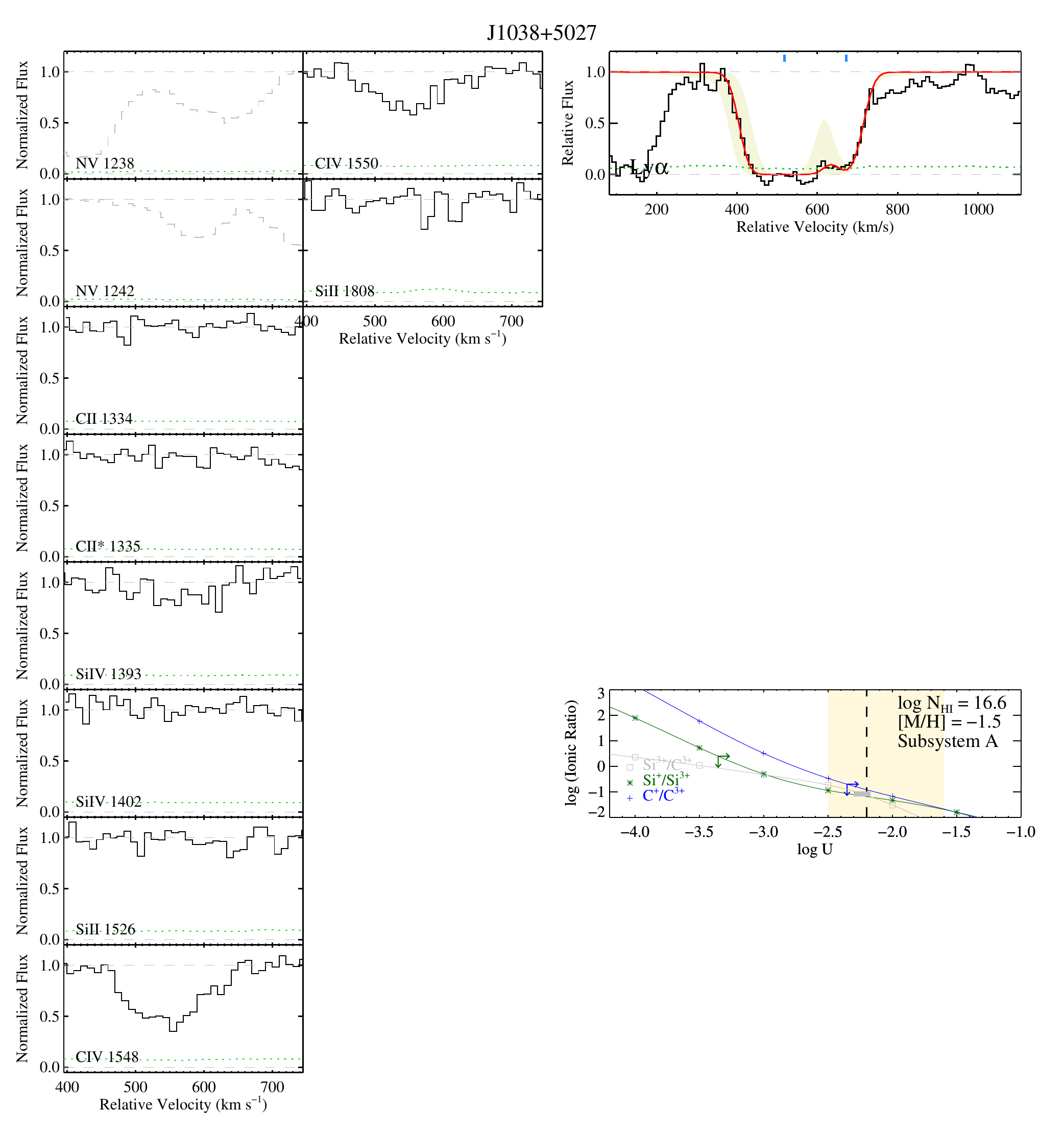} 
\caption{Similar to Figure~\ref{fig:j0225} but for J1038+5027 at $\mzfg = 3.1322$.
}
\label{fig:j1038}
\end{figure*}

{\it J1038+5027}---
The ions show absorption spanning a total of $\Delta v\approx260\;{\rm km\;s}^{-1}$ 
(Figure~\ref{fig:j1038}). We present the ionic column densities in Table~\ref{tab:J1038+5027_colm}. 
There is strong \ion{C}{4} absorption and moderate \ion{Si}{4} absorption, together with weak low 
ions suggesting a highly ionized gas. 

We present the Ly$\alpha$ profile of this absorption system. The Ly$\beta$ profile is blended with 
a strong absorption system is not useful for $N_{\rm HI}$ modeling. The asymmetric 
Ly$\alpha$ profile is well modeled by least two \ion{H}{1} components, with the weaker one 
unsaturated at Ly$\beta$. The lack of obvious damping wings restricts 
$N_{\rm HI}<10^{18.4}\;{\rm cm}^{-2}$. Assuming a maximum $b$-value of $60\;{\rm km\;s}^{-1}$, the 
Ly$\alpha$ equivalent width requires $N_{\rm HI}>10^{15.3}\;{\rm cm}^{-2}$. We adopt a total 
$N_{\rm HI}=10^{16.6\pm0.5}\;{\rm cm}^{-2}$, where the large errors are due to line saturation.

The Si$^+$/Si$^{3+}$ and C$^+$/C$^{3+}$ ratios place lower limits on $U$. To better constrain the 
$U$ value, we also consider the Si$^+$/C$^{3+}$ ratio under the assumption of solar relative 
abundances. 
Adopting $\log U=-2.2$, we recovered $x_{\rm HI} = 0.0005$, corresponding to 
$N_{\rm H}=19.9\;{\rm cm}^{-2}$. 

\begin{figure*}
\includegraphics[width=7in]{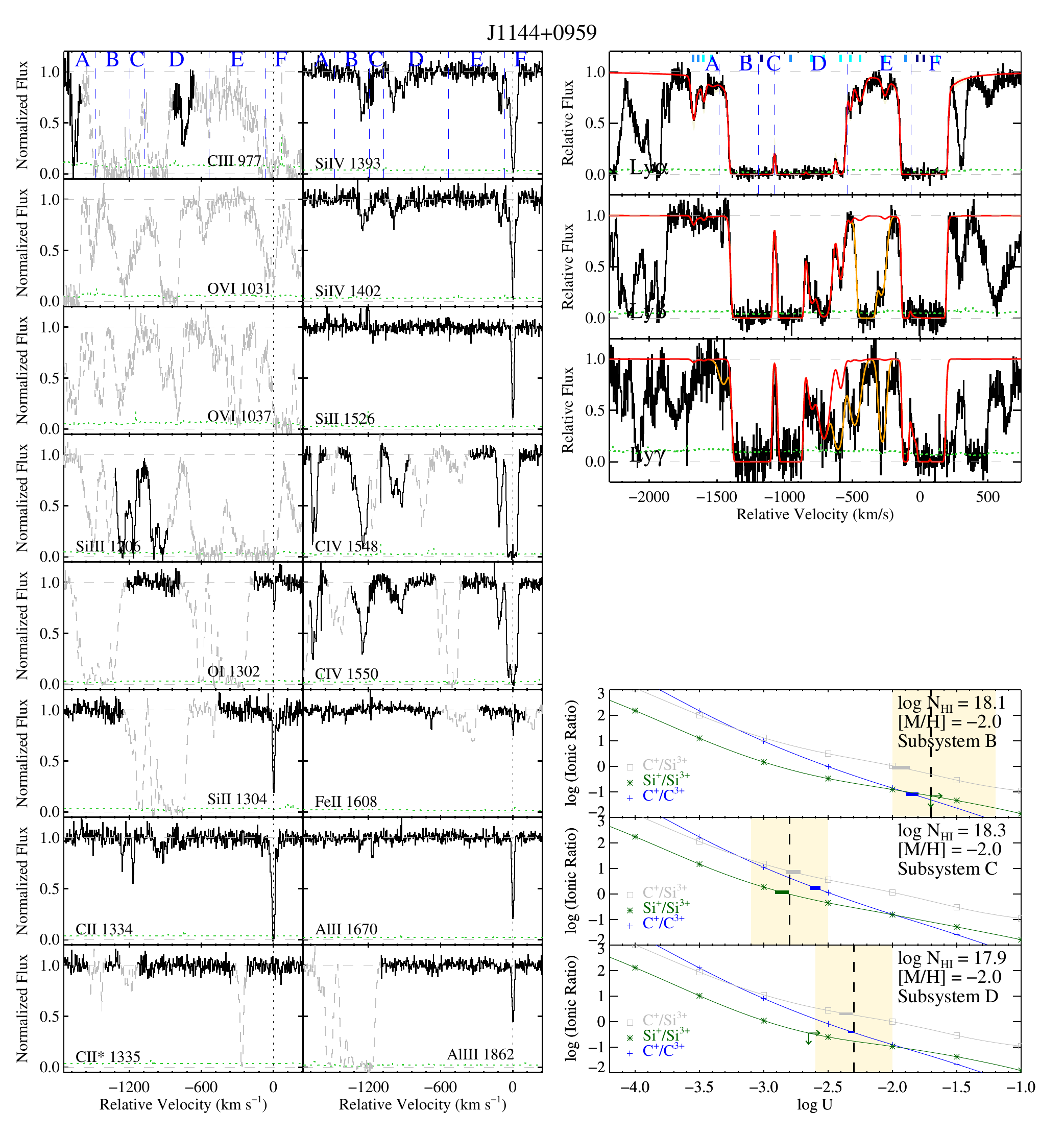} 
\caption{Similar to Figure~\ref{fig:j0225} but for J1144+0959 at $\mzfg = 2.9731$.
}
\label{fig:j1144}
\end{figure*}

{\it J1144+0959}---
This very complex absorption system exhibits a velocity spread of nearly $2000\;{\rm km\;s}^{-1}$ 
(Figure~\ref{fig:j1144}), which we divide into six subsystems. We present the ionic column 
densities in Table~\ref{tab:J1144+0959_colm}. 
Given the large velocity separation of the subsystems, we examined the possibility that some of 
the gas is unassociated to the foreground quasar. The total equivalent width of 
\ion{C}{4}~1548 of subsystems A, B, C, and D, $W_{1548}=0.86\;{\rm\AA}$, is large. The \ion{C}{4} 
survey conducted by \cite{Cooksey+13} reported the incidence of strong \ion{C}{4} absorbers of 
equivalent width $>0.6\;{\rm\AA}$ at $z\approx2.97$ to be 
$\frac{dN_{\rm CIV}^{>0.6{\rm\AA}}}{dz}=0.83$. Thus in a $\pm1500\;{\rm km\;s}^{-1}$ window 
around $z\approx2.97$, the probability of finding at least one strong \ion{C}{4} absorber is 3\%.  
According to the QSO-\ion{C}{4} clustering analysis in QPQ7 and \cite{Vikas+13}, clustering would 
at most quadrupole this probability. 
We consider it unlikely that subsystems A, B, C, and D are not physically associated to the 
foreground quasar. We note further that the positive detections of \ion{C}{2}~1334 and 
\ion{Al}{2}~1670, which have an even a smaller random incidence, strongly imply the physical 
association of all the gas to the environment of J1144+0959FG.

Subsystem A shows strong \ion{C}{4}, strong \ion{C}{3} absorption in an apparently clean region of 
the \lya\ forest, and the absence of low ion absorption. 
Subsystem B shows moderate absorption from C$^+$ and strong absorption from C$^{3+}$ and 
Si$^{3+}$. Subsystem C shows moderate absorption from C$^+$ and weak absorption from C$^{3+}$ and 
Si$^{3+}$. Subsystem D shows moderate absorption from C$^+$ and moderate absorption from C$^{3+}$ 
and Si$^{3+}$. As a group subsystems A, B, C, and D trace a highly ionized gas. 
Subsystem E shows moderate absorption from high ions C$^{3+}$ and Si$^{3+}$ and no corresponding 
low ion absorption. Lastly, subsystem F shows strong absorption from high ions C$^{3+}$ and 
Si$^{3+}$ and strong absorption from low ions O$^0$, C$^+$, Si$^+$, Al$^+$ and Fe$^+$. 

We present the Ly$\alpha$, Ly$\beta$, and Ly$\gamma$ velocity profiles of this complex absorption 
system. The data also cover Ly$\delta$, but the majority of the complex is blended strongly with a 
damped \lya\ system with $N_{\rm HI}\approx10^{20.3}\;{\rm cm}^{-2}$ at $z\approx2.0933$. Two 
groups of absorbers with associated metal lines separated by $\sim1000\;{\rm km\;s^{-1}}$ are 
found in a $\pm1500\;{\rm km\;s^{-1}}$ window around $z_{\rm fg}$. 
Subsystem A has four weak components, among which two are associated with high ion absorption, 
e.g. \ion{C}{4}, and possibly \ion{O}{6}. Its total 
$N_{\rm HI}^{\rm A}=10^{13.5\pm0.2}\;{\rm cm}^{-2}$ is well constrained. 
In the Lyman series, subsystems B and C are blended together, however their centroid velocities 
can be precisely constrained by the unblended absorption profiles of low ions. the absence of 
strong Ly$\alpha$ damping wings demand a total $N_{\rm HI}^{\rm B+C}<10^{18.7}\;{\rm cm}^{-2}$. 
Assuming a maximum $b_{\rm B,C}$ value of $60\;{\rm km\;s}^{-1}$, the large Ly$\alpha$ equivalent 
width demands a total $N_{\rm HI}^{\rm B+C}>10^{16.1}\;{\rm cm}^{-2}$. Our best fit solution gives 
$N_{\rm HI}^{\rm B}=10^{18.1_{-0.4}^{+0.2}}\;{\rm cm}^{-2}$ and 
$N_{\rm HI}^{\rm C}=10^{18.3_{-0.4}^{+0.2}}\;{\rm cm}^{-2}$, where the errors are dominated by 
line blending and line saturation. 
Subsystem D shows multiple components, one associated with ions 
and three weaker components that lack any metal ion detection. For the stronger component, the 
lack of strong Ly$\alpha$ damping wings demand a total 
$N_{\rm HI}^{\rm D}<10^{19.2}\;{\rm cm}^{-2}$, while the large Ly$\alpha$ equivalent width 
requires $N_{\rm HI}^{\rm D}>10^{15.6}\;{\rm cm}^{-2}$, assuming $b_{\rm D}<60\;{\rm km\;s^{-1}}$. 
The three weaker components at $v\approx-700\;{\rm km\;s}^{-1}$ have Ly$\alpha$ and Ly$\beta$ 
equivalent widths that lie on the linear part of the curve of growth and hence are tightly 
constrained. Summing up the four components, our best fit solution gives a total 
$N_{\rm HI}^{\rm D}=10^{17.9\pm0.5}\;{\rm cm}^{-2}$, where the errors are dominated by line 
blending and saturation. For subsystem E, the unsaturated Lyman lines yield precise constraint of 
$N_{\rm HI}^{\rm E}=10^{15.6\pm0.2}\;{\rm cm}^{-2}$. Subsystem F contains two components 
associated with metal absorption and one component that is not associated with metals.
The lack of strong Ly$\alpha$ damping wings restricts 
$N_{\rm HI}^{\rm F}<10^{18.5}\;{\rm cm}^{-2}$, while the large Ly$\alpha$ equivalent width 
demands a total $N_{\rm HI}^{\rm F}>10^{16.1}\;{\rm cm}^{-2}$. Our best fit solution gives a total 
$N_{\rm HI}^{\rm F}=10^{18.4_{-0.4}^{+0.2}}\;{\rm cm}^{-2}$, where the errors are dominated by 
blending in the Ly$\alpha$, Ly$\beta$ and Ly$\gamma$ profiles and line saturation. Altogether, we 
adopt $N_{\rm HI}^{\rm total}=10^{{18.8}_{-0.4}^{+0.2}}\;{\rm cm}^{-2}$ with the upper bound a 
strict limit given the absence of \lya damping wings.

For subsystem A, the C$^+$/C$^{3+}$ and C$^{2+}$/C$^{3+}$ ratios constrain 
$-2.2<\log U_{\rm A}<-1.5$. We expect the $U$ value to lie closer to the upper value because 
\ion{C}{3} is only mildly saturated and the Si$^{3+}$/C$^{3+}$ ratio is consistent with this 
estimate. 
For subsystems B and C, the observed Si$^+$/Si$^{3+}$ ratios put $U_{\rm B,C}$ at a different 
values than C$^+$/C$^{3+}$ does, indicating a multiphase absorber. We give stronger weight to the 
constraint from Si$^+$/Si$^{3+}$ regarding the lower ionization gas phase. 
For subsystem D, the observed Si$^+$/Si$^{3+}$ and C$^+$/C$^{3+}$ ratios well constrain 
$U_{\rm D}$. 
For subsystem E, the observed C$^+$/C$^{3+}$ ratio gives a precise value for $U_{\rm E}$ that is 
fully consistent with the observed Si$^+$/Si$^{3+}$ and Al$^+$/Al$^{2+}$ ratios. 
Subsystem F exhibits positive detections from a wide range of ions.
Despite the significant low ion absorption, the measurements imply a highly ionized system. This 
is argued from the Si$^+$/Si$^{3+}$, Al$^+$/Al$^{2+}$, and C$^+$/C$^{3+}$ ratios. It is further 
implied by the very low O$^0$/Fe$^+$ ratio, as discussed in Section~\ref{sec:abundance}. 
Unfortunately none of these is highly constraining because a number of the measurements are 
formally lower limits. Adopting the C$^+$, C$^{3+}$, and Si$^{3+}$ values as measurements instead 
of limits, their measured ratios suggest 
$\log U_{\rm F} \approx -2$. We adopt this value and a large uncertainty.

\begin{figure*}
\includegraphics[width=7in]{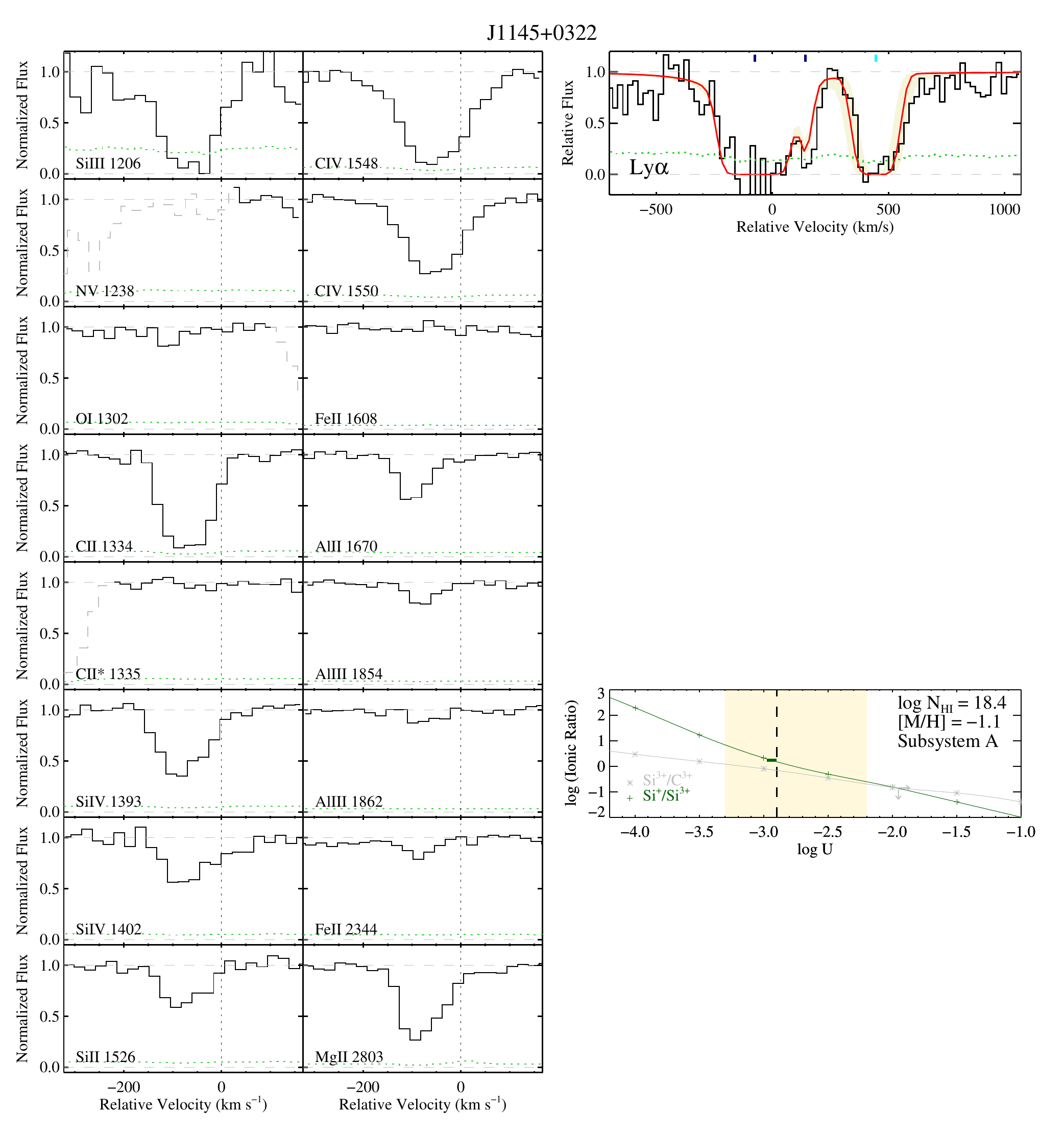} 
\caption{Similar to Figure~\ref{fig:j0225} but for J1145+0322 at $\mzfg = 1.7652$.
}
\label{fig:j1145}
\end{figure*}

{\it J1145+0322}---
The ions show absorption spanning a total of $\Delta v\approx300\;{\rm km\;s}^{-1}$ (Figure~
\ref{fig:j1145}). We present the ionic column densities in Table~\ref{tab:J1145+0322_colm}. 
Absorption from low ions is strong for C$^+$, Si$^+$, Al$^+$ and Mg$^+$ and 
moderate for Fe$^+$ and Mg$^0$. There is also strong absorption from high ions Si$^{3+}$ and 
C$^{3+}$. Together the data suggests a partially ionized gas characteristic of Lyman limit 
systems. 

Figure~\ref{fig:j1145} presents the Magellan/MagE spectrum at Ly$\alpha$. Given the relatively low 
S/N of these data, we also included a lower resolution Keck/LRIS spectrum in the profile fits.
The dominant absorber is asymmetric, suggesting blending with a weaker, unresolved component. The 
absorption at $v\approx+500\;{\rm km\;s}^{-1}$ is not associated with any metal ion detection. 
Assuming a single component with a $b$ value ranging from $15\text{--}60\;{\rm km\;s}^{-1}$, the 
large Ly$\alpha$ equivalent width and the lack of strong damping wings together restrict the range 
of $N_{\rm HI}$ to be $10^{18.0}\text{--}10^{18.6}\;{\rm cm}^{-2}$. Our best fit solution gives a 
total $N_{\rm HI}=10^{18.4\pm0.4}\;{\rm cm}^{-2}$, where the errors are dominated by the lack of 
higher Lyman series lines.

The high Si$^+$/Si$^{3+}$ ratio suggests a lower ionization state with $\log U \approx -3$. On the 
other hand the Al$^+$/Al$^{3+}$ and highly saturated \ion{C}{4} doublet implies higher $U$ values. 
These conflicting constraints suggest the profile is a blend of material with varying ionization 
state, although there is no obvious evidence for such a blend in the line profiles. We proceeded 
by adopting $\log U=-2.9$ with a larger uncertainty towards higher values. 

\begin{figure*}
\includegraphics[width=7in]{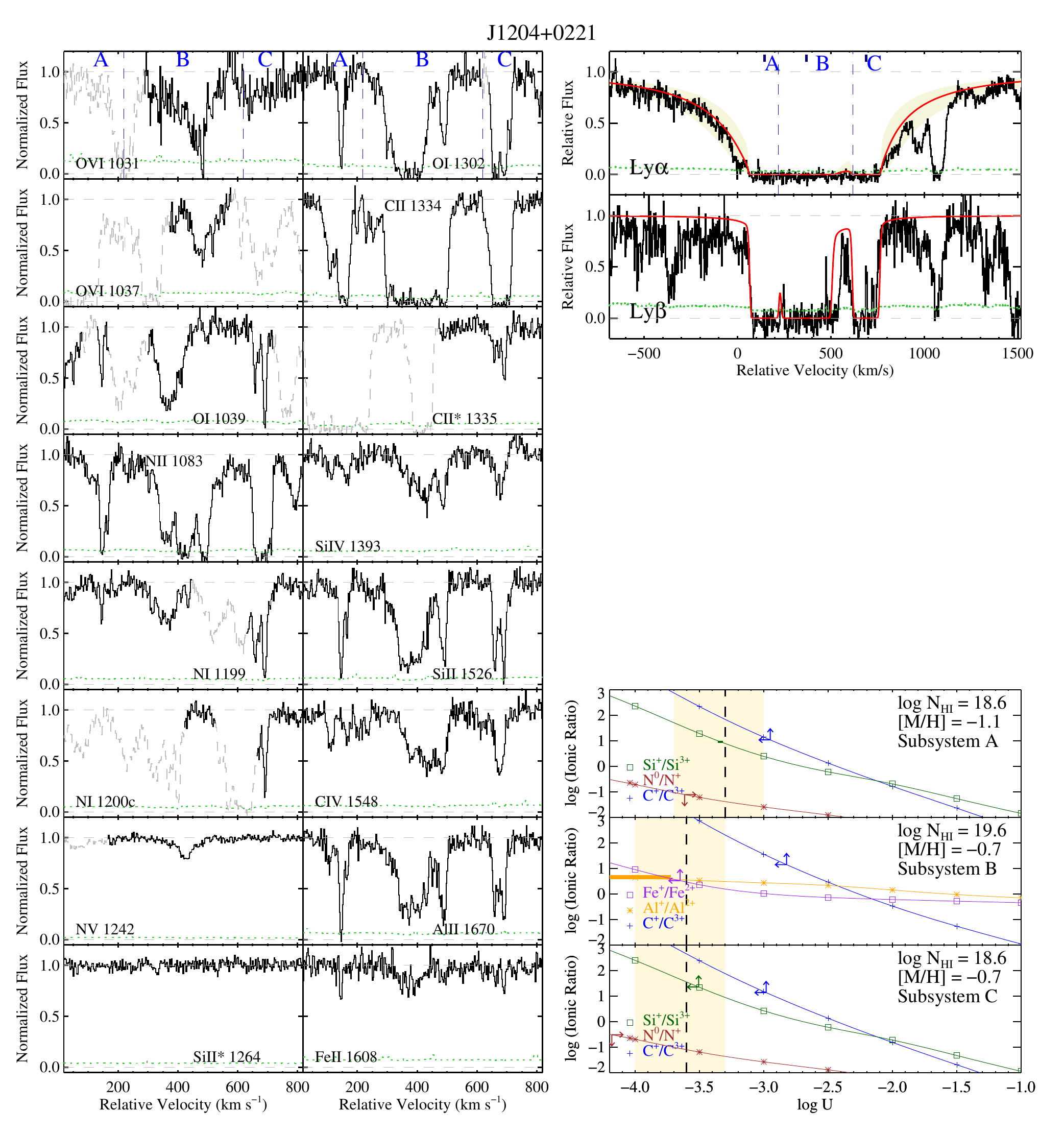} 
\caption{Similar to Figure~\ref{fig:j0225} but for J1204+0221 at $\mzfg = 2.4358$. 
}
\label{fig:j1204}
\end{figure*}

{\it J1204+0221}---
As reported previously in QPQ3, the ions show absorption spanning a total of 
$\Delta v\approx760\;{\rm km\;s}^{-1}$ (Figure~\ref{fig:j1204}, or see Figure~3 of QPQ3) that we 
separate into three subsystems. We present ionic column densities in 
Table~\ref{tab:J1204+0221_colm}. We refer the reader to QPQ3 for details on the \ion{H}{1} and 
photoionization modeling. 
Summarizing the previous findings, there is 
absorption from a series of low ions O$^0$, C$^+$, Si$^+$, N$^0$ , N$^+$, Al$^+$ and Fe$+$, 
characteristic of optically thick absorbers. Weak \ion{Si}{4} and \ion{C}{4} absorption indicate 
the ionization state of the gas is not extreme. The absence of strong \ion{N}{5} and \ion{O}{6} 
limits the flux of photons with energies $h\nu\gtrsim4$~Ryd and rules out a collisionally ionized 
gas with $T\approx10^5$~K. The strong \ion{N}{2} absorption traces the \ion{N}{1} profile for 
subsystems A and C, but the ionic ratio N$^+$/N$^0$ varies significantly across subsystem B.  
In QPQ3, the \ion{N}{2} column density of subsystem A is obtained by Voigt profile modeling, while 
in this study it is obtained by AODM. In our AODM treatment, if a line is resolved, as in the case 
for this echelle sightline, we consider a component saturated if the absorption trough minimum is 
below 0.5 times the 1-$\sigma$ error. Hence, while a good measurement for the \ion{N}{2} column 
density is reported in QPQ3, a lower limit is reported in QPQ8 for consistency across the whole 
sample. 

We present the Ly$\alpha$ and Ly$\beta$ profiles of this absorption system and reproduced the 
results shown in Figure~2 of QPQ3. For subsystems A and C, the absence of strong Ly$\alpha$ 
damping wings restricts $N_{\rm HI}^{\rm A,C}<10^{19}\;{\rm cm}^{-2}$, while the Ly$\beta$ profile 
demands $b_{\rm A,C}<25\;{\rm km\;s^{-1}}$. Our best estimates give 
$N_{\rm HI}^{\rm A,C}=10^{18.6\pm0.4}\;{\rm cm}^{-2}$. With the constraints on subsystems A and C, 
we estimated $N_{\rm HI}^{\rm B}=10^{19.6\pm0.2}\;{\rm cm}^{-2}$ that is insensitive to the 
$b_{\rm B}$ value. 

In QPQ3, we adopted $\log U\lesssim-3$ for all three subsystems. We have revised our estimate in 
this study. 
This absorption system shows a varying $U$ parameter. For subsystems A and C, the observed 
Fe$^+$/Fe$^{2+}$ ratio requires a lower $U$ parameter than other ionic ratios. Nonetheless the 
Si$^+$/Si$^{3+}$, Al$^+$/Al$^{2+}$, N$^0$/N$^+$ and C$^+$/C$^{3+}$ ratios are roughly consistent 
with a single $U$ value for each of subsystems A and C. For subsystem B, due to the high 
$N_{\rm HI}$ column the Cloudy predicted ionic ratios are rather insensitive to $U$. Other than 
the Si$^+$/Si$^{3+}$ ratio, other ionic ratios are roughly consistent with a single $U$. There is 
compelling evidence that the foreground quasar is not shining on the gas for any reasonable gas 
density. The observed ionic ratios are also consistent with the EUVB being dominant if 
$n_{\rm H}\lesssim10^{-3}\;{\rm cm}^{-3}$. 
Detailed component-by-component fitting of the echelle 
spectrum reveals that the low and high ions do not have the same velocity structure. The 
low-to-high ion ratios, in particular C$^+$/C$^{3+}$, should therefore be considered lower limits 
for the low ionization phase.
We adopted $\log U_{\rm A}=-3.3$, $\log U_{\rm B}=-3.6$ and $\log U_{\rm C}=-3.6$. 

\begin{figure*}
\includegraphics[width=7in]{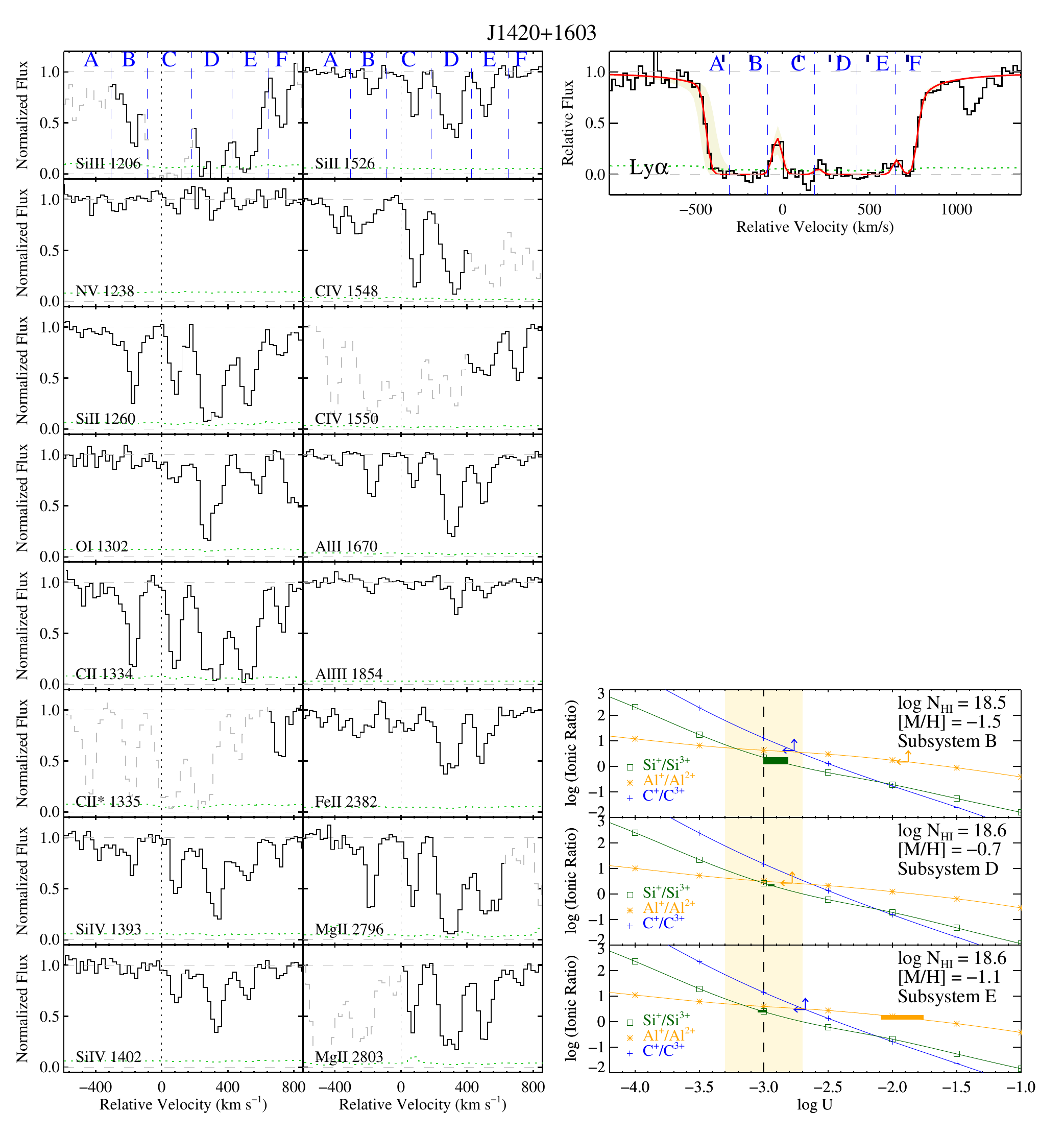} 
\caption{Similar to Figure~\ref{fig:j0225} but for J1420+1603 at $\mzfg = 2.0197$.
}
\label{fig:j1420}
\end{figure*}

{\it J1420+1603}---
The ions show complex absorption in a series of components spanning 
$\Delta v\approx1350\;{\rm km\;s}^{-1}$ (Figure~\ref{fig:j1420}) that we separate into six 
subsystems. We present ionic column densities in Table~\ref{tab:J1420+1603_colm}. 
There is strong absorption throughout the interval from low-ions, e.g. O$^0$, Si$^+$, C$^+$, 
Al$^+$, Fe$^+$,Mg$^+$ and Mg$^0$. The absorption consistent with \ion{C}{2*}~1335 from 
subsystem F cannot be associated to the \ion{C}{2}~1334 transition at another velocity. 
Together with modest absorption from high ions Si$^{3+}$ and C$^{3+}$, the data suggests a 
partially ionized gas. The spectral resolution of FWHM $\approx 51\;{\rm km\;s}^{-1}$ implies 
\ion{C}{2}~1334 and \ion{C}{4}~1548 are heavily saturated in subsystems D and E. This system is one 
of the few cases where the \ion{N}{5} doublet lies redward of the background quasar's Ly$\alpha$ 
forest. 
We report no positive detections in any of the subsystems. 

Our data covers only the Ly$\alpha$ transition of this complex absorption system. The high S/N 
spectrum exhibits no evidence for strong damping wings which constrains the total 
$N_{\rm HI}<10^{19}\;{\rm cm}^{-2}$ assuming the majority of \ion{H}{1} gas traces the low ion 
metal absorption. Tighter limits may be placed on the subsystems at the ends of the interval, i.e. 
for subsystems A and F, $N_{\rm HI}^{\rm A,F}<10^{18.5}\;{\rm cm}^{-2}$. The large \lya\ equivalent 
widths for subsystems D and E imply $N_{\rm HI}>10^{18}\;{\rm cm}^{-2}$ for 
$b_{\rm D,E}<60\;{\rm km\;s}^{-1}$ provided line blending is not severe. We took the best 
estimates from ALIS and adopted a 0.4 dex uncertainty.


With the exception of the weakly absorbing subsystem A, we found the gas throughout the system is 
consistent with a single ionization parameter of $\log U \approx -3$. Subsystems B, C, D, E and F 
have a series of observed ionic ratios that together impose tight and consistent constraints on 
$U$. The relatively low $U$ value reflects the strong low ion absorption observed throughout the 
complex. The $U$ parameter for subsystem A is not as well constrained, but a higher value is 
preferred. 

\begin{figure*}
\includegraphics[width=7in]{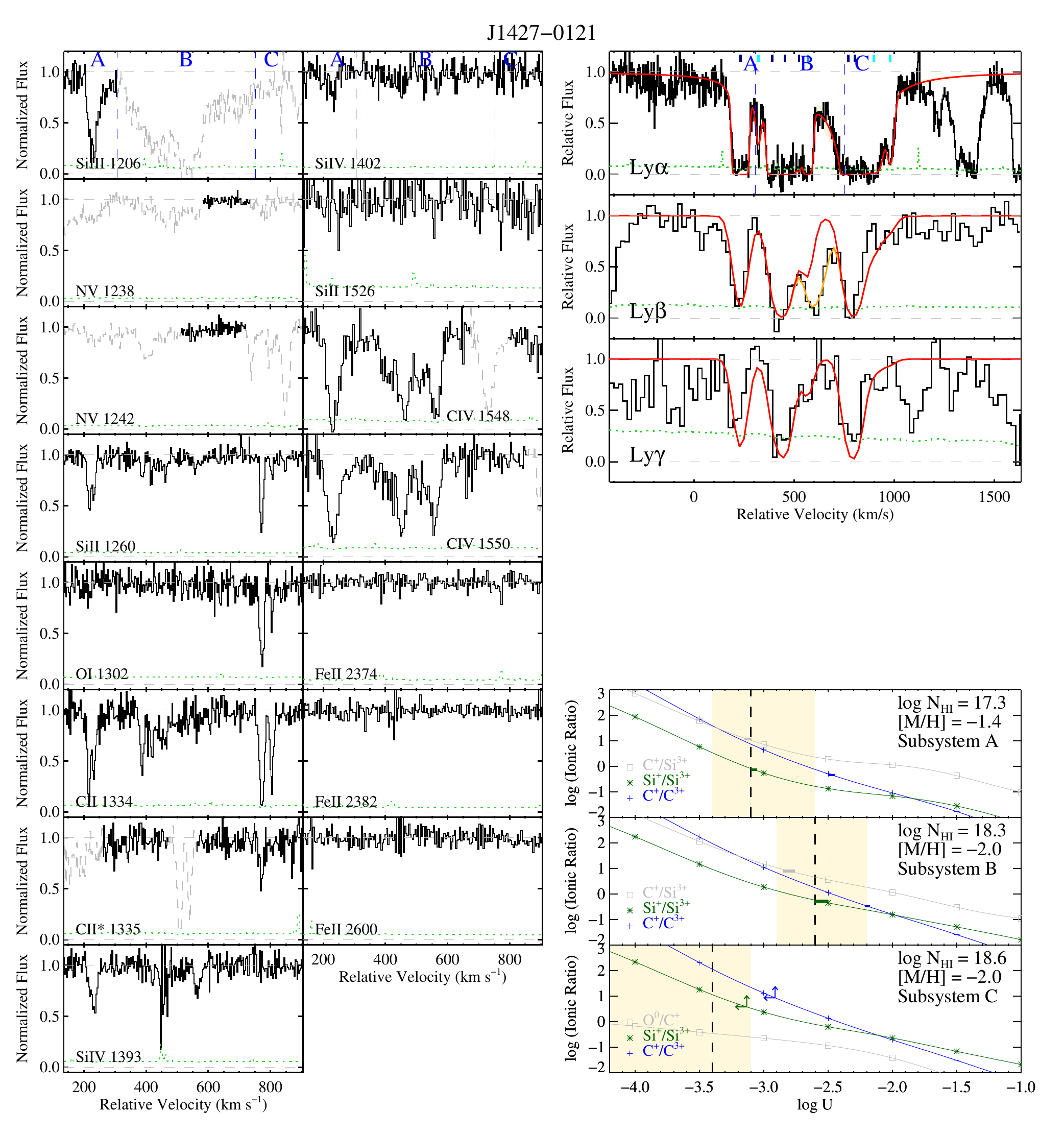} 
\caption{Similar to Figure~\ref{fig:j0225} but for J1427-0121 at $\mzfg = 2.2736$.
}
\label{fig:j1427}
\end{figure*}

{\it J1427-0121}---
The ions show absorption in roughly three distinct velocity intervals spanning a total of 
$\Delta v\approx670\;{\rm km\;s}^{-1}$ (Figure~\ref{fig:j1427}) that we divide into three 
subsystems. We present the ionic column densities in Table~\ref{tab:J1427-0121_colm}. 
Low ions are detected throughout the complex, including moderate absorption from \ion{O}{1}~1302 in 
subsystem C. In addition, weak \ion{C}{2*}~1335 absorption is detected in one of the components, 
implying a relatively dense gas. High ion absorption from C$^{3+}$ and Si$^{3+}$ is detected in 
subsystems A and B and remarkably absent in subsystem C. The \ion{C}{2}, \ion{Si}{2} and 
\ion{Si}{4} profiles are similar throughout the complex, whereas the \ion{C}{4} profile differs 
from the low ions in subsystem B. This suggests a contribution to \ion{C}{4} absorption from a 
different phase along the sightline. 

We present the Ly$\alpha$, Ly$\beta$, and Ly$\gamma$ transitions of this absorption system 
measured using Magellan/MIKE and Magellan/MagE. These data offer a terrific puzzle:  the \lya\ 
profile of subsystem A, and to a lesser extent subsystem B, is unsaturated despite the presence of 
strong low ion absorption and strong \lyb\ and \lyg\ absorption in the lower resolution MagE data. 
In turn, the \lya\ profile would require $N_{\rm HI}<10^{14.5}\;{\rm cm}^{-2}$ while the higher
order lines demand much larger $N_{\rm HI}$ values. We identify three scenarios that could 
resolve this apparent conundrum:
(i) the gas is optically thin at \lya, represents the first such case reported with corresponding 
\ion{C}{2} and \ion{Si}{2} absorption, and the absorption at \lyb\ and \lyg\ is unrelated IGM 
absorption;
(ii) we have performed poor sky subtraction at these wavelengths; 
(iii) there is unresolved \lya\ emission that is `filling in' the \lya\ absorption \citep[e.g. 
from fluorescence of the quasar ionizing flux;][]{QPQ4,Cantalupo+12,HennawiProchaska+09,Finley+13,
Cai+14}. Of these three, the second is the least extraordinary. We have carefully inspected the 
data reduction process for this spectrum and cannot identify any error and further note that other 
lines close by in wavelength, e.g. subsystem C, exhibits complete absorption. We consider the 
first option to the be the most improbable and therefore proceeded by fitting the Lyman series 
lines with a revised zero level that gives complete absorption at \lya. Presently, we interpret 
the non-zero flux as unresolved \lya\ emission along the sightline.
We estimated a total $N_{\rm HI}^{\rm A}=10^{17.3_{-1.0}^{+0.5}}\;{\rm cm}^{-2}$, a total 
$N_{\rm HI}^{\rm B}=10^{18.3_{-1.0}^{+0.2}}\;{\rm cm}^{-2}$ and a total
$N_{\rm HI}^{\rm C}=10^{18.6_{-1.0}^{+0.2}}\;{\rm cm}^{-2}$, where the errors are dominated by 
blending of the absorption profiles and degeneracy between $N_{\rm HI}$ and $b$ values.  

Because the \ion{C}{4} profile does not closely track other ions, including \ion{Si}{4}, we give 
stronger weight to the Si$^+$/Si$^{3+}$ ratio for constraining the $U$ parameter. This approach is 
further supported by the observed C$^+$/Si$^{3+}$ ratio. The only significant ionization gradient 
within a subsystem is in \ion{C}{4}, hence we consider the \ion{C}{4} column densities measured as 
upper limits when we perform ionization modeling.
The gas in subsystems A and B, 
which exhibit high ions, is well modeled by $\log U =-3.1$ and $-2.6$ respectively. 
The absence of high ion absorption in subsystem~C together with stronger low ion absorption 
requires a much lower $U$ value. We adopted $\log U < -3.4$.
Subsystem C shows \ion{C}{2}* which allows us to constrain its electron density 
$n_e=16\;{\rm cm^{-3}}$, as discussed in Section~\ref{sec:volumedensity}. This $n_e$ implies the 
gas receives an ionizing flux that is $\approx0.25$ of that expected should the quasar shine 
isotropically. Should the gas be illuminated only by the EUVB, the $n_e$ would imply 
$\log\;U=-7.1$. 

\begin{figure*}
\includegraphics[width=7in]{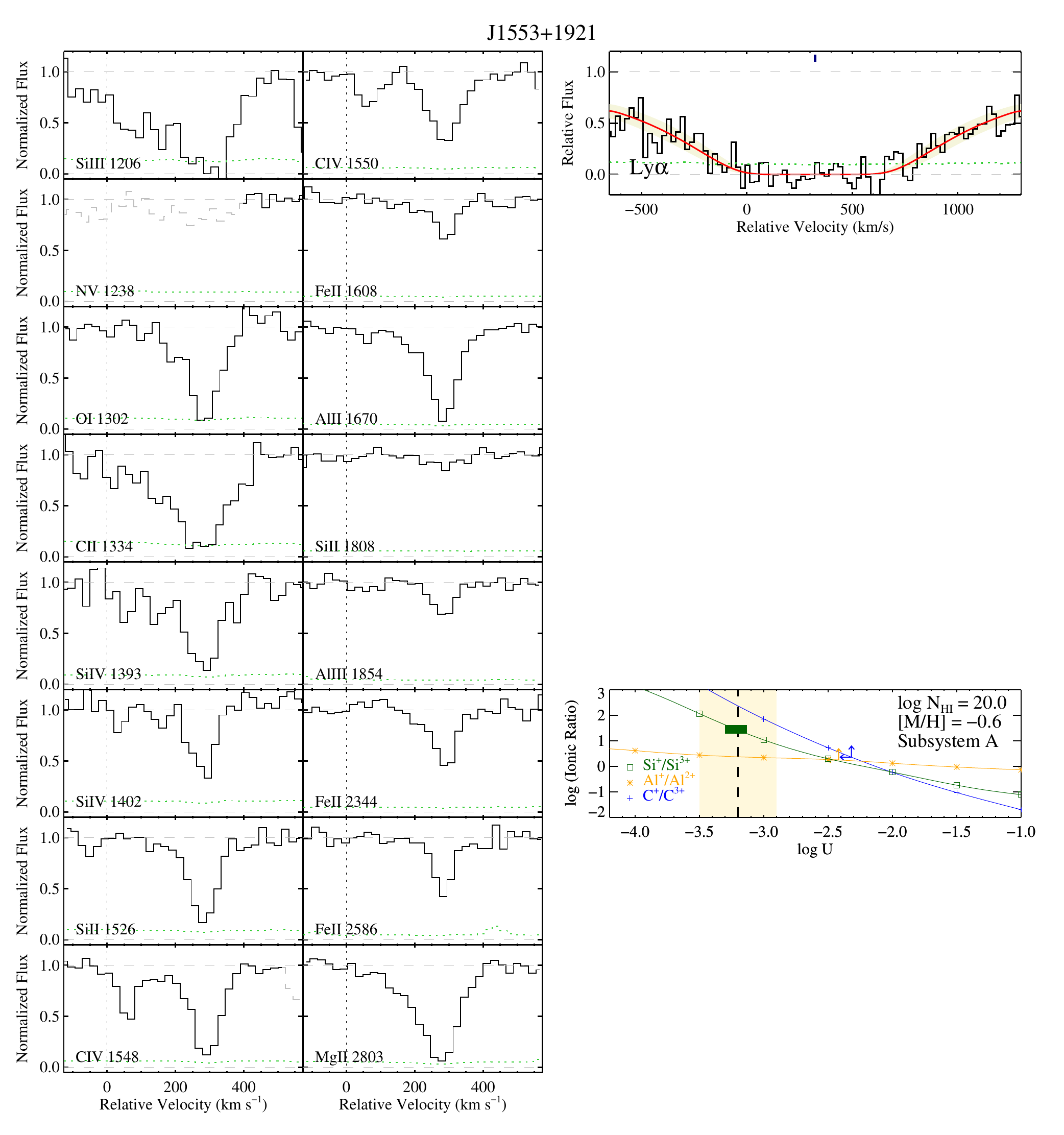} 
\caption{Similar to Figure~\ref{fig:j0225} but for J1553+1921 at $\mzfg = 2.0098$.
}
\label{fig:j1553}
\end{figure*}

{\it J1553+1921}---
The ions show absorption spanning a total of $\Delta v\approx500\;{\rm km\;s}^{-1}$, with the 
majority of low ion absorption confined to $\approx 100\;\mkms$ (Figure~\ref{fig:j1553}). We 
present the ionic column densities in Table~\ref{tab:J1553+1921_colm}. This is 
a damped \lya\ system and we observe strong absorption from low ion transitions of O$^0$, Si$^+$, 
C$^+$, Fe$^+$, Al$^+$, Mg$^+$ and Mg$^0$. Interestingly, we find corresponding high ion absorption 
at the same velocities. 

The Ly$\alpha$ profile shows strong damping wings that constrain $N_{\rm HI}$ to be 
$10^{20.2\pm0.1}\;{\rm cm}^{-2}$, insensitive to the Doppler $b$.  The line centroid of \lya\ is 
consistent with the peak optical depth of the low ion transitions.

The observed Si$^+$/Si$^{3+}$ ratio suggests $\log U \approx -3.2$, if these ions trace the same 
phase of gas. In damped Lyman alpha systems this is rarely the case 
\citep[e.g.][]{WolfeProchaska00b}, but we do find close kinematic alignment between the 
\ion{Si}{2} and \ion{Si}{4} transitions, 
and the absorption profiles are all narrow. This $\log U$ value is also consistent with the upper 
limits placed by the observed C$^+$/C$^{3+}$ and Al$^+$/Al$^{2+}$ ratios. Since the absorption at 
\ion{C}{2}~1334 is strongly saturated, the upper limit given by C$^+$/C$^{3+}$ is a generous limit 
even if some of the C$^{3+}$ comes from another phase of gas. 
We adopted $\log U=-3.2$ and allow for much lower values. At this $U$ value, the neutral fraction 
is approximately 50\%. 

\begin{figure*}
\includegraphics[width=7in]{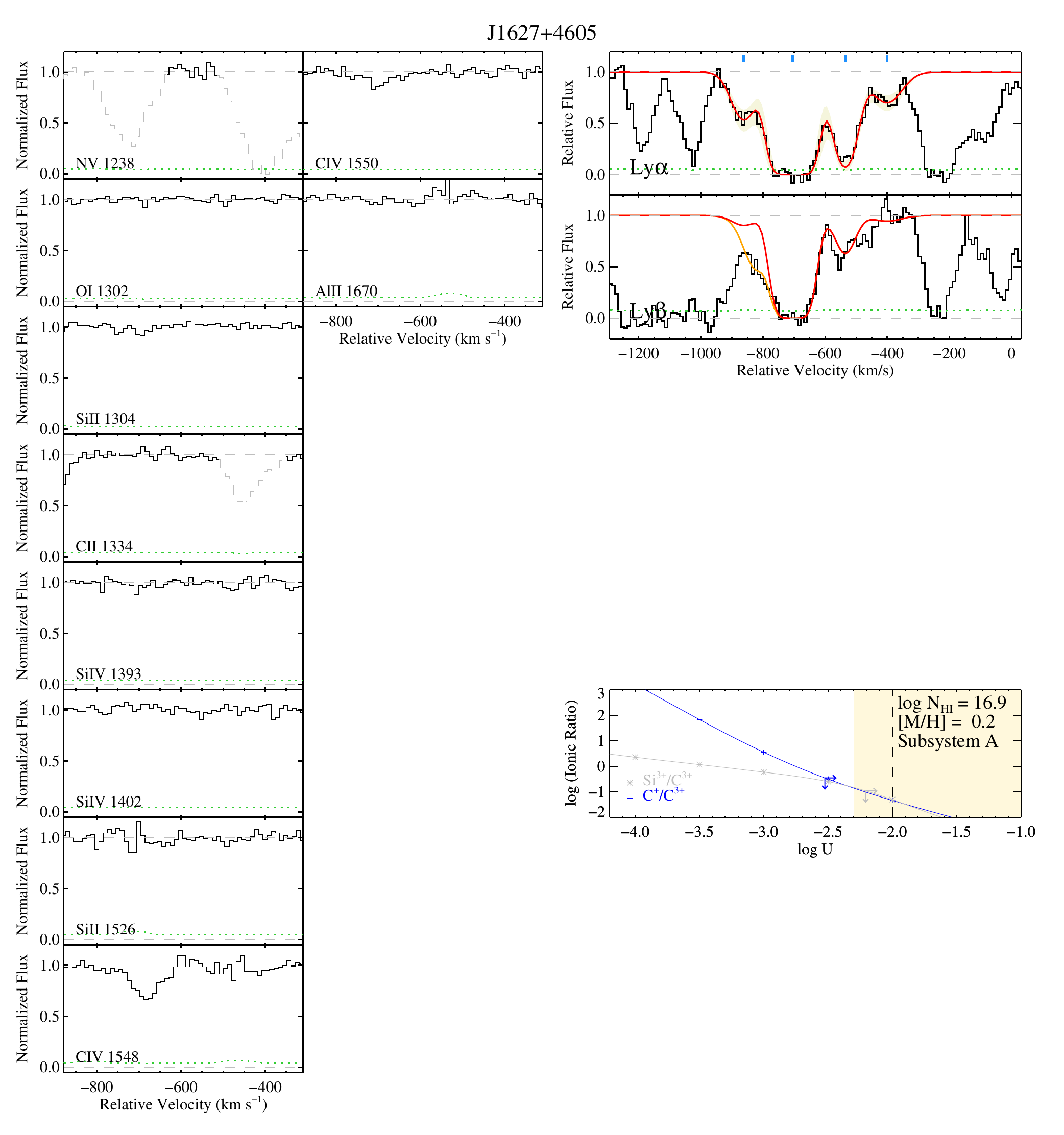} 
\caption{Similar to Figure~\ref{fig:j0225} but for J1627+4605 at $\mzfg = 3.8137$.
}
\label{fig:j1627}
\end{figure*}

{\it J1627+4605}---
Only moderate absorption from C$^{3+}$ is detected in the system we associated to J1627+4605FG. 
This gas spans a velocity interval of $\Delta v\approx150\;{\rm km\;s}^{-1}$ 
(Figure~\ref{fig:j1627}). We present the ionic column densities in 
Table~\ref{tab:J1627+4605_colm}. Together with an absence of any low ions, the data suggests a 
highly ionized gas. 

We present the Ly$\alpha$ and Ly$\beta$ velocity profiles of this system. Higher order lines are 
compromised by a Lyman limit system at lower redshift. Restricting to $b$-values ranging from 
$15\text{--}60\;{\rm km\;s}^{-1}$, the Ly$\alpha$ equivalent width and the lack of Ly$\alpha$ 
damping wings of the dominant absorber requires 
$10^{15.1}\;{\rm cm}^{-2}<N_{\rm HI}<10^{18.0}\;{\rm cm}^{-2}$. There are three additional weak 
absorbers included in our model. 
We recovered $\mnhi = 10^{16.9} \cm{-2}$ as the best fitted value and adopted a large uncertainty 
of 0.8~dex owing to line saturation.

C$^{3+}$ is the only ion detected and our ionization constraints include limits to 
C$^+$/C$^{3+}$ and Si$^{3+}$/C$^{3+}$. These require $\log U$ greater than $-2.5$ and $-2.2$ 
respectively. We adopted $\log U = -2.0$ with a large uncertainty to higher values.

\clearpage
\LongTables
\begin{deluxetable}{lcccccccc}
\tablewidth{0pc}
\tablecaption{Ionic Column Densities for J0225+0048FG \label{tab:J0225+0048_colm}}
\tabletypesize{\scriptsize}
\setlength{\tabcolsep}{0.03in}
\tablehead{\colhead{Ion}
& \colhead{$\lambda_{\rm rest}$}
& \colhead{$v_{\rm int}^a$} 
& \colhead{$\log N_{\rm AODM}$}
& \colhead{Rest EW} \\
& (\AA) & (${\rm km~s^{-1}}$) & & (\AA) }
\startdata
\cutinhead{A}
\ion{C}{2}&&&$< 13.66$\\
&1334.5323 &$[ -363,   88]$&$< 13.66$&$ 0.000 \pm  0.030$&\\
\ion{C}{2}*&&&$< 13.70$\\
&1335.7077 &$[ -363,   88]$&$< 13.70$&$ 0.005 \pm  0.030$&\\
\ion{C}{4}&&&$  14.07 \pm 0.03$\\
&1548.1950 &$[ -363,   88]$&$  14.07 \pm 0.03$&$ 0.419 \pm  0.025$&\\
\ion{Si}{2}&&&$< 12.55$\\
&1260.4221 &$[ -363,   88]$&$< 12.55$&$ 0.028 \pm  0.016$&\\
\enddata
\tablenotetext{a}{Velocity interval for the AODM
column density measurement.
Velocities are relative to the redshift of the foreground quasar J0225+0048FG, $z_{\rm fg}=2.7265$.}
\end{deluxetable}
\LongTables
\begin{deluxetable}{lcccccccc}
\tablewidth{0pc}
\tablecaption{Ionic Column Densities for J0341+0000FG \label{tab:J0341+0000_colm}}
\tabletypesize{\scriptsize}
\setlength{\tabcolsep}{0.03in}
\tablehead{\colhead{Ion}
& \colhead{$\lambda_{\rm rest}$}
& \colhead{$v_{\rm int}^a$} 
& \colhead{$\log N_{\rm AODM}$}
& \colhead{Rest EW} \\
& (\AA) & (${\rm km~s^{-1}}$) & & (\AA) }
\startdata
\cutinhead{A}
\ion{C}{2}&&&$< 14.00$\\
&1334.5323 &$[  120,  470]$&$< 14.00$&$ 0.146 \pm  0.028$&\\
\ion{C}{2}*&&&$< 13.69$\\
&1335.7077 &$[  120,  470]$&$< 13.69$&$ 0.006 \pm  0.029$&\\
\ion{C}{4}&&&$< 13.28$\\
&1548.1950 &$[  120,  470]$&$< 13.28$&$-0.003 \pm  0.026$&\\
&1550.7700 &$[  120,  470]$&$< 13.59$&$ 0.029 \pm  0.026$&\\
\ion{N}{5}&&&$< 13.54$\\
&1238.8210 &$[  120,  470]$&$< 13.54$&$ 0.038 \pm  0.024$&\\
\ion{Al}{2}&&&$< 12.18$\\
&1670.7874 &$[  120,  470]$&$< 12.18$&$ 0.031 \pm  0.023$&\\
\ion{Si}{2}&&&$< 13.42$\\
&1526.7066 &$[  120,  470]$&$< 13.42$&$-0.033 \pm  0.023$&\\
\ion{Fe}{2}&&&$< 13.62$\\
&1608.4511 &$[  120,  470]$&$< 13.62$&$ 0.035 \pm  0.018$&\\
\enddata
\tablenotetext{a}{Velocity interval for the AODM
column density measurement.
Velocities are relative to the redshift of the foreground quasar J0341+0000FG, $z_{\rm fg}=2.1233$.}
\end{deluxetable}
\LongTables
\begin{deluxetable}{lcccccccc}
\tablewidth{0pc}
\tablecaption{Ionic Column Densities for J0409-0411FG \label{tab:J0409-0411_colm}}
\tabletypesize{\scriptsize}
\setlength{\tabcolsep}{0.03in}
\tablehead{\colhead{Ion}
& \colhead{$\lambda_{\rm rest}$}
& \colhead{$v_{\rm int}^a$} 
& \colhead{$\log N_{\rm AODM}$}
& \colhead{Rest EW} \\
& (\AA) & (${\rm km~s^{-1}}$) & & (\AA) }
\startdata
\cutinhead{A}
\ion{C}{2}&&&$< 13.97$\\
&1334.5323 &$[-1458,-1238]$&$< 13.97$&$ 0.156 \pm  0.038$&\\
\ion{C}{4}&&&$< 13.31$\\
&1548.1950 &$[-1458,-1238]$&$< 13.31$&$-0.033 \pm  0.028$&\\
&1550.7700 &$[-1458,-1238]$&$< 13.64$&$ 0.024 \pm  0.029$&\\
\ion{Al}{2}&&&$< 12.26$\\
&1670.7874 &$[-1458,-1238]$&$< 12.26$&$-0.050 \pm  0.029$&\\
\ion{Si}{2}&&&$< 13.48$\\
&1526.7066 &$[-1458,-1238]$&$< 13.48$&$-0.005 \pm  0.026$&\\
\ion{Fe}{2}&&&$< 14.12$\\
&1608.4511 &$[-1458,-1238]$&$< 14.12$&$ 0.128 \pm  0.027$&\\
\enddata
\tablenotetext{a}{Velocity interval for the AODM
column density measurement.
Velocities are relative to the redshift of the foreground quasar J0409-0411FG, $z_{\rm fg}=1.7155$.}
\end{deluxetable}
\LongTables
\begin{deluxetable}{lcccccccc}
\tablewidth{0pc}
\tablecaption{Ionic Column Densities for J0853-0011FG \label{tab:J0853-0011_colm}}
\tabletypesize{\scriptsize}
\setlength{\tabcolsep}{0.03in}
\tablehead{\colhead{Ion}
& \colhead{$\lambda_{\rm rest}$}
& \colhead{$v_{\rm int}^a$} 
& \colhead{$\log N_{\rm AODM}$}
& \colhead{Rest EW} \\
& (\AA) & (${\rm km~s^{-1}}$) & & (\AA) }
\startdata
\cutinhead{A}
\ion{C}{2}&&&$  13.89 \pm 0.03$\\
&1334.5323 &$[ -271,  -51]$&$  13.89 \pm 0.03$&$ 0.139 \pm  0.010$&\\
\ion{C}{4}&&&$  13.45 \pm 0.04$\\
&1548.1950 &$[ -271,  -51]$&$  13.49 \pm 0.04$&$ 0.118 \pm  0.010$&\\
&1550.7700 &$[ -271,  -51]$&$  13.26 \pm 0.13$&$ 0.035 \pm  0.010$&\\
\ion{N}{5}&&&$< 13.90$\\
&1242.8040 &$[ -271,  -51]$&$< 13.90$&$ 0.080 \pm  0.010$&\\
\ion{O}{1}&&&$< 13.36$\\
\enddata
\tablenotetext{a}{Velocity interval for the AODM
column density measurement.
Velocities are relative to the redshift of the foreground quasar J0853-0011FG, $z_{\rm fg}=2.4014$.}
\end{deluxetable}
\LongTables
\begin{deluxetable}{lcccccccc}
\tablewidth{0pc}
\tablecaption{Ionic Column Densities for J0932+0925FG \label{tab:J0932+0925_colm}}
\tabletypesize{\scriptsize}
\setlength{\tabcolsep}{0.03in}
\tablehead{\colhead{Ion}
& \colhead{$\lambda_{\rm rest}$}
& \colhead{$v_{\rm int}^a$} 
& \colhead{$\log N_{\rm AODM}$}
& \colhead{Rest EW} \\
& (\AA) & (${\rm km~s^{-1}}$) & & (\AA) }
\startdata
\cutinhead{A}
\ion{C}{2}&&&$  13.72 \pm 0.11$\\
&1334.5323 &$[ -237,  316]$&$  13.72 \pm 0.11$&$ 0.100 \pm  0.025$&\\
\ion{C}{2}*&&&$< 13.62$\\
&1335.7077 &$[ -190,  316]$&$< 13.62$&$ 0.069 \pm  0.025$&\\
\ion{C}{4}&&&$  13.62 \pm 0.06$\\
&1548.1950 &$[ -237,  316]$&$  13.62 \pm 0.06$&$ 0.159 \pm  0.023$&\\
&1550.7700 &$[ -237,  316]$&$< 14.83$&$ 0.786 \pm  0.022$&\\
\ion{N}{5}&&&$< 15.36$\\
\enddata
\tablenotetext{a}{Velocity interval for the AODM
column density measurement.
Velocities are relative to the redshift of the foreground quasar J0932+0925FG, $z_{\rm fg}=2.4170$.}
\end{deluxetable}
\LongTables
\begin{deluxetable}{lcccccccc}
\tablewidth{0pc}
\tablecaption{Ionic Column Densities for J1026+4614FG \label{tab:J1026+4614_colm}}
\tabletypesize{\scriptsize}
\setlength{\tabcolsep}{0.03in}
\tablehead{\colhead{Ion}
& \colhead{$\lambda_{\rm rest}$}
& \colhead{$v_{\rm int}^a$} 
& \colhead{$\log N_{\rm AODM}$}
& \colhead{Rest EW} \\
& (\AA) & (${\rm km~s^{-1}}$) & & (\AA) }
\startdata
\cutinhead{A}
\ion{C}{2}&&&$< 13.02$\\
&1334.5323 &$[  240,  360]$&$< 13.02$&$ 0.008 \pm  0.007$&\\
\ion{C}{2}*&&&$< 13.06$\\
&1335.7077 &$[  240,  360]$&$< 13.06$&$-0.016 \pm  0.007$&\\
\ion{C}{4}&&&$  13.98 \pm 0.03$\\
&1548.1950 &$[  240,  360]$&$> 13.76$&$ 0.182 \pm  0.006$&\\
&1550.7700 &$[  240,  360]$&$  13.98 \pm 0.03$&$ 0.156 \pm  0.006$&\\
\ion{N}{2}&&&$< 13.32$\\
\enddata
\tablenotetext{a}{Velocity interval for the AODM
column density measurement.
Velocities are relative to the redshift of the foreground quasar J1026+4614FG, $z_{\rm fg}=3.3401$.}
\end{deluxetable}
\LongTables
\begin{deluxetable}{lcccccccc}
\tablewidth{0pc}
\tablecaption{Ionic Column Densities for J1038+5027FG \label{tab:J1038+5027_colm}}
\tabletypesize{\scriptsize}
\setlength{\tabcolsep}{0.03in}
\tablehead{\colhead{Ion}
& \colhead{$\lambda_{\rm rest}$}
& \colhead{$v_{\rm int}^a$} 
& \colhead{$\log N_{\rm AODM}$}
& \colhead{Rest EW} \\
& (\AA) & (${\rm km~s^{-1}}$) & & (\AA) }
\startdata
\cutinhead{A}
\ion{C}{2}&&&$< 13.38$\\
&1334.5323 &$[  466,  676]$&$< 13.38$&$-0.011 \pm  0.016$&\\
\ion{C}{2}*&&&$< 13.35$\\
&1335.7077 &$[  536,  676]$&$< 13.35$&$ 0.016 \pm  0.013$&\\
\ion{C}{4}&&&$  14.08 \pm 0.03$\\
&1548.1950 &$[  466,  676]$&$  14.08 \pm 0.03$&$ 0.364 \pm  0.019$&\\
&1550.7700 &$[  466,  676]$&$  14.09 \pm 0.04$&$ 0.211 \pm  0.019$&\\
\ion{Si}{2}&&&$< 13.39$\\
&1526.7066 &$[  466,  676]$&$< 13.39$&$ 0.029 \pm  0.021$&\\
&1808.0130 &$[  466,  676]$&$< 15.17$&$ 0.044 \pm  0.029$&\\
\ion{Si}{4}&&&$  13.00 \pm 0.10$\\
&1393.7550 &$[  466,  676]$&$  13.00 \pm 0.10$&$ 0.079 \pm  0.019$&\\
&1402.7700 &$[  466,  676]$&$< 13.13$&$-0.023 \pm  0.021$&\\
\enddata
\tablenotetext{a}{Velocity interval for the AODM
column density measurement.
Velocities are relative to the redshift of the foreground quasar J1038+5027FG, $z_{\rm fg}=3.1323$.}
\end{deluxetable}
\LongTables
\begin{deluxetable}{lcccccccc}
\tablewidth{0pc}
\tablecaption{Ionic Column Densities for J1144+0959FG \label{tab:J1144+0959_colm}}
\tabletypesize{\scriptsize}
\setlength{\tabcolsep}{0.03in}
\tablehead{\colhead{Ion}
& \colhead{$\lambda_{\rm rest}$}
& \colhead{$v_{\rm int}^a$} 
& \colhead{$\log N_{\rm AODM}$}
& \colhead{Rest EW} \\
& (\AA) & (${\rm km~s^{-1}}$) & & (\AA) }
\startdata
\cutinhead{A}
\ion{C}{2}&&&$< 12.72$\\
&1334.5323 &$[-1690,-1605]$&$< 12.72$&$-0.004 \pm  0.004$&\\
\ion{C}{3}&&&$> 13.66$\\
& 977.0200 &$[-1690,-1605]$&$> 13.66$&$ 0.157 \pm  0.005$&\\
\ion{C}{4}&&&$  13.98 \pm 0.03$\\
&1548.1950 &$[-1690,-1605]$&$  13.92 \pm 0.03$&$ 0.196 \pm  0.002$&\\
&1550.7700 &$[-1690,-1605]$&$  14.09 \pm 0.03$&$ 0.169 \pm  0.003$&\\
\ion{Si}{2}&&&$< 12.53$\\
\enddata
\tablenotetext{a}{Velocity interval for the AODM
column density measurement.
Velocities are relative to the redshift of the foreground quasar J1144+0959FG, $z_{\rm fg}=2.9731$.}
\end{deluxetable}
\LongTables
\begin{deluxetable}{lcccccccc}
\tablewidth{0pc}
\tablecaption{Ionic Column Densities for J1145+0322FG \label{tab:J1145+0322_colm}}
\tabletypesize{\scriptsize}
\setlength{\tabcolsep}{0.03in}
\tablehead{\colhead{Ion}
& \colhead{$\lambda_{\rm rest}$}
& \colhead{$v_{\rm int}^a$} 
& \colhead{$\log N_{\rm AODM}$}
& \colhead{Rest EW} \\
& (\AA) & (${\rm km~s^{-1}}$) & & (\AA) }
\startdata
\cutinhead{A}
\ion{C}{2}&&&$> 14.71$\\
&1334.5323 &$[ -278,  122]$&$> 14.71$&$ 0.504 \pm  0.022$&\\
\ion{C}{2}*&&&$< 13.61$\\
&1335.7077 &$[ -258,  122]$&$< 13.61$&$ 0.055 \pm  0.024$&\\
\ion{C}{4}&&&$> 14.59$\\
&1548.1950 &$[ -278,  122]$&$> 14.55$&$ 0.777 \pm  0.028$&\\
&1550.7700 &$[ -278,  122]$&$> 14.59$&$ 0.523 \pm  0.029$&\\
\ion{O}{1}&&&$  14.08 \pm 0.14$\\
\enddata
\tablenotetext{a}{Velocity interval for the AODM
column density measurement.
Velocities are relative to the redshift of the foreground quasar J1145+0322FG, $z_{\rm fg}=1.7652$.}
\end{deluxetable}
\LongTables
\begin{deluxetable}{lcccccccc}
\tablewidth{0pc}
\tablecaption{Ionic Column Densities for J1204+0221FG \label{tab:J1204+0221_colm}}
\tabletypesize{\scriptsize}
\setlength{\tabcolsep}{0.03in}
\tablehead{\colhead{Ion}
& \colhead{$\lambda_{\rm rest}$}
& \colhead{$v_{\rm int}^a$} 
& \colhead{$\log N_{\rm AODM}$}
& \colhead{Rest EW} \\
& (\AA) & (${\rm km~s^{-1}}$) & & (\AA) }
\startdata
\cutinhead{A}
\ion{C}{2}&&&$> 14.63$\\
&1036.3367 &$[   57,  207]$&$> 14.63$&$ 0.248 \pm  0.009$&\\
&1334.5323 &$[   77,  247]$&$> 14.58$&$ 0.363 \pm  0.005$&\\
\ion{C}{4}&&&$  13.60 \pm 0.03$\\
&1548.1950 &$[   77,  247]$&$  13.60 \pm 0.03$&$ 0.141 \pm  0.007$&\\
\ion{N}{1}&&&$  13.34 \pm 0.05$\\
&1199.5496 &$[  107,  207]$&$  13.33 \pm 0.05$&$ 0.032 \pm  0.004$&\\
\ion{N}{2}&&&$> 14.45$\\
\enddata
\tablenotetext{a}{Velocity interval for the AODM
column density measurement.
Velocities are relative to the redshift of the foreground quasar J1204+0221FG, $z_{\rm fg}=2.4358$.}
\end{deluxetable}
\LongTables
\begin{deluxetable}{lcccccccc}
\tablewidth{0pc}
\tablecaption{Ionic Column Densities for J1420+1603FG \label{tab:J1420+1603_colm}}
\tabletypesize{\scriptsize}
\setlength{\tabcolsep}{0.03in}
\tablehead{\colhead{Ion}
& \colhead{$\lambda_{\rm rest}$}
& \colhead{$v_{\rm int}^a$} 
& \colhead{$\log N_{\rm AODM}$}
& \colhead{Rest EW} \\
& (\AA) & (${\rm km~s^{-1}}$) & & (\AA) }
\startdata
\cutinhead{A}
\ion{C}{2}&&&$< 13.65$\\
&1334.5323 &$[ -543, -305]$&$< 13.65$&$ 0.083 \pm  0.027$&\\
\ion{C}{4}&&&$  13.63 \pm 0.04$\\
&1548.1950 &$[ -543, -305]$&$  13.63 \pm 0.04$&$ 0.155 \pm  0.014$&\\
\ion{N}{5}&&&$< 13.56$\\
&1238.8210 &$[ -543, -305]$&$< 13.56$&$ 0.008 \pm  0.026$&\\
&1242.8040 &$[ -543, -305]$&$< 13.92$&$ 0.011 \pm  0.030$&\\
\ion{O}{1}&&&$< 14.01$\\
\enddata
\tablenotetext{a}{Velocity interval for the AODM
column density measurement.
Velocities are relative to the redshift of the foreground quasar J1420+1603FG, $z_{\rm fg}=2.0197$.}
\end{deluxetable}
\LongTables
\begin{deluxetable}{lcccccccc}
\tablewidth{0pc}
\tablecaption{Ionic Column Densities for J1427-0121FG \label{tab:J1427-0121_colm}}
\tabletypesize{\scriptsize}
\setlength{\tabcolsep}{0.03in}
\tablehead{\colhead{Ion}
& \colhead{$\lambda_{\rm rest}$}
& \colhead{$v_{\rm int}^a$} 
& \colhead{$\log N_{\rm AODM}$}
& \colhead{Rest EW} \\
& (\AA) & (${\rm km~s^{-1}}$) & & (\AA) }
\startdata
\cutinhead{A}
\ion{C}{2}&&&$  13.91 \pm 0.03$\\
&1334.5323 &$[  183,  313]$&$  13.91 \pm 0.03$&$ 0.106 \pm  0.006$&\\
\ion{C}{4}&&&$  14.26 \pm 0.03$\\
&1548.1950 &$[  183,  313]$&$> 14.12$&$ 0.280 \pm  0.011$&\\
&1550.7700 &$[  183,  313]$&$  14.26 \pm 0.03$&$ 0.236 \pm  0.011$&\\
\ion{O}{1}&&&$< 13.42$\\
&1302.1685 &$[  183,  313]$&$< 13.42$&$ 0.005 \pm  0.006$&\\
\ion{Si}{2}&&&$  12.73 \pm 0.03$\\
\enddata
\tablenotetext{a}{Velocity interval for the AODM
column density measurement.
Velocities are relative to the redshift of the foreground quasar J1427-0121FG, $z_{\rm fg}=2.2736$.}
\end{deluxetable}
\LongTables
\begin{deluxetable}{lcccccccc}
\tablewidth{0pc}
\tablecaption{Ionic Column Densities for J1553+1921FG \label{tab:J1553+1921_colm}}
\tabletypesize{\scriptsize}
\setlength{\tabcolsep}{0.03in}
\tablehead{\colhead{Ion}
& \colhead{$\lambda_{\rm rest}$}
& \colhead{$v_{\rm int}^a$} 
& \colhead{$\log N_{\rm AODM}$}
& \colhead{Rest EW} \\
& (\AA) & (${\rm km~s^{-1}}$) & & (\AA) }
\startdata
\cutinhead{A}
\ion{C}{2}&&&$> 14.92$\\
&1334.5323 &$[  -62,  535]$&$> 14.92$&$ 0.950 \pm  0.067$&\\
\ion{C}{2}*&&&$< 13.93$\\
&1335.7077 &$[  166,  535]$&$< 13.93$&$-0.034 \pm  0.052$&\\
\ion{C}{4}&&&$  14.57 \pm 0.03$\\
&1548.1950 &$[  -62,  535]$&$> 14.47$&$ 0.757 \pm  0.036$&\\
&1550.7700 &$[  -62,  535]$&$  14.57 \pm 0.03$&$ 0.566 \pm  0.036$&\\
\ion{O}{1}&&&$> 15.05$\\
\enddata
\tablenotetext{a}{Velocity interval for the AODM
column density measurement.
Velocities are relative to the redshift of the foreground quasar J1553+1921FG, $z_{\rm fg}=2.0098$.}
\end{deluxetable}
\LongTables
\begin{deluxetable}{lcccccccc}
\tablewidth{0pc}
\tablecaption{Ionic Column Densities for J1627+4605FG \label{tab:J1627+4605_colm}}
\tabletypesize{\scriptsize}
\setlength{\tabcolsep}{0.03in}
\tablehead{\colhead{Ion}
& \colhead{$\lambda_{\rm rest}$}
& \colhead{$v_{\rm int}^a$} 
& \colhead{$\log N_{\rm AODM}$}
& \colhead{Rest EW} \\
& (\AA) & (${\rm km~s^{-1}}$) & & (\AA) }
\startdata
\cutinhead{A}
\ion{C}{2}&&&$< 13.17$\\
&1334.5323 &$[ -828, -508]$&$< 13.17$&$ 0.008 \pm  0.010$&\\
\ion{C}{4}&&&$  13.63 \pm 0.04$\\
&1548.1950 &$[ -828, -358]$&$  13.67 \pm 0.04$&$ 0.168 \pm  0.017$&\\
&1550.7700 &$[ -828, -358]$&$  13.47 \pm 0.11$&$ 0.056 \pm  0.015$&\\
\ion{O}{1}&&&$< 13.53$\\
&1302.1685 &$[ -828, -358]$&$< 13.53$&$-0.002 \pm  0.008$&\\
\ion{Al}{2}&&&$< 12.06$\\
&1670.7874 &$[ -828, -358]$&$< 12.06$&$-0.040 \pm  0.018$&\\
\ion{Si}{2}&&&$< 13.25$\\
&1304.3702 &$[ -828, -358]$&$< 13.25$&$-0.007 \pm  0.008$&\\
&1526.7066 &$[ -828, -358]$&$< 13.36$&$ 0.021 \pm  0.020$&\\
\ion{Si}{4}&&&$< 12.66$\\
&1393.7550 &$[ -828, -358]$&$< 12.66$&$ 0.020 \pm  0.014$&\\
&1402.7700 &$[ -828, -358]$&$< 12.96$&$-0.009 \pm  0.014$&\\
\enddata
\tablenotetext{a}{Velocity interval for the AODM
column density measurement.
Velocities are relative to the redshift of the foreground quasar J1627+4605FG, $z_{\rm fg}=3.8137$.}
\end{deluxetable}

\LongTables
\begin{deluxetable*}{llllllllll}
\tablewidth{0pc}
\tablecaption{\ion{H}{1} Column Density Measurements \label{tab:NHI}}
\tabletypesize{\scriptsize}
\setlength{\tabcolsep}{0.0in}
\tablehead{\colhead{Subsystem} & \colhead{Velocity Range} & \colhead{Adopted $\log\,N_{\rm HI}^{a}$} & \colhead{Component Centroid} & \colhead{Component $\log\,N_{\rm HI}$} & \colhead{Doppler $b$} \\
 & (${\rm km\;s^{-1}}$) & & (${\rm km\;s^{-1}}$) & & (${\rm km\;s^{-1}}$) }
\startdata
\cutinhead{J0225+0048  $\mzfg = 2.7265$  $\mnhi^{\rm total} = 17.9$}
A& $[-362,88)$&$17.5^{+1.0}_{-1.0}$\\
 & & & $ -217\pm    2$ & $17.5$ & $ 34$ \\
 & & & $  -59\pm    5$ & $14.0\pm0.2$ & $ 15$ \\
 & & & $+   33\pm    4$ & $14.1\pm0.1$ & $ 60$ \\
B& $[88,498)$&$16.9^{+1.0}_{-1.0}$\\
 & & & $+  303\pm    2$ & $16.9\pm0.2$ & $ 60$ \\
C& $[498,827)$&$17.5^{+0.8}_{-0.8}$\\
 & & & $+  536\pm    7$ & $13.4\pm0.1$ & $ 15$ \\
\enddata
\tablenotetext{a}{The adopted $N_{\rm HI}$ values match approximately the middle of lower bounds from equivalent widths and upper bounds from wings of the Lyman series absorptions. The $N_{\rm HI}$ values of the individual components of a subsystem come from ALIS voigt profile modeling outputs, which serve as guides to our adopted $N_{\rm HI}$ for a subsystem. The two need not agree and ALIS tends to give biased high values.}
\end{deluxetable*}

\end{document}